    \documentclass[twoside,twocolumn,9pt,]{article}
    
    \usepackage[T1]{fontenc}
    \usepackage[utf8]{inputenc}
    
    \usepackage{extsizes}
    \usepackage[super,sort&compress,comma]{natbib} 
    \usepackage[version=3]{mhchem}
    \usepackage[left=1.5cm, right=1.5cm, top=1.785cm, bottom=2.0cm]{geometry}
    \usepackage{balance}
    \usepackage{mathptmx}
    \usepackage{sectsty}
    \usepackage{graphicx} 
    \usepackage{lastpage}
    \usepackage[format=plain,justification=justified,singlelinecheck=false,font={stretch=1.125,sf},labelfont=bf,labelsep=space]{caption}
    \usepackage{float}
    \usepackage{fancyhdr}
    \usepackage{fnpos}
    \usepackage[english]{babel}
    \addto{\captionsenglish}{%
      
    }
    \usepackage{array}
    \usepackage{droidsans}
    \usepackage{charter}
    \usepackage[T1]{fontenc}
    \usepackage[usenames,dvipsnames]{xcolor}
    \usepackage{setspace}
    \usepackage[compact]{titlesec}
    \usepackage{hyperref}
    \usepackage{xspace}
    
    \usepackage{amsmath,amssymb}
    \usepackage[format=plain,justification=justified,singlelinecheck=false,font={stretch=1.125,sf},labelfont=bf,labelsep=space]{subcaption}

    \usepackage{dblfloatfix}  

    \definecolor{cream}{RGB}{222,217,201}
    
    \newcommand{\pmg}{pymatgen\xspace}
    \newcommand{\Pmg}{Pymatgen\xspace}
    
    \providecommand{\pmgORCID}[1]{%
      \href{https://orcid.org/#1}{%
        \textcolor[HTML]{A6CE39}{%
          \raisebox{0.15ex}{\scriptsize\sffamily\bfseries iD}%
        }%
      }%
    }
    
\begin{document}
    
    \pagestyle{fancy}
    \thispagestyle{plain}
    \fancypagestyle{plain}{
    \renewcommand{\headrulewidth}{0pt}
    }
    
    \makeFNbottom
    \makeatletter
    \renewcommand\LARGE{\@setfontsize\LARGE{15pt}{17}}
    \renewcommand\Large{\@setfontsize\Large{12pt}{14}}
    \renewcommand\large{\@setfontsize\large{10pt}{12}}
    \renewcommand\footnotesize{\@setfontsize\footnotesize{7pt}{10}}
    \makeatother
    
    \renewcommand{\thefootnote}{\fnsymbol{footnote}}
    \renewcommand\footnoterule{\vspace*{1pt}%
    \color{cream}\hrule width 3.5in height 0.4pt \color{black}\vspace*{5pt}} 
    \setcounter{secnumdepth}{5}
    
    \makeatletter 
    \renewcommand\@biblabel[1]{#1}            
    \renewcommand\@makefntext[1]%
    {\noindent\makebox[0pt][r]{\@thefnmark\,}#1}
    \makeatother 
    \renewcommand{\figurename}{\small{Fig.}~}
    \sectionfont{\sffamily\Large}
    \subsectionfont{\normalsize}
    \subsubsectionfont{\bf}
    \setstretch{1.125} 
    \setlength{\skip\footins}{0.8cm}
    \setlength{\footnotesep}{0.25cm}
    \setlength{\jot}{10pt}
    \titlespacing*{\section}{0pt}{4pt}{4pt}
    \titlespacing*{\subsection}{0pt}{15pt}{1pt}
    
\fancyfoot{}
\fancyfoot[LO,RE]{\vspace{-7.1pt}\includegraphics[height=9pt]{head_foot/LF_anonymized}}
\fancyfoot[CO]{\vspace{-7.1pt}\hspace{13.2cm}\includegraphics{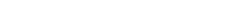}}
\fancyfoot[CE]{\vspace{-7.2pt}\hspace{-14.2cm}\includegraphics{head_foot/LF_anonymized.pdf}}
\fancyfoot[RO]{\footnotesize{\sffamily{1--\pageref{LastPage} ~\textbar  \hspace{2pt}\thepage}}}
\fancyfoot[LE]{\footnotesize{\sffamily{\thepage~\textbar\hspace{3.45cm} 1--\pageref{LastPage}}}}
\fancyhead{}
\renewcommand{\headrulewidth}{0pt} 
\renewcommand{\footrulewidth}{0pt}
\setlength{\arrayrulewidth}{1pt}
\setlength{\columnsep}{6.5mm}
\setlength\bibsep{1pt}

    \makeatletter 
    \newlength{\figrulesep} 
    \setlength{\figrulesep}{0.5\textfloatsep} 
    
    \newcommand{\topfigrule}{\vspace*{-1pt}%
    \noindent{\color{cream}\rule[-\figrulesep]{\columnwidth}{1.5pt}} }
    
    \newcommand{\botfigrule}{\vspace*{-2pt}%
    \noindent{\color{cream}\rule[\figrulesep]{\columnwidth}{1.5pt}} }
    
    \newcommand{\dblfigrule}{\vspace*{-1pt}%
    \noindent{\color{cream}\rule[-\figrulesep]{\textwidth}{1.5pt}} }
    
    \makeatother

\twocolumn[
  \begin{@twocolumnfalse}
{
}\par
\vspace{1em}
\sffamily

    
    
    \noindent\LARGE{\textbf{pymatgen: 15 years of community growth, new functionality, and future
      prospects}} \\
    \vspace{0.3cm} 
    
    \noindent\small{%
      Janine George\,\pmgORCID{0000-0001-8907-0336}\textit{$^{a,b,\dagger}$},\allowbreak\ 
      Matthew K. Horton\,\pmgORCID{0000-0001-7777-8871}\textit{$^{c,\dagger}$},\allowbreak\ 
      Aaron D. Kaplan\,\pmgORCID{0000-0003-3439-4856}\textit{$^{d,\dagger}$},\allowbreak\ 
      Seán R. Kavanagh\,\pmgORCID{0000-0003-4577-9647}\textit{$^{e,\dagger}$},\allowbreak\ 
      Janosh Riebesell\,\pmgORCID{0000-0001-5233-3462}\textit{$^{d,f,\dagger}$},\allowbreak\ 
      Andrew S. Rosen\,\pmgORCID{0000-0002-0141-7006}\textit{$^{g,\dagger}$},\allowbreak\ 
      Haoyu Yang\textit{$^{h,\dagger}$},\allowbreak\ 
      Felix Adams\,\pmgORCID{0000-0001-5802-1072}\textit{$^{i}$},\allowbreak\ 
      Mahdi Amachraa\textit{$^{j}$},\allowbreak\ 
      Oxana Andriuc\,\pmgORCID{0000-0002-4011-3339}\textit{$^{k,l,m}$},\allowbreak\ 
      Omar A. Ashour\,\pmgORCID{0000-0002-7830-9281}\textit{$^{d,n}$},\allowbreak\ 
      Saurabh Bajaj\textit{$^{o}$},\allowbreak\ 
      Ella Banyas\,\pmgORCID{0000-0002-3449-2846}\textit{$^{d}$},\allowbreak\ 
      Luis Barroso-Luque\textit{$^{d}$},\allowbreak\ 
      Karlo Berket\,\pmgORCID{0009-0008-9646-0226}\textit{$^{p}$},\allowbreak\ 
      Noam Bernstein\,\pmgORCID{0000-0002-6532-1337}\textit{$^{q}$},\allowbreak\ 
      Arpit Bhardwaj\,\pmgORCID{0009-0002-2007-1511}\textit{$^{r}$},\allowbreak\ 
      Ali Hussain Umar Bhatti\,\pmgORCID{0000-0003-4342-0219}\textit{$^{s,t}$},\allowbreak\ 
      Samuel M. Blau\,\pmgORCID{0000-0003-3132-3032}\textit{$^{u}$},\allowbreak\ 
      Brandon Bocklund\,\pmgORCID{0000-0002-3264-8413}\textit{$^{v}$},\allowbreak\ 
      Alexander Bonkowski\,\pmgORCID{0000-0002-0525-4742}\textit{$^{w}$},\allowbreak\ 
      Danny Broberg\textit{$^{x}$},\allowbreak\ 
      Guillaume Brunin\,\pmgORCID{0000-0002-5685-4668}\textit{$^{y}$},\allowbreak\ 
      Kyle Bystrom\,\pmgORCID{0000-0003-1342-4972}\textit{$^{z}$},\allowbreak\ 
      Charles A. Cardot\,\pmgORCID{0000-0001-6914-8650}\textit{$^{aa}$},\allowbreak\ 
      Rees Chang\,\pmgORCID{0000-0001-5773-7183}\textit{$^{ab}$},\allowbreak\ 
      Chi Chen\,\pmgORCID{0000-0001-8008-7043}\textit{$^{ac}$},\allowbreak\ 
      Wei Chen\,\pmgORCID{0000-0002-1135-7721}\textit{$^{ad}$},\allowbreak\ 
      Xin Chen\,\pmgORCID{0000-0001-9643-0870}\textit{$^{ae}$},\allowbreak\ 
      Yuan Chiang\,\pmgORCID{0000-0002-4017-7084}\textit{$^{d,x}$},\allowbreak\ 
      Kamal Choudhary\,\pmgORCID{0000-0001-9737-8074}\textit{$^{af}$},\allowbreak\ 
      Orion A. Cohen\,\pmgORCID{0000-0003-3940-2456}\textit{$^{ag}$},\allowbreak\ 
      Miguel D. Costa\,\pmgORCID{0000-0001-8859-5763}\textit{$^{ah}$},\allowbreak\ 
      Stephen T. Dacek III\textit{$^{ai}$},\allowbreak\ 
      Maarten de Jong\textit{$^{x}$},\allowbreak\ 
      Tom Demeyere\,\pmgORCID{0000-0002-5023-6156}\textit{$^{a}$},\allowbreak\ 
      Alan K. Dozier\textit{$^{aj,ak}$},\allowbreak\ 
      Alexander Dunn\,\pmgORCID{0000-0002-8567-1879}\textit{$^{u}$},\allowbreak\ 
      Shyam Dwaraknath\,\pmgORCID{0000-0003-0289-2607}\textit{$^{al}$},\allowbreak\ 
      Maxwell T. Dylla\,\pmgORCID{0000-0001-8461-5122}\textit{$^{am}$},\allowbreak\ 
      Kristjan Eimre\,\pmgORCID{0000-0002-3444-3286}\textit{$^{an}$},\allowbreak\ 
      Alexander R. Epstein\,\pmgORCID{0000-0002-9914-2388}\textit{$^{x}$},\allowbreak\ 
      Christina Ertural\,\pmgORCID{0000-0002-7696-5824}\textit{$^{a}$},\allowbreak\ 
      Marco Esters\,\pmgORCID{0000-0002-8793-2200}\textit{$^{ao}$},\allowbreak\ 
      Matthew L. Evans\,\pmgORCID{0000-0002-1182-9098}\textit{$^{e}$},\allowbreak\ 
      Johannes Falke\,\pmgORCID{0000-0003-4122-9381}\textit{$^{ap}$},\allowbreak\ 
      Adam Fekete\textit{$^{h}$},\allowbreak\ 
      Bernard Field\,\pmgORCID{0000-0003-0352-9156}\textit{$^{d,n}$},\allowbreak\ 
      Alexandr Fonari\textit{$^{aq}$},\allowbreak\ 
      Nathan C. Frey\,\pmgORCID{0000-0001-5291-6131}\textit{$^{ar}$},\allowbreak\ 
      Dale W. Gaines II\,\pmgORCID{0000-0001-9280-1240}\textit{$^{am}$},\allowbreak\ 
      Alex M. Ganose\,\pmgORCID{0000-0002-4486-3321}\textit{$^{as}$},\allowbreak\ 
      Matteo Giantomassi\,\pmgORCID{0000-0002-7007-9813}\textit{$^{at}$},\allowbreak\ 
      Rhys E. A. Goodall\,\pmgORCID{0000-0002-6589-1700}\textit{$^{ai}$},\allowbreak\ 
      Petre-Flaviu Gostin\,\pmgORCID{0000-0002-0528-7093}\textit{$^{au}$},\allowbreak\ 
      Aayush Gupta\,\pmgORCID{0000-0001-6208-9889}\textit{$^{av}$},\allowbreak\ 
      Viet-Anh Ha\,\pmgORCID{0000-0002-6665-1274}\textit{$^{aw}$},\allowbreak\ 
      Geoffroy Hautier\,\pmgORCID{0000-0003-1754-2220}\textit{$^{ax,ay,az}$},\allowbreak\ 
      Niclas Heinsdorf\,\pmgORCID{0009-0009-5188-2852}\textit{$^{ba}$},\allowbreak\ 
      Michael F. Herbst\,\pmgORCID{0000-0003-0378-7921}\textit{$^{bb,bc}$},\allowbreak\ 
      Tingzheng Hou\,\pmgORCID{0000-0002-7163-2561}\textit{$^{bd}$},\allowbreak\ 
      Patrick Huck\,\pmgORCID{0000-0001-8237-2227}\textit{$^{d}$},\allowbreak\ 
      Yuji Ikeda\,\pmgORCID{0000-0001-9176-3270}\textit{$^{be}$},\allowbreak\ 
      Adam J. Jackson\,\pmgORCID{0000-0001-5272-6530}\textit{$^{bf}$},\allowbreak\ 
      Saivenkataraman Jayaraman\textit{$^{bg}$},\allowbreak\ 
      Finn R. Kearns\textit{$^{bh}$},\allowbreak\ 
      Sunghyun Kim\,\pmgORCID{0000-0001-5072-6801}\textit{$^{bi}$},\allowbreak\ 
      Ryan S. Kingsbury\,\pmgORCID{0000-0002-7168-3967}\textit{$^{bj}$},\allowbreak\ 
      Adam M. Krajewski\,\pmgORCID{0000-0002-2266-0099}\textit{$^{bk}$},\allowbreak\ 
      Yu Kumagai\,\pmgORCID{0000-0003-0489-8148}\textit{$^{bl}$},\allowbreak\ 
      Matthew C. Kuner\,\pmgORCID{0000-0002-8218-8558}\textit{$^{d,x}$},\allowbreak\ 
      William T. Laderer\,\pmgORCID{0009-0004-5079-8644}\textit{$^{bm}$},\allowbreak\ 
      Weitang Li\,\pmgORCID{0000-0002-8739-641X}\textit{$^{bn}$},\allowbreak\ 
      Xiang-Guo Li\,\pmgORCID{0000-0002-2062-3809}\textit{$^{bo}$},\allowbreak\ 
      Yu Hsuan Liang\textit{$^{bp}$},\allowbreak\ 
      Hanyu Liu\textit{$^{bq}$},\allowbreak\ 
      Ke Liu\,\pmgORCID{0000-0003-3604-1026}\textit{$^{br}$},\allowbreak\ 
      Miao Liu\,\pmgORCID{0000-0002-1843-9519}\textit{$^{bs,bt}$},\allowbreak\ 
      Ahmad Lutfi\,\pmgORCID{0000-0003-0213-075X}\textit{$^{bu}$},\allowbreak\ 
      Olivier Marchand\,\pmgORCID{0009-0008-6745-3801}\textit{$^{bv}$},\allowbreak\ 
      Matthew J. McDermott\,\pmgORCID{0000-0002-4071-3000}\textit{$^{x,bw}$},\allowbreak\ 
      Bharat K. Medasani\,\pmgORCID{0000-0002-2073-4162}\textit{$^{bx}$},\allowbreak\ 
      Fanchen Meng\,\pmgORCID{0000-0001-9869-9005}\textit{$^{by}$},\allowbreak\ 
      Kyle D. Miller\,\pmgORCID{0000-0001-6083-6502}\textit{$^{am,bz}$},\allowbreak\ 
      Henrique Miranda\,\pmgORCID{0000-0002-2843-0876}\textit{$^{ca}$},\allowbreak\ 
      Rohith Srinivaas Mohanakrishnan\,\pmgORCID{0000-0003-1898-0660}\textit{$^{d,x}$},\allowbreak\ 
      Joseph H. Montoya\,\pmgORCID{0000-0001-5760-2860}\textit{$^{cb}$},\allowbreak\ 
      Guy C. Moore\,\pmgORCID{0000-0002-8346-6871}\textit{$^{d}$},\allowbreak\ 
      Benjamin J. Morgan\,\pmgORCID{0000-0002-3056-8233}\textit{$^{cc}$},\allowbreak\ 
      Jason M. Munro\textit{$^{cd}$},\allowbreak\ 
      Koki Muraoka\,\pmgORCID{0000-0003-1830-7978}\textit{$^{u,ce,cf}$},\allowbreak\ 
      Aakash A. Naik\,\pmgORCID{0000-0002-6071-6786}\textit{$^{a,b}$},\allowbreak\ 
      Willis O'Leary\,\pmgORCID{0000-0003-1126-0214}\textit{$^{cg}$},\allowbreak\ 
      Anthony Onwuli\,\pmgORCID{0000-0003-2107-153X}\textit{$^{r,bh}$},\allowbreak\ 
      Jaesuk Park\,\pmgORCID{0009-0005-1584-5385}\textit{$^{aw}$},\allowbreak\ 
      Drew F. Parsons\,\pmgORCID{0000-0002-3956-6031}\textit{$^{ch}$},\allowbreak\ 
      Kai Pei\,\pmgORCID{0000-0001-7378-9987}\textit{$^{ci}$},\allowbreak\ 
      Guido Petretto\,\pmgORCID{0000-0001-7220-5417}\textit{$^{y,at}$},\allowbreak\ 
      Thomas A. R. Purcell\,\pmgORCID{0000-0003-4564-7206}\textit{$^{cj}$},\allowbreak\ 
      Ji Qi\,\pmgORCID{0000-0001-5808-9931}\textit{$^{ck}$},\allowbreak\ 
      Xiaohui Qu\,\pmgORCID{0000-0001-5651-8405}\textit{$^{cl}$},\allowbreak\ 
      Ihor Radchenko\,\pmgORCID{0000-0002-0498-7100}\textit{$^{cm}$},\allowbreak\ 
      Nav Nidhi Rajput\,\pmgORCID{0000-0003-4740-8217}\textit{$^{cn}$},\allowbreak\ 
      Francesco Ricci\,\pmgORCID{0000-0002-2677-7227}\textit{$^{at}$},\allowbreak\ 
      Benjamin S. Rich\,\pmgORCID{0000-0003-2887-5944}\textit{$^{co}$},\allowbreak\ 
      William D. Richards\textit{$^{cg}$},\allowbreak\ 
      Gian-Marco Rignanese\,\pmgORCID{0000-0002-1422-1205}\textit{$^{at,cp}$},\allowbreak\ 
      Xiaoxu Ruan\,\pmgORCID{0009-0009-9897-4105}\textit{$^{cq}$},\allowbreak\ 
      Germain Salvato Vallverdu\,\pmgORCID{0000-0003-1116-8776}\textit{$^{cr}$},\allowbreak\ 
      Jonathan Schmidt\,\pmgORCID{0000-0001-5685-6404}\textit{$^{bc}$},\allowbreak\ 
      Jimmy-Xuan Shen\,\pmgORCID{0000-0002-2743-7531}\textit{$^{cs}$},\allowbreak\ 
      Ryuichi Shimogawa\,\pmgORCID{0000-0002-6507-1876}\textit{$^{cn}$},\allowbreak\ 
      Kohei Shinohara\,\pmgORCID{0000-0002-5907-2549}\textit{$^{ct}$},\allowbreak\ 
      Arunima K. Singh\,\pmgORCID{0000-0002-7212-6310}\textit{$^{cu}$},\allowbreak\ 
      Eric Sivonxay\,\pmgORCID{0000-0002-6408-1255}\textit{$^{u}$},\allowbreak\ 
      Stef Smeets\,\pmgORCID{0000-0002-5413-9038}\textit{$^{cv}$},\allowbreak\ 
      Tess E. Smidt\,\pmgORCID{0000-0001-5581-5344}\textit{$^{av,cw}$},\allowbreak\ 
      Andrei Sobolev\,\pmgORCID{0000-0001-5086-6601}\textit{$^{cx}$},\allowbreak\ 
      Evan Walter Clark Spotte-Smith\,\pmgORCID{0000-0003-1554-197X}\textit{$^{cy,cz}$},\allowbreak\ 
      Christopher Stihl\textit{$^{da}$},\allowbreak\ 
      Wenhao Sun\,\pmgORCID{0000-0002-8416-455X}\textit{$^{db}$},\allowbreak\ 
      Jack D. Sundberg\,\pmgORCID{0000-0001-5739-8919}\textit{$^{dc}$},\allowbreak\ 
      Yuta Suzuki\,\pmgORCID{0000-0002-0019-4832}\textit{$^{dd}$},\allowbreak\ 
      Leopold Talirz\,\pmgORCID{0000-0002-1524-5903}\textit{$^{de}$},\allowbreak\ 
      Hanmei Tang\,\pmgORCID{0000-0003-2659-7768}\textit{$^{cq}$},\allowbreak\ 
      Eric Taw\textit{$^{df}$},\allowbreak\ 
      Cooper Tezak\,\pmgORCID{0000-0001-7512-7792}\textit{$^{dg}$},\allowbreak\ 
      Félix Therrien\,\pmgORCID{0000-0003-1074-7805}\textit{$^{dh}$},\allowbreak\ 
      Scott D. Thiel\,\pmgORCID{0000-0002-9947-0277}\textit{$^{di}$},\allowbreak\ 
      Steven B. Torrisi\,\pmgORCID{0000-0002-4283-8077}\textit{$^{dj}$},\allowbreak\ 
      Mark E. Turiansky\,\pmgORCID{0000-0002-9154-3582}\textit{$^{q}$},\allowbreak\ 
      Katharina Ueltzen\,\pmgORCID{0009-0003-2967-1182}\textit{$^{a,b}$},\allowbreak\ 
      Michiel J. van Setten\,\pmgORCID{0000-0003-0557-5260}\textit{$^{dk}$},\allowbreak\ 
      Josua Vieten\,\pmgORCID{0000-0002-2060-2039}\textit{$^{dl}$},\allowbreak\ 
      Sudarshan Vijay\,\pmgORCID{0000-0001-8242-0161}\textit{$^{dm}$},\allowbreak\ 
      Nicholas Wagner\textit{$^{dn}$},\allowbreak\ 
      Frank Wan\textit{$^{d}$},\allowbreak\ 
      Amanda Wang\,\pmgORCID{0000-0002-5084-4039}\textit{$^{db}$},\allowbreak\ 
      Jingyang Wang\,\pmgORCID{0000-0003-3307-5132}\textit{$^{do}$},\allowbreak\ 
      Zhenbin Wang\,\pmgORCID{0000-0002-7016-9245}\textit{$^{dp}$},\allowbreak\ 
      Logan Ward\,\pmgORCID{0000-0002-1323-5939}\textit{$^{dj}$},\allowbreak\ 
      Thomas Warford\,\pmgORCID{0009-0000-6039-2083}\textit{$^{dq}$},\allowbreak\ 
      David Waroquiers\,\pmgORCID{0000-0001-8943-9762}\textit{$^{y}$},\allowbreak\ 
      Mingjian Wen\,\pmgORCID{0000-0003-0013-575X}\textit{$^{dr}$},\allowbreak\ 
      Oskar Weser\,\pmgORCID{0000-0001-5503-1195}\textit{$^{ds,dt}$},\allowbreak\ 
      Nicholas Winner\textit{$^{x}$},\allowbreak\ 
      Donald Winston\,\pmgORCID{0000-0002-8424-0604}\textit{$^{du}$},\allowbreak\ 
      Michael Wolloch\,\pmgORCID{0000-0002-3419-5526}\textit{$^{ca}$},\allowbreak\ 
      Brandon M. Wood\textit{$^{dv}$},\allowbreak\ 
      Rachel Woods-Robinson\,\pmgORCID{0000-0001-5009-9510}\textit{$^{dw}$},\allowbreak\ 
      Yihan Xiao\,\pmgORCID{0000-0002-4409-6898}\textit{$^{x}$},\allowbreak\ 
      Yaoshu Xie\textit{$^{bd}$},\allowbreak\ 
      Firat Yalcin\,\pmgORCID{0000-0001-7809-1094}\textit{$^{dx}$},\allowbreak\ 
      Ruo Xi Yang\,\pmgORCID{0000-0001-8225-5856}\textit{$^{d}$},\allowbreak\ 
      Weike Ye\,\pmgORCID{0000-0002-9541-7006}\textit{$^{cb}$},\allowbreak\ 
      Pezhman Zarabadi-Poor\textit{$^{dy}$},\allowbreak\ 
      Jinzhe Zeng\,\pmgORCID{0000-0002-1515-8172}\textit{$^{dz,ea,eb,ec}$},\allowbreak\ 
      Hongsheng Zhao\,\pmgORCID{0009-0003-9131-7539}\textit{$^{ed}$},\allowbreak\ 
      Kaifeng Zheng\,\pmgORCID{0000-0001-6766-1805}\textit{$^{cn}$},\allowbreak\ 
      He Zhu\textit{$^{ee}$},\allowbreak\ 
      Zhuoying Zhu\,\pmgORCID{0000-0003-1775-7651}\textit{$^{ea}$},\allowbreak\ 
      Nils E. R. Zimmermann\,\pmgORCID{0000-0003-1063-5926}\textit{$^{ai}$},\allowbreak\ 
      Daniel Zügner\,\pmgORCID{0000-0003-1626-5065}\textit{$^{ef}$},\allowbreak\ 
      Kristin A. Persson\,\pmgORCID{0000-0003-2495-5509}\textit{$^{d,x}$},\allowbreak\ 
      Anubhav Jain\,\pmgORCID{0000-0001-5893-9967}\textit{$^{eg}$}, and\allowbreak\ 
      Shyue Ping Ong\,\pmgORCID{0000-0001-5726-2587}\textit{$^{cq,eh,\ast}$}
    } \\
   
    \includegraphics{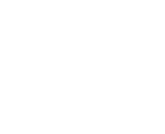}
    \noindent\normalsize{
    
    \textbf{Abstract}
    \large{
    We present a retrospective on the Python Materials Genomics (\pmg) library, reflecting on 15 years of community growth and established best practices.
    The current version of \pmg builds on its robust, open-source foundation, emphasising its collaborative nature.
    Over the past decade, \pmg has thrived as one of the largest open-source materials science codebases.
    We detail how \pmg supports advanced computational materials science workflows, how it adapts to changing demands, and the lessons learned from its growing community, providing guidance beyond \pmg's open-source community.
    \Pmg will continue to evolve in response to the current rapidly changing landscape of computational materials science and materials informatics.
    We discuss the challenges and potential solutions to managing the community-driven evolution of \pmg and provide an outlook on its future development.}
    } \\
    


     \end{@twocolumnfalse} \vspace{0.6cm}
    
      ]
    {\footnotesize\textit{$^{a}$} Materials Chemistry, Federal Institute for Materials Research and Testing (BAM), Berlin, 12205, Germany.}
    {\footnotesize\textit{$^{b}$} Institute of Condensed Matter Theory and Optics, Friedrich Schiller University Jena, Jena, 07743, Germany.}
    {\footnotesize\textit{$^{c}$} Isotropy, LLC, Seattle, WA, 98112, United States.}
    {\footnotesize\textit{$^{d}$} Materials Sciences Division, Lawrence Berkeley National Laboratory, Berkeley, CA, 94720, United States.}
    {\footnotesize\textit{$^{e}$} Yusuf Hamied Department of Chemistry, University of Cambridge, Cambridge, CB2 1EW, United Kingdom.}
    {\footnotesize\textit{$^{f}$} Cavendish Laboratory, University of Cambridge, Cambridge, CB3 0US, United Kingdom.}
    {\footnotesize\textit{$^{g}$} Department of Chemical and Biological Engineering, Princeton University, Princeton, NJ, 08544, United States.}
    {\footnotesize\textit{$^{h}$} Institut für Physik, Humboldt-Universität zu Berlin, Berlin, Germany.}
    {\footnotesize\textit{$^{i}$} Applied Research Lab for Intelligence and Security, University of Maryland, College Park, MD, 20740, United States.}
    {\footnotesize\textit{$^{j}$} R\&D, Elemynt Pte. Ltd., Singapore.}
    {\footnotesize\textit{$^{k}$} Department of Chemistry, University of California, Berkeley, Berkeley, CA, 94720, United States.}
    {\footnotesize\textit{$^{l}$} Liquid Sunlight Alliance, Lawrence Berkeley National Laboratory, Berkeley, CA, 94720, United States.}
    {\footnotesize\textit{$^{m}$} Chemical Sciences Division, Lawrence Berkeley National Laboratory, Berkeley, CA, 94720, United States.}
    {\footnotesize\textit{$^{n}$} Molecular Foundry, Lawrence Berkeley National Laboratory, Berkeley, CA, 94720, United States.}
    {\footnotesize\textit{$^{o}$} Smart Manufacturing and Artificial Intelligence, Micron Technology Inc., Manassas, VA, 20110, United States.}
    {\footnotesize\textit{$^{p}$} Scientific Data Division, Lawrence Berkeley National Laboratory, Berkeley, CA, 94720, United States.}
    {\footnotesize\textit{$^{q}$} Center for Materials Physics and Technology, U.S. Naval Research Laboratory, Washington, DC, 20375, United States.}
    {\footnotesize\textit{$^{r}$} MatNex, Cambridge, CB1 2LA, United Kingdom.}
    {\footnotesize\textit{$^{s}$} Energy Storage Research Center, Korea Institute of Science and Technology, Seoul 02792, Republic of Korea.}
    {\footnotesize\textit{$^{t}$} Division of Energy and Environment Technology, KIST School, Korea University of Science and Technology, Seoul, 02792, Republic of Korea.}
    {\footnotesize\textit{$^{u}$} Energy Technologies Area, Lawrence Berkeley National Laboratory, Berkeley, CA, 94720, United States.}
    {\footnotesize\textit{$^{v}$} Materials Science Division, Lawrence Livermore National Laboratory, Livermore, CA, 94550, United States.}
    {\footnotesize\textit{$^{w}$} Institute of Physical Chemistry, RWTH Aachen University, Aachen, 52074, Germany.}
    {\footnotesize\textit{$^{x}$} Department of Materials Science and Engineering, University of California, Berkeley, Berkeley, CA, 94720, United States.}
    {\footnotesize\textit{$^{y}$} Research and Development, Matgenix, Charleroi, Belgium.}
    {\footnotesize\textit{$^{z}$} Initiative for Computational Catalysis, Flatiron Institute, New York, NY, 10010, United States.}
    {\footnotesize\textit{$^{aa}$} Physics, University of Washington, Seattle, WA, 98195, United States.}
    {\footnotesize\textit{$^{ab}$} Department of Materials Science and Engineering, University of Illinois at Urbana-Champaign, Urbana, IL, 61801, United States.}
    {\footnotesize\textit{$^{ac}$} Quantum Applications, IonQ Inc, College Park, MD, 20740, United States.}
    {\footnotesize\textit{$^{ad}$} Department of Materials Design and Innovation, University at Buffalo, Buffalo, NY, 14260, United States.}
    {\footnotesize\textit{$^{ae}$} National Key Laboratory of Computational Physics, Institute of Applied Physics and Computational Mathematics, Beijing, China.}
    {\footnotesize\textit{$^{af}$} Department of Materials Science and Engineering, Johns Hopkins University, Baltimore, MD, 21218, United States.}
    {\footnotesize\textit{$^{ag}$} Riven Systems, New York, NY, 10027, United States.}
    {\footnotesize\textit{$^{ah}$} Department of Physics, University of Coimbra, Coimbra, Portugal.}
    {\footnotesize\textit{$^{ai}$} Independent Researcher.}
    {\footnotesize\textit{$^{aj}$} Center for Computational Science (CCS), University of Kentucky, Lexington, KY, 40506, United States.}
    {\footnotesize\textit{$^{ak}$} Health Effects Laboratory Division (HELD), National Institute for Occupational Health and Safety (NIOSH), Cincinnati, OH, 45226, United States.}
    {\footnotesize\textit{$^{al}$} Barrier Labs, San Francisco, CA, United States.}
    {\footnotesize\textit{$^{am}$} Department of Materials Science and Engineering, Northwestern University, Evanston, IL, 60208, United States.}
    {\footnotesize\textit{$^{an}$} PSI Center for Scientific Computing, Theory and Data, Paul Scherrer Institute, Villigen, Switzerland.}
    {\footnotesize\textit{$^{ao}$} Department of Chemistry and Biochemistry, University of Oregon, Eugene, OR, 97403, United States.}
    {\footnotesize\textit{$^{ap}$} Physics of Correlated Matter, Max Planck Institute for Chemical Physics of Solids, Dresden, 01187, Germany.}
    {\footnotesize\textit{$^{aq}$} Schrödinger Inc., New York, NY, 10036, United States.}
    {\footnotesize\textit{$^{ar}$} Life Sciences, Anthropic, New York, NY, United States.}
    {\footnotesize\textit{$^{as}$} Department of Chemistry, Imperial College London, London, W12 0BZ, United Kingdom.}
    {\footnotesize\textit{$^{at}$} Institute of Condensed Matter and Nanosciences (IMCN), UCLouvain, Louvain-la-Neuve, B-1348, Belgium.}
    {\footnotesize\textit{$^{au}$} CoaST, Department of Chemical and Biochemical Engineering, Technical University of Denmark, DTU, Kongens Lyngby, 2800, Denmark.}
    {\footnotesize\textit{$^{av}$} Electrical Engineering and Computer Science, Massachusetts Institute of Technology, Cambridge, MA, 02139, United States.}
    {\footnotesize\textit{$^{aw}$} Oden Institute for Computational Engineering and Sciences, The University of Texas at Austin, Austin, TX, 78712, United States.}
    {\footnotesize\textit{$^{ax}$} Materials Science and NanoEngineering, Rice University, Houston, TX, 77005, United States.}
    {\footnotesize\textit{$^{ay}$} Rice Advanced Materials Institute, Rice University, Houston, TX, 77005, United States.}
    {\footnotesize\textit{$^{az}$} Thayer School of Engineering, Dartmouth College, Hanover, NH, 03755, United States.}
    {\footnotesize\textit{$^{ba}$} Department of Physics, California Institute of Technology, Pasadena, CA, 91125, United States.}
    {\footnotesize\textit{$^{bb}$} Institute of Mathematics, École Polytechnique Fédérale de Lausanne, Lausanne, 1015, Switzerland.}
    {\footnotesize\textit{$^{bc}$} Institute of Materials, École Polytechnique Fédérale de Lausanne, Lausanne, 1015, Switzerland.}
    {\footnotesize\textit{$^{bd}$} Institute of Materials Research (iMR), Tsinghua Shenzhen International Graduate School, Tsinghua University, Shenzhen, 518055, China.}
    {\footnotesize\textit{$^{be}$} Institute for Materials Science, University of Stuttgart, Stuttgart, 70569, Germany.}
    {\footnotesize\textit{$^{bf}$} Scientific Computing, UKRI Science and Technology Facilities Council, Didcot, OX11 0QX, United Kingdom.}
    {\footnotesize\textit{$^{bg}$} Drug Product Development, Bristol Myers Squibb, New Brunswick, NJ, 08901, United States.}
    {\footnotesize\textit{$^{bh}$} Department of Materials, Imperial College London, London, SW7 2AZ, United Kingdom.}
    {\footnotesize\textit{$^{bi}$} Samsung Advanced Institute of Technology, Samsung Electronics Co., Ltd., Suwon, Republic of Korea.}
    {\footnotesize\textit{$^{bj}$} Civil and Environmental Engineering and the Andlinger Center for Energy and the Environment, Princeton University, Princeton, NJ, 08544, United States.}
    {\footnotesize\textit{$^{bk}$} Department of Materials Science and Engineering, The Pennsylvania State University, University Park, PA, 16802, United States.}
    {\footnotesize\textit{$^{bl}$} Institute for Materials Research, Tohoku University, Sendai, Miyagi, 980-8577, Japan.}
    {\footnotesize\textit{$^{bm}$} Chemistry and Biochemistry, University of California, Los Angeles, Los Angeles, CA, 90095, United States.}
    {\footnotesize\textit{$^{bn}$} School of Science and Engineering, The Chinese University of Hong Kong (Shenzhen), Shenzhen, China.}
    {\footnotesize\textit{$^{bo}$} School of Materials, Sun Yat-sen University, Shenzhen, 518107, China.}
    {\footnotesize\textit{$^{bp}$} Chemistry, Columbia University, New York, NY, 10027, United States.}
    {\footnotesize\textit{$^{bq}$} School of Materials Science and Engineering, Beihang University, Beijing, 100191, China.}
    {\footnotesize\textit{$^{br}$} Shanghai Jiao Tong University Global College, Shanghai Jiao Tong University, Shanghai, China.}
    {\footnotesize\textit{$^{bs}$} Institute of Physics, Chinese Academy of Sciences, Beijing, 100190, China.}
    {\footnotesize\textit{$^{bt}$} Dongguan Institute of Materials Science and Technology, Chinese Academy of Sciences, Dongguan, Guangdong, 523808, China.}
    {\footnotesize\textit{$^{bu}$} Institute of Biotechnology-Bioanalytics, Technische Universität Berlin, Berlin, 10623, Germany.}
    {\footnotesize\textit{$^{bv}$} Department of Chemistry and Biomolecular Sciences, University of Ottawa, Ottawa, Ontario, K1N 6N5, Canada.}
    {\footnotesize\textit{$^{bw}$} Newfound Materials, Inc., Houston, TX, 77004, United States.}
    {\footnotesize\textit{$^{bx}$} Computational Sciences Department, Princeton Plasma Physics Laboratory, Princeton, NJ, 08540, United States.}
    {\footnotesize\textit{$^{by}$} Watt Family Innovation Center, Clemson University, Clemson, SC, 29634, United States.}
    {\footnotesize\textit{$^{bz}$} External Research, Citrine Informatics, Redwood City, CA, 94063, United States.}
    {\footnotesize\textit{$^{ca}$} VASP Software GmbH, Vienna, A-1090, Austria.}
    {\footnotesize\textit{$^{cb}$} Energy and Materials, Toyota Research Institute, Los Altos, CA, 94022, United States.}
    {\footnotesize\textit{$^{cc}$} Department of Chemistry, University of Bath, Bath, BA2 7AY, United Kingdom.}
    {\footnotesize\textit{$^{cd}$} Atomscale, Boston, MA, 02110, United States.}
    {\footnotesize\textit{$^{ce}$} Institute of Engineering Innovation, The University of Tokyo, Bunkyo-ku, Tokyo, 113-8656, Japan.}
    {\footnotesize\textit{$^{cf}$} Department of Chemical System Engineering, The University of Tokyo, Bunkyo-ku, Tokyo, 113-8656, Japan.}
    {\footnotesize\textit{$^{cg}$} Department of Materials Science and Engineering, Massachusetts Institute of Technology, Cambridge, MA, 02139, United States.}
    {\footnotesize\textit{$^{ch}$} Department of Chemical and Geological Sciences and CSGI, University of Cagliari, Monserrato, CA, 09042, Italy.}
    {\footnotesize\textit{$^{ci}$} Frontier Institute of Science and Technology, Xi'an Jiaotong University, Xi'an, 712046, China.}
    {\footnotesize\textit{$^{cj}$} Chemistry and Biochemistry, University of Arizona, Tucson, AZ, 85719, United States.}
    {\footnotesize\textit{$^{ck}$} Program of Materials Science and Engineering, University of California San Diego, La Jolla, CA, 92093, United States.}
    {\footnotesize\textit{$^{cl}$} Center for Functional Nanomaterials, Brookhaven National Laboratory, Upton, NY, 11973-5000, United States.}
    {\footnotesize\textit{$^{cm}$} Ensemble3 sp. z o.o., Warsaw, 01-919, Poland.}
    {\footnotesize\textit{$^{cn}$} Department of Materials Science and Chemical Engineering, Stony Brook University, Stony Brook, NY, 11794, United States.}
    {\footnotesize\textit{$^{co}$} Chemistry, University of Colorado, Boulder, Boulder, CO, 80309, United States.}
    {\footnotesize\textit{$^{cp}$} WEL Research Institute, Wavre, Belgium.}
    {\footnotesize\textit{$^{cq}$} Aiiso Yufeng Li Family Department of Chemical and Nano Engineering, University of California San Diego, La Jolla, CA, 92093, United States.}
    {\footnotesize\textit{$^{cr}$} CNRS, IPREM, Université de Pau et des Pays de l'Adour, Pau, France.}
    {\footnotesize\textit{$^{cs}$} Physical Sciences, Lila Sciences, San Francisco, CA, United States.}
    {\footnotesize\textit{$^{ct}$} Materials and Drug Discovery, Preferred Networks, Inc., Tokyo, 1000004, Japan.}
    {\footnotesize\textit{$^{cu}$} Department of Physics, Arizona State University, Tempe, AZ, 85287, United States.}
    {\footnotesize\textit{$^{cv}$} Natural Sciences and Engineering, Netherlands eScience Center, Amsterdam, 1098 XH, Netherlands.}
    {\footnotesize\textit{$^{cw}$} Research Laboratory of Electronics, Massachusetts Institute of Technology, Cambridge, MA, 02139, United States.}
    {\footnotesize\textit{$^{cx}$} Molecular Simulations from First Principles e.V., Berlin, Germany.}
    {\footnotesize\textit{$^{cy}$} School of Chemistry, University College Dublin, Dublin, County Dublin, D04 V1W8, Ireland.}
    {\footnotesize\textit{$^{cz}$} Department of Chemical Engineering, Carnegie Mellon University, Pittsburgh, PA, 15289, United States.}
    {\footnotesize\textit{$^{da}$} Institute for Applied Materials - Applied Materials Physics, Karlsruhe Institute of Technology, Karlsruhe, Germany.}
    {\footnotesize\textit{$^{db}$} Department of Materials Science and Engineering, University of Michigan, Ann Arbor, MI, 48109, United States.}
    {\footnotesize\textit{$^{dc}$} Chemistry, University of North Carolina (Chapel Hill), Chapel Hill, NC, 27599, United States.}
    {\footnotesize\textit{$^{dd}$} Strategic R\&D Planning Department, Toyota Motor Corporation, Tokyo, 103-0022, Japan.}
    {\footnotesize\textit{$^{de}$} Central R\&D, SCHOTT AG, Mainz, Germany.}
    {\footnotesize\textit{$^{df}$} MIT Lincoln Laboratory, Lexington, MA, United States.}
    {\footnotesize\textit{$^{dg}$} Chemical Engineering, Massachusetts Institute of Technology, Cambridge, MA, 02139, United States.}
    {\footnotesize\textit{$^{dh}$} Applied Machine Learning Research Team, Mila, Montréal, Quebec, H2S 3H1, Canada.}
    {\footnotesize\textit{$^{di}$} Department of Chemistry, University of Massachusetts Amherst, Amherst, MA, 01003, United States.}
    {\footnotesize\textit{$^{dj}$} NVIDIA, Santa Clara, CA, 95051, United States.}
    {\footnotesize\textit{$^{dk}$} MSP, imec, Leuven, 3001, Belgium.}
    {\footnotesize\textit{$^{dl}$} Founding Team, ExoMatter GmbH, Munich, Germany.}
    {\footnotesize\textit{$^{dm}$} Chemical Engineering, Indian Institute of Technology Bombay, Mumbai, Maharashtra, 400076, India.}
    {\footnotesize\textit{$^{dn}$} Learning Journey AI Inc, Arlington, VA, 22202, United States.}
    {\footnotesize\textit{$^{do}$} Key Laboratory of Theoretical and Computational Photochemistry, Ministry of Education, College of Chemistry, Beijing Normal University, Beijing, 100875, China.}
    {\footnotesize\textit{$^{dp}$} Department of Materials Science and Engineering, City University of Hong Kong, Hong Kong SAR, 999077, China.}
    {\footnotesize\textit{$^{dq}$} Department of Engineering, University of Cambridge, Cambridge, CB2 1PZ, United Kingdom.}
    {\footnotesize\textit{$^{dr}$} Institute of Fundamental and Frontier Sciences, University of Electronic Science and Technology of China, Chengdu, 611731, China.}
    {\footnotesize\textit{$^{ds}$} Institute of Physical Chemistry, University of Göttingen, Göttingen, 37077, Germany.}
    {\footnotesize\textit{$^{dt}$} Forecast and Services, European Centre for Medium Range Weather Forecast, Bonn, North Rhine-Westphalia, 53175, Germany.}
    {\footnotesize\textit{$^{du}$} Research and Development, Polyneme LLC, New York, NY, United States.}
    {\footnotesize\textit{$^{dv}$} Applied Science and Technology, University of California, Berkeley, Berkeley, CA, 94720, United States.}
    {\footnotesize\textit{$^{dw}$} Clean Energy Institute, University of Washington, Seattle, WA, 98195, United States.}
    {\footnotesize\textit{$^{dx}$} Physics, University of Vienna, Vienna, 1010, Austria.}
    {\footnotesize\textit{$^{dy}$} Department of Materials, University of Oxford, Oxford, OX1 3PH, United Kingdom.}
    {\footnotesize\textit{$^{dz}$} School of Artificial Intelligence and Data Science, University of Science and Technology of China, Hefei, Anhui, 230026, China.}
    {\footnotesize\textit{$^{ea}$} State Key Laboratory of Precision and Intelligent Chemistry, University of Science and Technology of China, Hefei, Anhui, 230026, China.}
    {\footnotesize\textit{$^{eb}$} Suzhou Institute for Advanced Research, University of Science and Technology of China, Suzhou, Jiangsu, 215123, China.}
    {\footnotesize\textit{$^{ec}$} Suzhou Big Data \& AI Research and Engineering Center, Suzhou, Jiangsu, 215123, China.}
    {\footnotesize\textit{$^{ed}$} Department of Automotive Engineering, Hebei Vocational University of Technology and Engineering, Xingtai, Hebei, 054000, China.}
    {\footnotesize\textit{$^{ee}$} Department of Chemical Engineering and Materials Science, University of Minnesota, Minneapolis, MN, 55455, United States.}
    {\footnotesize\textit{$^{ef}$} AI for Science, Microsoft Research, Berlin, 10178, Germany.}
    {\footnotesize\textit{$^{eg}$} Energy Storage and Distributed Resources Division, Lawrence Berkeley National Laboratory, Berkeley, CA, 94720, United States.}
    {\footnotesize\textit{$^{eh}$} Department of Materials Science and Engineering, National University of Singapore, Singapore.}
    {\footnotesize\textit{$^\dagger$} These authors contributed equally.}
    {\footnotesize\textit{$^\ast$} Corresponding author: Shyue Ping Ong (\href{mailto:shyue@nus.edu.sg}{shyue@nus.edu.sg}).}
    \clearpage

    \renewcommand*\rmdefault{bch}\normalfont\upshape
    \rmfamily
    \section*{}
    \vspace{-1cm}

    
    

    
    
    \section{Introduction} 
    
\begin{figure*}[h]
        \centering
        \includegraphics[width=0.75\textwidth]{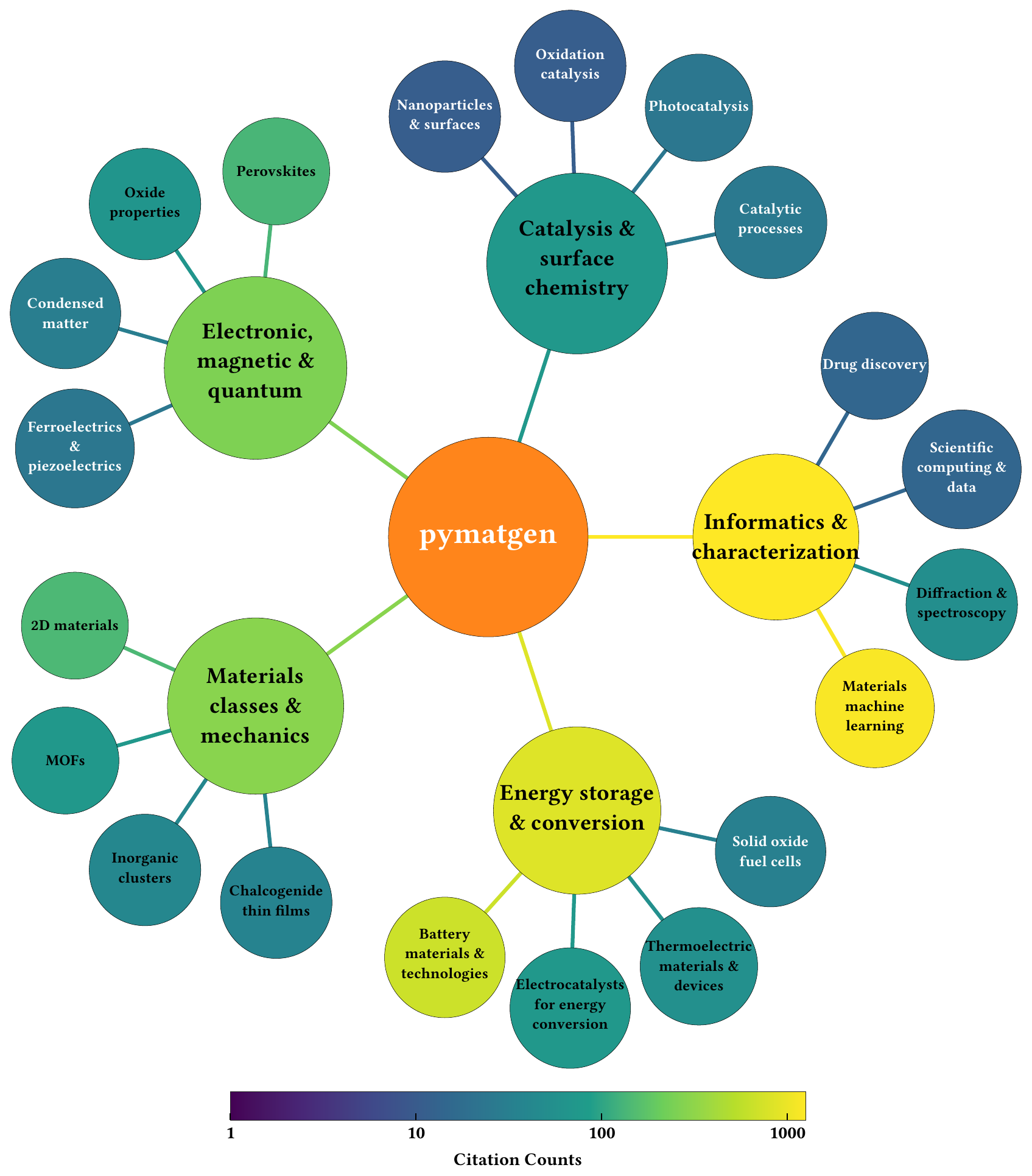}
        \caption{Selected research topics among works citing the original \pmg\ paper, grouped from OpenAlex primary topics into five thematic branches with LLM assistance. Up to four topics are displayed per branch. Node colours and the colour bar share a logarithmic count scale; branch counts sum their displayed children, while the root node colour is fixed for visual distinction.}
        \label{fig:mindmap}
    \end{figure*}

    Python Materials Genomics (\pmg) \cite{ong_python_2013} was launched in 2011 to provide a robust, reusable software foundation for materials analysis in high-throughput density functional theory (DFT) workflows. The predecessor infrastructure, built on Java \cite{jain_high-throughput_2011}, was showing its age. The ubiquity of Python, its mix of functional and object-oriented design, rapid development cycle, and interoperability established it as an ideal platform for modern computational materials science software. What emerged from that practical need has, over 15 years, grown into one of the most widely used open-source libraries in the field \cite{open_comp_matsci}. Today, \pmg is a cornerstone of the Materials Project \cite{jain_commentary_2013,horton_2025_mp}, a dependency for hundreds of downstream packages, and a standard toolkit across research areas spanning battery materials, defect modelling, machine learning interatomic potentials (MLIPs), and materials informatics\cite{butler_machine_2018} (Figs. \ref{fig:mindmap} and \ref{fig:shells-of-influence}).
    
    The trajectory of \pmg offers an opportunity for reflection. Scientific software rarely achieves broad, cross-disciplinary adoption, as exemplified by Fig. \ref{fig:mindmap}, combined with persistent long-term development. \Pmg has sustained active development across more than a decade, and the original software publication now receives 1000 citations per year. It has catalysed an ecosystem of tools \cite{mathew_atomate_2017, ganose_atomate2_2025, horton_2025_mp} in computational materials science. Understanding how and why this happened carries lessons that extend well beyond \pmg itself.
    
    In this retrospective, we review \pmg's design principles and the philosophy of interoperability that enabled its broad adoption. We trace the growth of its contributor community and downstream package ecosystem and examine the technical and organisational challenges that accompany maintaining infrastructure at this scale. Lastly, we outline the directions we believe will matter most in the next phase of its development. Throughout, we aim to extract lessons applicable to the broader challenge of building and sustaining community-driven scientific software --- a challenge that grows more consequential as computational methods become increasingly central to materials discovery.
    
    
    
    \section{Overview and Design Principles}
    
    \subsection{Background}
    
    \Pmg was originally developed in 2011 with the goal of supporting the Materials Project \cite{ong_python_2013,jain_commentary_2013,horton_2025_mp}. In this capacity, \pmg provided essential tools for crystallographic operations,
    symmetry analysis, integration with the Vienna \emph{Ab Initio} Simulation
    Package (VASP) density functional theory (DFT) code \cite{kresse_efficient_1996}
    and basic electronic-structure and phase-diagram analysis.
    These foundational capabilities enabled researchers to automate and standardise
    many aspects of computational materials science, laying the groundwork for
    high-throughput materials discovery.
    
    Since then, \pmg has expanded significantly, incorporating new features and
    adapting to the evolving landscape of materials science and informatics
    \cite{butler_machine_2018}.
    Key trends shaping its development include:
    \begin{enumerate}
      \item The growing importance of research data management\cite{himanen_data-driven_2019}
      \item The rise of automated workflows for reproducible simulations and
            experiments \cite{himanen_data-driven_2019, flores-leonar_materials_2020, stier_materials_2024}
      \item Increasing use of machine learning for materials property prediction\cite{butler_machine_2018, schmidt_recent_2019}
      \item Greater availability of computational resources, enabling more complex analyses, machine learning tasks, and simulations
    \end{enumerate}
    
    Today, \pmg aims to accelerate materials discovery by providing a
    comprehensive toolkit for researchers at all levels
    \cite{curtarolo_highthroughput_2013}.
    While \pmg's original goals---such as crystallographic operations and
    electronic structure analysis---remain central, its capabilities have expanded substantially in response to evolving community needs. \Pmg nowadays offers tools for crystallographic and structural analysis,
    including coordination environment analysis, electronic structure parsing,
    thermodynamic and phase diagram construction, analysis of phonon properties,
    and interfaces to various electronic structure codes.
    Additionally, it offers reusable utilities for machine learning applications in
    materials property prediction that underpin materials informatics packages,\cite{ward_matminer_2018} reflecting its evolution from a materials analysis library into a broader platform for materials informatics. By implementing these core materials analysis functions in a modular and reusable software framework, \pmg removes the need to repeatedly reimplement common tasks and allows both users and downstream developers to focus on specialised domain-specific analyses and applications. The same modularity and reuse may also prove increasingly valuable for AI-driven workflows, where LLM-based agents can build on established \pmg functionality as trusted building blocks for scientific reasoning and analysis.
    \Pmg therefore aims to be a powerful, flexible, and reliable software tool
    \cite{horton_promises_2021} for increasingly complex computational materials
    science and materials informatics research workflows.
    
    A key lesson learned from the wide community adoption and continued growth of
    the \pmg codebase is that scientific software must continuously evolve to
    meet emerging demands in its field; without this adaptability, it risks becoming
    obsolete.
    It also must continuously onboard and support new developers; without a welcoming environment, we believe that open-source development is not sustainable. \Pmg's growth has been supported by a low barrier to contribution and responsive maintainership, which facilitate community participation and the timely integration of new functionality.
    
    \subsection{Interoperability and Comparative Tools}
    
    One of \pmg's strengths is its interoperability, having the capability to
    read various molecule and structure file formats and to automatically retrieve
    molecules and structures from databases.
    Furthermore, \pmg provides interfaces to a number of electronic structure /
    DFT codes (e.g., VASP \cite{kresse_efficient_1996}, Quantum ESPRESSO
    \cite{giannozzi_qe_2017}, FHI-aims \cite{blum_fhiaims_2009}, CP2K
    \cite{kuhne_cp2k_2020}, ABINIT \cite{gonze_abinit_2020}, Q-Chem \cite{epifanovsky2021software}) or post-processing
    tools, such as phonopy \cite{togo_firstprinciples_2023,phonopy-phono3py-JPCM} or
    LOBSTER \cite{nelson_lobster_2020}.
    
    A crucial challenge in orchestrating heterogeneous computational workflows is
    normalisation of the data structures used to transfer information between codes.
    Separate codes typically have bespoke input and output data structures, which
    hinder interoperability between related simulation tools: for example, nearly
    all molecular DFT codes accept \texttt{.xyz} files as input to represent
    molecular geometries, but the precise formats used to configure calculation parameters (self-consistent
    field convergence criteria, force convergence criteria, basis sets, etc.) and
    store calculation outputs are unique to each code.
    
    By normalising the representation of this data, \pmg facilitates seamlessly
    chaining the outputs of different simulation tools when building complex workflows.
    Critically, when running complex workflows, the workflow provenance, metadata,
    and progress are typically stored in an external database.
    \Pmg was built with these requirements in mind, and thus supports JSON-compliant object
    de-/serialisation for interoperability with databases that can store JSON data.
    
    \Pmg also complements other materials science tools, most notably the Atomic
    Simulation Environment (ASE) \cite{larsen_atomic_2017}.
    While both software packages offer parsers for DFT and quantum chemistry codes,
    as well as tools for structure and molecule manipulation, their focuses differ.
    \Pmg centres on complex materials data analyses \cite{jain_computational_2016},
    supporting tasks from oxidation state prediction to thermal conductivity modelling.
    ASE emphasises atomistic simulation using normalisation of crystalline and
    molecular data with similar basic functionality as their analogues in \pmg.
    Its calculator framework integrates quantum-chemical interoperability
    and in-built geometry optimisation and molecular dynamics functionality.
    
    \Pmg emphasises materials-analysis objects and Monty-based serialisation, while ASE centres on atomistic simulation through its calculator and dynamics interfaces. Both provide persistence facilities, and conversion between their structure representations enables complementary workflows.
    
    This distinction means that \pmg is often used in conjunction with ASE to
    orchestrate workflows where certain features (e.g., molecular dynamics driven
    by an interatomic potential) are only available in ASE.
    To streamline the usage of both codes, \pmg makes it possible to convert
    between the \pmg \texttt{Structure} object representation of a material and
    the ASE \texttt{Atoms} object representation.
    In a diverse and rapidly evolving research ecosystem, it is unrealistic and undesirable for a single code to dominate community adoption.
    Rather than striving for exclusivity, the focus should be on interoperability
    between codes, allowing users to combine the strengths of both software
    packages.
    In this way, it is on the code maintainers to develop and maintain good
    relationships with their fellow developers, and ensure this interoperability is as smooth as engineering constraints allow.
    
    
    This high degree of interoperability enables the seamless
    integration of computational materials analyses within diverse research
    workflows.
    To illustrate, one common materials science workflow for phonon calculations
    could involve:
    \begin{enumerate}
      \item Obtain a crystal structure from the ICSD \cite{zagorac2019recent} or
            Materials Project \cite{jain_commentary_2013, horton_2025_mp} database, either manually
            or automatically with the \pmg API tools.
      \item Use the symmetry and structure transformation tools in \pmg, built
            upon spglib \cite{togo2024spglib,spglibv2}, to reduce to the primitive
            unit cell.
      \item Write the primitive cell to the file format required by the chosen
            quantum chemistry code, as well as other calculation input files
            (optionally), using \pmg.
      \item Perform geometry relaxation with the chosen quantum chemistry code.
      \item Directly read in the relaxed structure to \pmg, and generate a set
            of supercells with displaced atoms using the
            \texttt{pymatgen.io.phonopy} interface.
      \item Calculate forces for each displaced supercell using the chosen quantum chemistry code, keeping the displaced atomic positions fixed.
      \item Parse the displaced-atom calculations with phonopy
            \cite{togo_firstprinciples_2023,phonopy-phono3py-JPCM} and compute atomic force constants.
      \item Plot the phonon dispersion with phonopy or the
            \texttt{pymatgen.io.phonopy} interface, which also supports phonon band-structure and density-of-states analysis. Analysis of Gr\"{u}neisen parameters additionally requires phonon calculations at multiple volumes.
    \end{enumerate}
    
    This is a simple example, but one that helps illustrate the utility of
    \pmg's comprehensive materials analysis functionality and wide-ranging
    interoperability.
    Computational materials research workflows now widely employ \pmg as both a
    core analysis toolkit and a seamless glue between quantum chemistry codes and
    advanced research software packages.
    
    While \pmg supports interfaces to many codes, it should be noted that these
    are driven in large part by the research routes pursued within the Materials
    Project.
    Thus, the interfaces to VASP and Q-Chem have been thoroughly developed, as they are regularly used in the Materials Project.
    Other interfaces to PACKMOL \cite{packmol}, ATAT \cite{atat}, or Zeo++ \cite{zeopp} have been developed to support workflows studying non-crystalline materials, configurationally disordered crystalline materials, or material porosity, respectively.
    These interfaces may overlap with those available in ASE or pyiron \cite{janssen_pyiron_2019}, but they are more often disjoint.
    We make no attempt to make a (necessarily outdated) list of codes to which
    \pmg natively interfaces, but we list a few which have been added or
    significantly enhanced since the first \pmg publication
    \cite{ong_python_2013} (alphabetically): ABINIT
    \cite{verstraete_Abinit2025_2025}, CP2K \cite{kuhne_cp2k_2020}, icet
    \cite{angqvist_icet_2019}, JDFTx \cite{sundararaman_jdftx_2017}, LAMMPS
    \cite{thompson_lammps_2022}, LOBSTER \cite{nelson_lobster_2020}, OpenFF
    \cite{wang_openff_2024}, phonopy \cite{togo_firstprinciples_2023,phonopy-phono3py-JPCM}, and Q-Chem
    \cite{krylov_qchem_2020}.
    The extensibility of \pmg allows for rapid development of interfaces to codes
    as needs arise.
    
    \section{Community Adoption and Impact}
    
\begin{figure*}[h]
    
        \centering
        \includegraphics[width=0.6\textwidth]{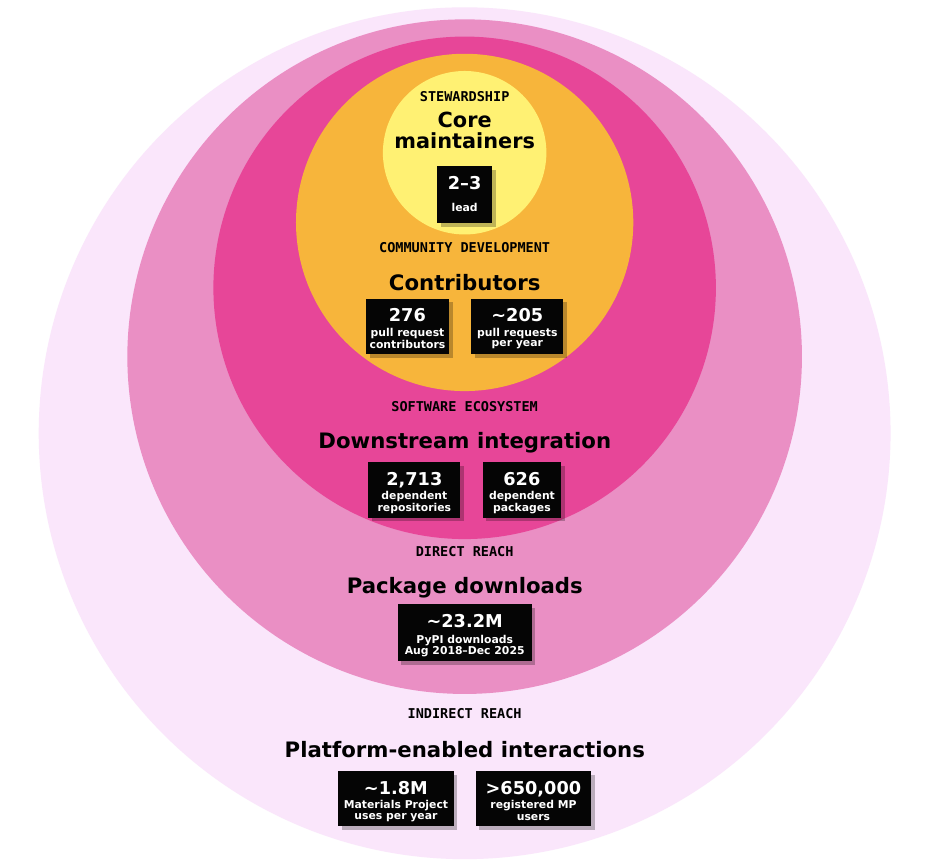} \caption{Different levels of influence within the \pmg ecosystem. From the center outward: maintainers, contributors, developers of downstream packages, direct users of \pmg, and indirect users via the Materials Project. See Supporting Information Section S2 for data sources and collection dates.} \label{fig:shells-of-influence} 
        \end{figure*}

    The growth of \pmg is reflected not only in its feature set but also in its expanding user base and community engagement across multiple materials sub-fields (Fig.~\ref{fig:mindmap}). Its applications span diverse fields, including energy storage and conversion,
    catalysis, defects, informatics, and materials discovery via machine learning
    \cite{jain_computational_2016}. 
    Further, the influence of \pmg extends across multiple layers of the materials informatics community, from maintainers and contributors to developers of downstream software packages, direct users of \pmg, and indirect users who benefit from its functionality through the Materials Project (Fig.~\ref{fig:shells-of-influence}).
    This broad impact is mirrored by sustained growth in adoption. The steadily increasing number of downloads and packages that require \pmg as a dependency provide evidence of widespread and increasing adoption across the materials science community.
    Since its publication in 2013, \pmg has been cited over 4000 times
    \cite{ong_python_2013}, reaching just under 1000 citations in 2025 alone (Google Scholar, accessed September 2026).
    \Pmg had approximately 23.2 million downloads from the Python Package
    Index (PyPI) between August 2018 and December 2025 and over 0.5 million per month in 2026, though we note that these numbers are likely inflated due to
    automated continuous integration (CI) testing.

    \subsection{Package Ecosystem}
    
\begin{figure*}[h]
    
        \centering
        \begin{subfigure}[b]{0.49\textwidth}
            \centering
            \includegraphics[width=\textwidth]{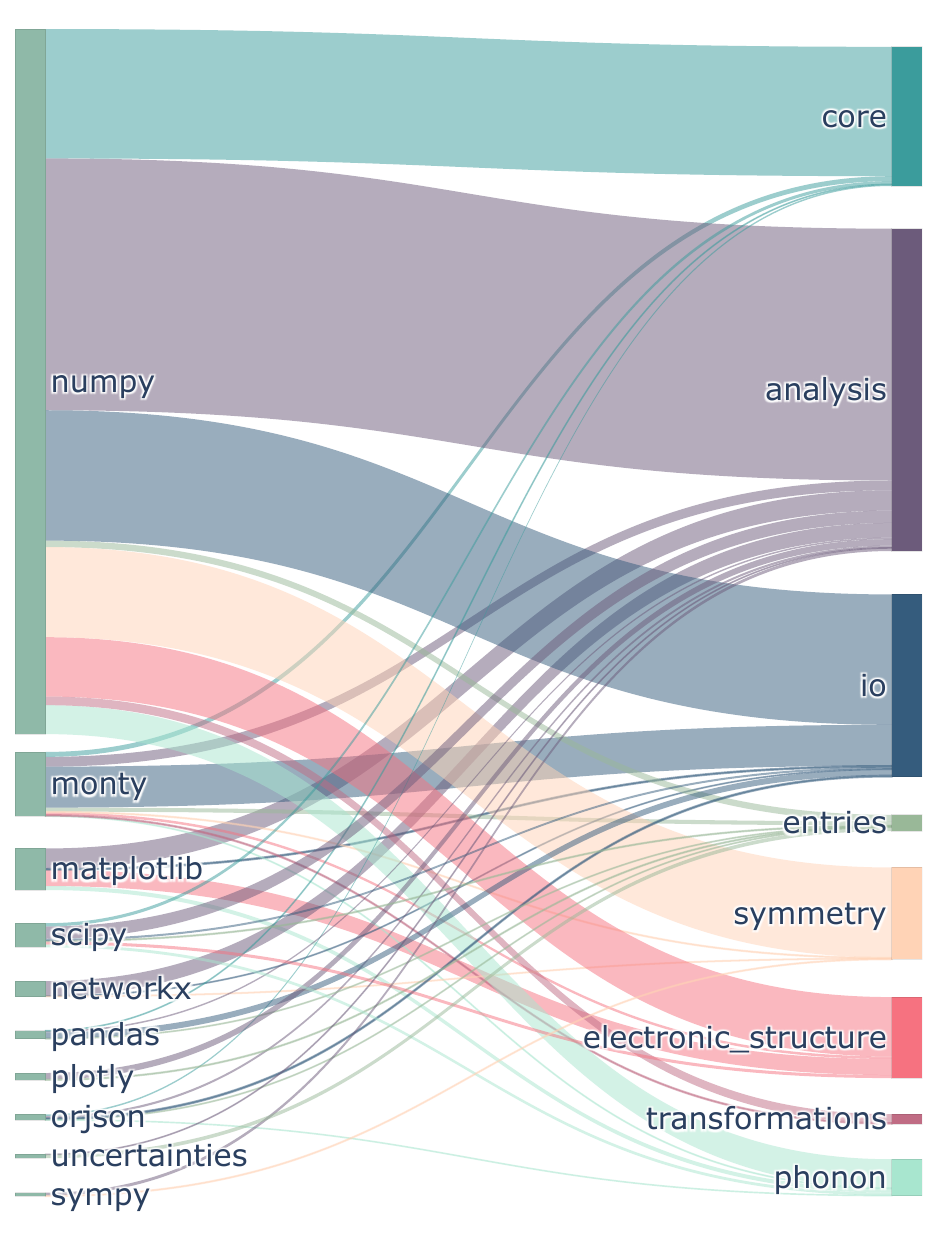}
            \caption{Upstream usage of third-party Python packages (left) by \pmg subpackages (middle).}
            \label{fig:pmg-dependency-usage}
        \end{subfigure}
        \hfill
        \begin{subfigure}[b]{0.49\textwidth}
            \centering
            \includegraphics[width=\textwidth]{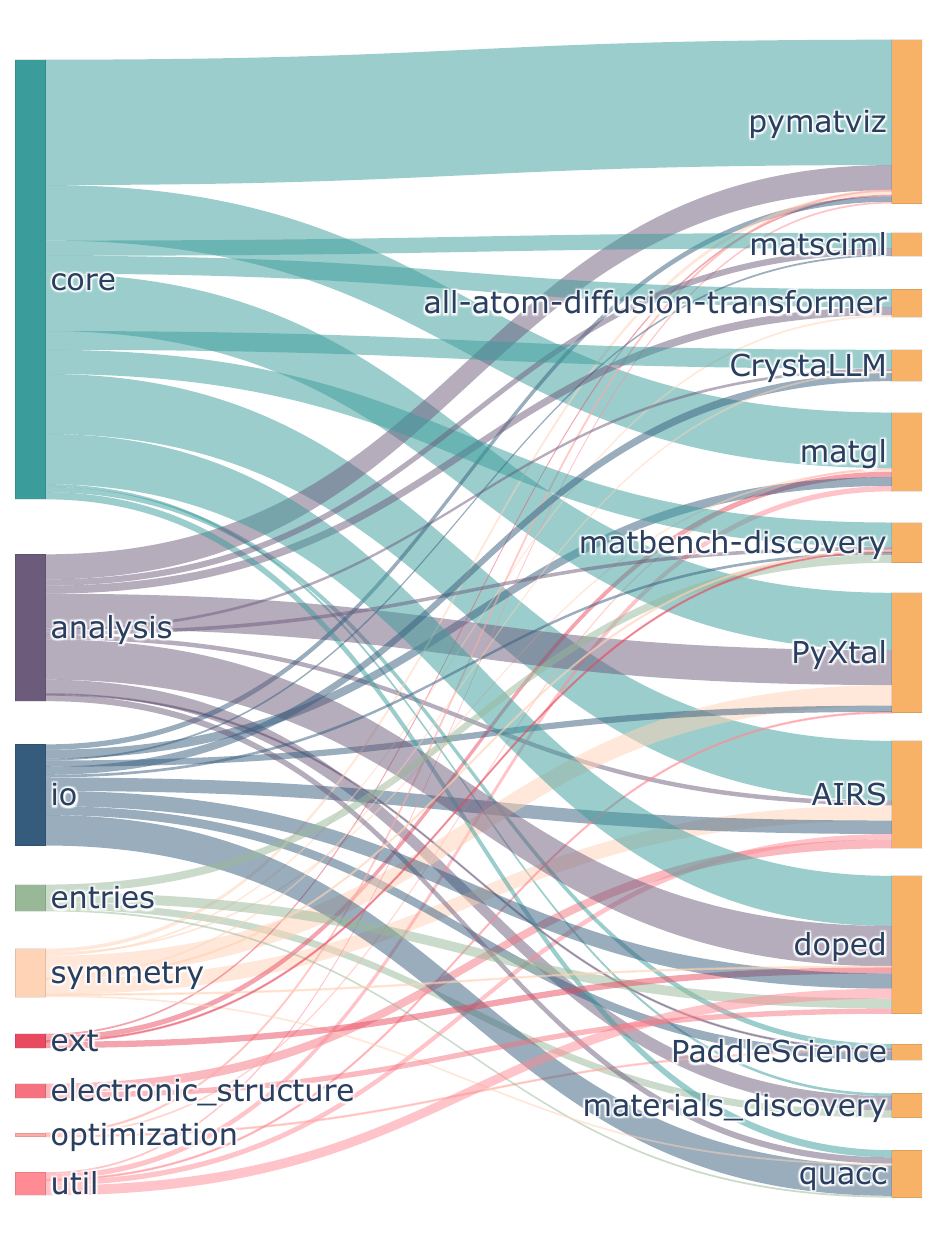}
            \caption{Downstream usage of \pmg subpackages (middle) by dependent packages (right).}
            \label{fig:dependent-usage-of-pmg}
        \end{subfigure}
        \caption{Upstream and downstream usage relationships of \pmg subpackages. Link width indicates the number of statically detected API call sites in the analysed source code. The panels show selected dependencies and downstream repositories; repository selection and filtering are described in Supporting Information Section S3. These subpackages broadly correspond to three common modes of \pmg usage: input/output (\texttt{io}), manipulation of materials objects (\texttt{core}), and analysis or simulation-enabling functionality (\texttt{analysis}).}
        \label{fig:package-ecosystem}
    \end{figure*}

    In addition to direct usage for common materials analyses as discussed in the previous section, many packages specialised for various computational materials science research areas have built upon the \pmg framework.
    Notable downstream packages, some of which are shown in Fig.~\ref{fig:dependent-usage-of-pmg}, include:
    
    \begin{enumerate}
      \item \href{https://github.com/materialsproject/atomate2}{atomate2}
            \cite{ganose_atomate2_2025},
            \href{https://github.com/materialsproject/custodian}{custodian},
            \href{https://quantum-accelerators.github.io/quacc/}{quacc}
            \cite{rosen_quacc_2025}: High-throughput computational materials science
            workflows.
      \item \href{https://github.com/materialsvirtuallab/pymatgen-analysis-diffusion}%
            {pymatgen-analysis-diffusion} \cite{deng_datadriven_2017}:
            Diffusion analysis suite.
      \item \href{https://doped.readthedocs.io/en/latest/}{doped}
            \cite{kavanagh_doped_2024},
            \href{https://shakenbreak.readthedocs.io/en/latest/}{ShakeNBreak}
            \cite{mosquera-lois_shakenbreak_2022}: Defect modelling and
            structure-searching, building on
            \href{https://materialsproject.github.io/pymatgen-analysis-defects}%
            {pymatgen-analysis-defects} \cite{shen_pymatgen-analysis-defects_2024}
            and \href{https://github.com/mbkumar/pycdt}{PyCDT}
            \cite{broberg_pycdt_2018}.
      \item \href{https://pymatviz.janosh.dev/}{pymatviz}
            \cite{riebesell_pymatviz_2022},
            \href{https://smact.readthedocs.io/en/latest/}{SMACT}
            \cite{davies_smact_2019}, \href{https://hackingmaterials.lbl.gov/matminer/}{matminer}
            \cite{ward_matminer_2018}: Materials informatics toolkits.
      \item \href{https://matbench-discovery.materialsproject.org/}%
            {matbench-discovery} \cite{riebesell_framework_2025},
            \href{https://www.air4.science/}{AIRS}
            \cite{zhangArtificialIntelligenceScience2025},
            \href{https://github.com/materialsvirtuallab/matgl}{MatGL}
            \cite{ko_materials_2025},
            \href{https://github.com/torchsim/torch-sim}{torch-sim}
            \cite{cohen_torchsim_2025}: Machine learning for materials science.
      \item \href{https://pyxtal.readthedocs.io/en/latest/}{PyXtal}
            \cite{fredericksPyXtalPythonLibrary2021}: Generation and manipulation
            of atomic structures with symmetry constraints.
      \item \href{https://jageo.github.io/LobsterPy/}{LobsterPy}
            \cite{naik_LobsterPypackage_2024}: Automatic analysis of
            LOBSTER outputs for summarised bonding information.
        \item \href{https://hackingmaterials.lbl.gov/amset/}{AMSET}
            \cite{Ganose2021}: Electronic transport property modelling.
    \end{enumerate}
    
    \vspace{1em}
    
    This diversity of downstream packages demonstrates \pmg's extensibility,
    serving a foundational role in the materials modelling and informatics
    ecosystems \cite{butler_machine_2018}.
    Of course, \pmg itself builds on foundational open-source Python packages,
    such as NumPy \cite{harris_array_2020}, SciPy
    \cite{virtanen_scipy_2020}, and pandas
    \cite{team_pandasdev_2025,mckinney_data_2010} for numerical operations,
    Matplotlib \cite{hunter_matplotlib_2007} and Plotly \cite{plotly} for plotting, spglib
    \cite{togo2024spglib,spglibv2} for symmetry analyses,  NetworkX \cite{hagberg2007exploring} for graph analysis, and monty for
    serialisation, as shown in Fig.~\ref{fig:package-ecosystem}.
    
    Cumulatively, \pmg's impact on the materials science community includes:
    \begin{enumerate}
      \item Accelerated research across various domains\cite{butler_machine_2018}, including the rapid development of specialised materials analysis toolkits (Fig.~\ref{fig:dependent-usage-of-pmg}).
      \item Standardisation of core materials analysis procedures.
      \item Improved reproducibility through user-friendly serialisation and analysis tools.
      \item Provision of accessible educational tools for students and early-career researchers \cite{ong_python_2013}, such as \url{https://github.com/materialsvirtuallab/matgenb} and \url{https://github.com/sp8rks/MaterialsInformatics}.
      \item Adoption in industrial research and development workflows \cite{jain_commentary_2013, horton_2025_mp}.
      \item Promotion of open science and collaborative development \cite{horton_promises_2021}.
    \end{enumerate}
    
    \subsection{Case Studies}
    
    As discussed above, many advanced materials analysis packages have been built upon the \pmg codebase to enable complex domain-specific research workflows.
    Below, we briefly discuss two such cases where the \pmg framework has facilitated advanced, complex materials analyses.
    
    \subsubsection{Defect Modelling Tools}
    
    Defect modelling is a rapidly growing field, driven by advances in computational power and methods which make these calculations increasingly tractable, along with the critical importance of defect species to diverse materials technologies, such as transistors, solar cells, transparent conducting materials, batteries, qubits and more.\cite{Yan2024}
    Defect simulation can be a long and complex process as it requires many different tasks;\cite{kumagai_insights_2021,kavanagh_doped_2024,freysoldt_firstprinciples_2014} including structure manipulations, symmetry analyses, input file generation and output parsing for electronic structure codes, lightweight metadata and serialisation for reproducibility, and interfaces with databases such as the Materials Project for phase diagram analysis.
    Moreover, many of these tasks have specific requirements for the specialised
    case of defects. 
    This includes efficient and appropriate supercell generation (to
    balance computational cost and accuracy), determination of defect point
    symmetries, configurational degeneracies, and site multiplicities (in
    symmetry-breaking supercells), targeted
    distortions for structure-searching, efficient algorithms for large and complex
    structure analyses, smart algorithms for sub-phase diagrams, and more \cite{squires_guidelines_2026}.
    
    Several open-source toolkits have now been developed, which make use of core
    functionalities in \pmg to implement stages of the defect modelling workflow,
    including doped \cite{kavanagh_doped_2024}
    (Fig.~\ref{fig:dependent-usage-of-pmg}), pydefect
    \cite{kumagai_insights_2021}, PyCDT \cite{broberg_pycdt_2018},
    pymatgen-analysis-defects \cite{shen_pymatgen-analysis-defects_2024},
    ShakeNBreak \cite{mosquera-lois_shakenbreak_2022}, AiiDA-defects
    \cite{muy_aiida-defects_2023} and pydecs \cite{ogawa_extended_2022}.
    These toolkits are seeing widespread usage in the defects modelling community, helping to
    accelerate research in this technologically crucial area, while reducing
    human errors and lowering the barrier to entry for new researchers.
    Here, the wide functionality and flexible, modular structure of \pmg has been
    a key enabler of these downstream developments, allowing the tailoring of \pmg
    functions for specialised use cases.
    
    One particularly promising aspect of these developments is that they have
    greatly reduced the burden for data sharing and reproducibility in defect
    simulations---a particularly challenging task given the many steps involved
    \cite{squires_guidelines_2026}.
    Through the object-oriented architecture and efficient serialisation tools
    provided by \pmg and its monty dependency for JSON (de)serialisation, the
    human effort to collate and export complete metadata and calculation provenance
    is significantly diminished.
    These developments are helping to establish clear community standards and
    expectations for reproducibility, strengthening the quality of defect modelling
    research while ensuring robust foundations for future database and machine
    learning efforts in this area \cite{squires_guidelines_2026}.
    
    \subsubsection{Data-Driven Heuristic Assessment and Machine Learning}
    
    \Pmg is a powerful toolkit that enables large-scale data analysis and machine
    learning studies in materials science.
    These capabilities are essential  for uncovering new design principles for specific material properties and for developing feature-based machine learning models, the latter frequently serving as a core component in Materials Acceleration Platforms (MAPs) and
    self-driving labs \cite{flores-leonar_materials_2020, stier_materials_2024}.
    
    Through \pmg's direct interfaces to material databases (Materials Project API,
    now spun off as its own package, or OPTIMADE API
    \cite{andersen2021optimade}), researchers can easily retrieve extensive datasets,
    including machine learning targets, which serve as the foundation for
    data-driven assessments of design principles and heuristics, as well as the
    development of machine learning models.
    When working with other crystal structure sources such as the Cambridge
    Structural Database \cite{groom2016cambridge}, Inorganic Crystal Structure
    Database \cite{zagorac2019recent} or MAGNDATA \cite{gallego2016magndata},
    \pmg's ability to parse (magnetic) Crystallographic Information Files (CIFs)
    directly into input objects needed for atomistic simulation facilitates the creation of large, structured datasets.
    \Pmg allows simplified filtering by composition or crystal symmetry, depending
    on requirements.
    Additionally, many post-processing features are available in \pmg, enabling
    the simpler generation of targets.
    
    Chemical heuristics have been frequently used to inspire design principles or
    features in machine learning models. Such features are extremely valuable, as they provide physically and chemically interpretable descriptors that not only improve model interpretability but can also serve as scientific consistency checks for machine-learning outputs.
    Many such descriptors and design principles---such as those related to stability, synthesisability, defect
    energetics, and magnetism---rely on accurate determinations of oxidation states
    and coordination environments, both of which can be assessed within \pmg.
    \Pmg implements several heuristics to determine oxidation states (e.g., based
    on bond valence sums or simple composition-based guessing methods) and to
    determine coordination environments (e.g., based on minimum distances or
    Hoppe's effective coordination numbers, and more advanced approaches implemented in the \texttt{local\_env} and \texttt{chemenv} modules) \cite{pan_benchmarking_2021,waroquiers_chemenv_2020}.
    Additionally, \pmg is deeply integrated with matminer
    \cite{ward_matminer_2018}, a library that streamlines the generation of popular
    features for machine learning workflows, further supporting the derivation and
    validation of heuristics.
    Here, \pmg formats are used directly as input for creating machine-learning
    features.
    
    In summary, \pmg provides a framework for materials data handling, feature
    extraction, and heuristic development, making it an essential tool for modern,
    data-driven materials informatics.
    By codifying chemical, crystallographic, and physical knowledge into reusable data structures, descriptors, and analysis workflows, \pmg transforms expert domain knowledge into a scalable resource for data-driven materials discovery.
    
    \subsubsection{Transferability and Lessons Learned \label{sec:addnl}}
    
    Numerous other case studies demonstrate the impact of \pmg across materials modelling sub-fields, such as in low-dimensional materials, surface slicing and analyses, high-throughput materials discovery and automated workflows, magnetic structure enumerations, among others.\\
    
    The success of subfield-specific modelling packages built on \pmg demonstrates that the provision of robust functionality for common tasks -- such as structure manipulations, oxidation state estimation or input/output to DFT codes -- allows the ready extension to more specialised research workflows by domain experts.
    By implementing these core materials analysis functions in a modular code structure, which can be readily imported into downstream packages, \pmg removes the need to re-implement common tasks and allows practitioners to focus on specialised domain-specific analyses (e.g., predicting defect concentrations). In this way, some functionalities have effectively become community reference points for specific tasks within the field (e.g., \texttt{StructureMatcher}), with methodological choices and default parameters adopted broadly through the software's widespread use. Packages such as \pmg, through their widespread use and expert oversight, can serve as important touchstones for the community, helping to promote consistent practices and methodologies. As a result, a natural form of standardisation can emerge, with widely adopted implementations shaping common approaches to computational materials analysis.
    Another aspect that has proven important to the success of these advanced simulation workflows is the use of computationally efficient implementations in core \pmg modules, through intelligent algorithms, vectorisation, Cython code, or other methods.
    This ensures efficiency and scalability in downstream uses, enabling complex workflows built on \pmg, such as high-throughput screening across immense materials search spaces or structural analyses of large defect supercells with low symmetry. At the same time, performance optimisations must be balanced against code readability, maintainability, and ease of understanding, as these are essential to encouraging community contributions and ensuring the project's long-term sustainability.
    
    Finally, clear communication about code deprecations and breaking changes is crucial for the smooth, uninterrupted development and maintenance of the many downstream scientific packages that rely on \pmg.
    
    \section{New Features}
    \label{sec:new-features}
    
    The growth of \pmg closely follows trends in materials science research,
    reflecting its critical role in advancing science.
    
    Modelling of Li- and Co-alternative battery electrodes has long driven atomistic
    analysis tools in \pmg.
    The pymatgen-analysis-diffusion extension \cite{shen_topological_2023}
    grew out of a smaller submodule of diffusivity analysis tools, which were
    eventually migrated to a separate namespace package.
    These tools enable faster estimates of ionic mobility from molecular dynamics
    trajectories.
    They also include tools to identify likely topotactic ion insertion defect
    sites based on electronic charge densities \cite{shen_charge_2020}, used when
    estimating ionic mobility from transition-state type mappings of the potential
    energy surface.
    
    More recently, machine learning (ML) methods \cite{butler_machine_2018},
    particularly ML interatomic potentials (MLIPs) \cite{chen_universal_2022}, have
    emerged as practical ways to rapidly estimate atomic structure without more
    costly recourse to electronic structure methods.
    \Pmg has integrated some of these structural prediction tools via the external
    MatGL Python package \cite{ko_materials_2025}.
    This allows users to, e.g., quickly relax crystal structures using common graph
    neural-network MLIPs such as M3GNet \cite{chen_universal_2022}, CHGNet
    \cite{deng_chgnet_2023}, TensorNet \cite{simeon_tensornet_2023}, and MatPES-derived models \cite{kaplan2025matpes}.
    MLIPs can also be directly used in \pmg to estimate small-cell ordered
    representations of disordered materials by enumeration of possible orderings,
    and selecting the predicted minimum energy configuration.
    
Analysis of local coordination environments has also been of critical importance in understanding the connections between atomic structure and materials characteristics.
These insights help identify structural motifs that lead to similar performance, e.g., frameworks which favour ionic mobility.
    \Pmg facilitates the analysis of electronic charge densities and electronic bonding by interfacing with LOBSTER \cite{nelson_lobster_2020}.
    \Pmg allows for computationally tractable analysis of local atomic coordination environments using, e.g., the ChemEnv \cite{waroquiers_chemenv_2020} and CrystalNN
    \cite{zimmermann_crystalnn_2020} methods, as discussed further below.
    CrystalNN can be further used to generate structural embeddings similar to the
    feature vectors of MLIPs, which in turn allow for rapid estimates of structural
    similarity \cite{zimmermann_crystalnn_2020} and automatic generation of
    human-readable descriptions of materials with the robocrys package
    \cite{ganose_robocrys_2019}.
    Local coordination environment analysis from ChemEnv, CrystalNN similarity
    estimates, and CrystalNN-based generated descriptions of materials are now
    standard analysis products of the Materials Project's build pipelines, and are currently distributed as part of its core data.
    
    Other extensions to \pmg, which can be separately installed under the
    \texttt{io} and \texttt{analysis} namespaces include, but are not limited to:
    validation of VASP \cite{kresse_ultrasoft_1999} calculations for compatibility
    with the Materials Project \cite{horton_2025_mp} via
    pymatgen-io-validation; orchestration and parsing of FHI-aims
    \cite{blum_fhiaims_2009}, Quantum ESPRESSO \cite{giannozzi_qe_2017}, and FLEUR
    \cite{wortmann_fleur_2023} electronic structure calculations via
    pymatgen-io-aims, pymatgen-io-espresso, and
    pymatgen-io-fleur, respectively; orchestration of classical molecular
    dynamics calculations with OpenMM \cite{eastman_openmm_2023} via
    pymatgen-io-openmm; core defect functions with
    pymatgen-analysis-defects \cite{shen_pymatgen-analysis-defects_2024}.
    A more complete list of extensions to \pmg is maintained in the add-ons
    section of the documentation.
    
\begin{figure*}[h!]
      \centering
      \begin{subfigure}[b]{0.5\linewidth}
        \centering
        \includegraphics[width=\linewidth]{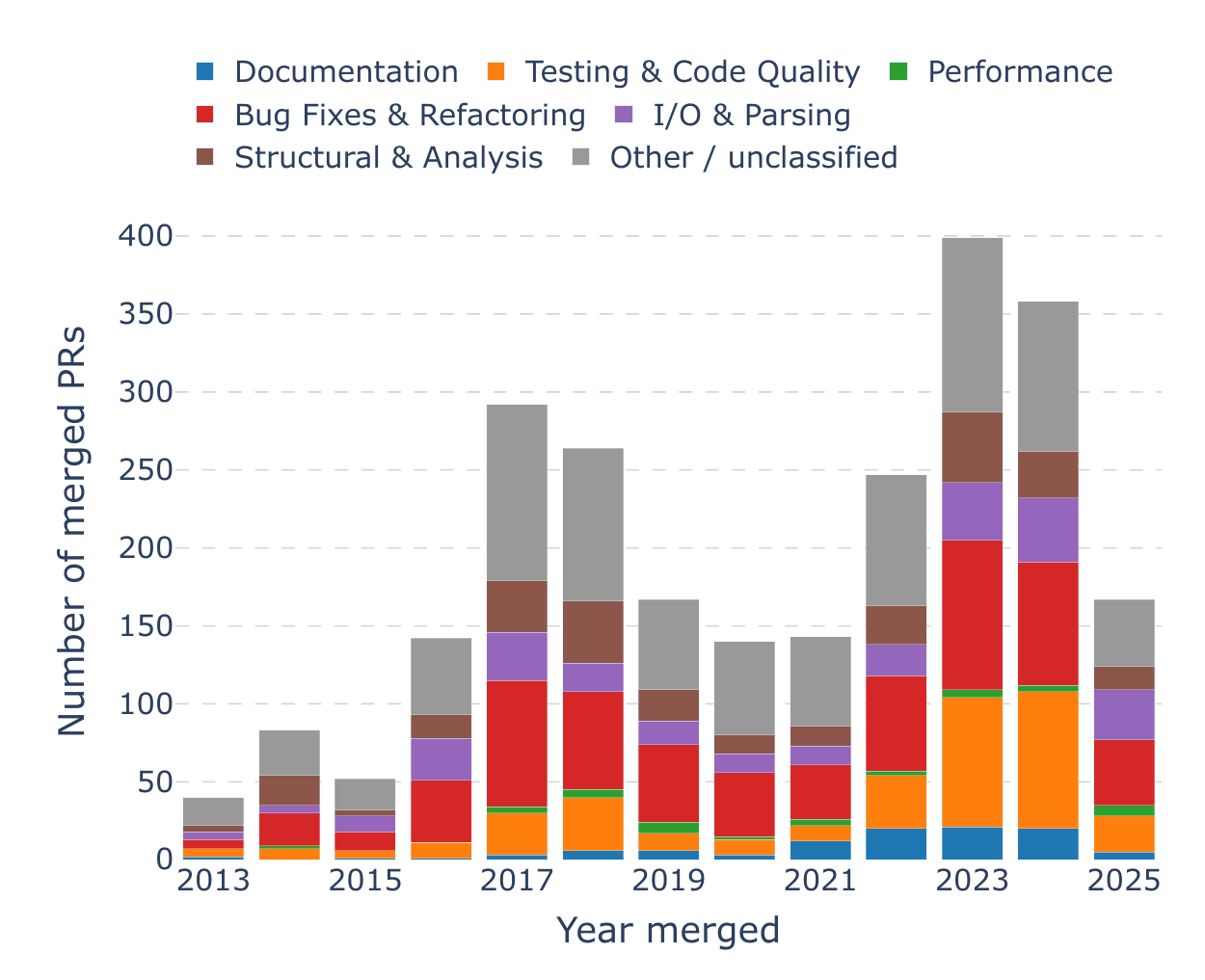}
        \caption{Merged non-bot pull requests by merger year and title-based theme.}
        \label{fig:pr-topics}
      \end{subfigure}
      \hfill
      \begin{subfigure}[b]{0.9\linewidth}
        \centering
        \includegraphics[width=\linewidth]{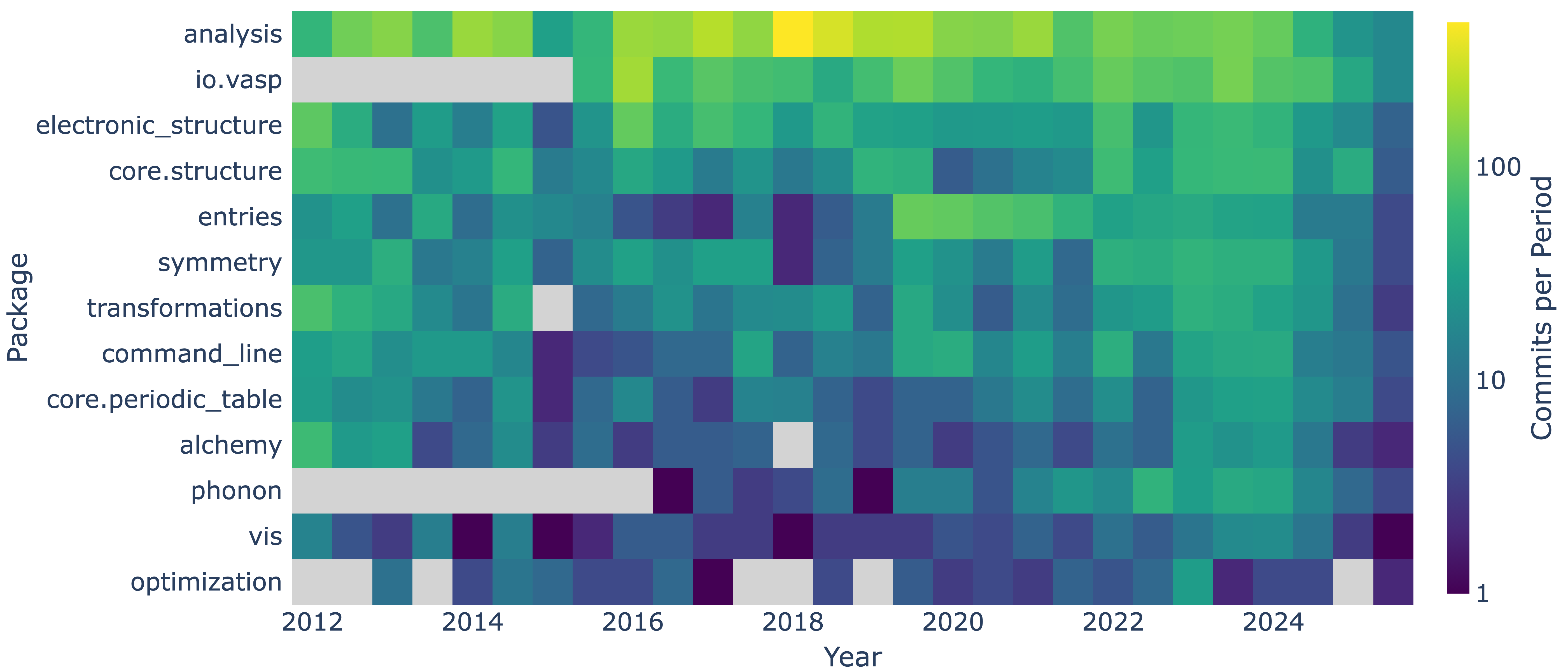}
        \caption{Number of commits per \pmg package over time, aggregated into calendar half-years and shown on a log-scaled colour scale.}
        \label{fig:commits-heatmap}
      \end{subfigure}
      \caption{Development activity in the \pmg repository. (a) All 2,494 non-bot pull requests merged through 31 December 2025, each assigned once by ordered keyword rules; unmatched titles are retained as Other / unclassified. Themes are approximate title-based categories. (b) Commits per package in calendar half-years (January--June and July--December) from 2012 through 2025.}
      \label{fig:development-activity}
    \end{figure*}

    As illustrated in Figs.~\ref{fig:pr-topics} and~\ref{fig:commits-heatmap},
    development activity has been sustained across multiple subpackages for over a
    decade, with contributions spanning bug fixes and refactoring, structural and I/O functionality, testing and code quality, documentation, and performance improvements.
    
    \section{Challenges and Solutions}
    
    Maintaining and evolving a project of \pmg's scale over more than a decade presents unique technical, organisational, and community challenges that offer valuable lessons for the broader scientific software community. Our experience highlights the importance of balancing sustainability, interoperability, governance, contributor onboarding, and community growth. We discuss key challenges encountered and possible approaches to addressing them.
    
    \subsection{Sustainability and Maintainer Resources}
    
    Perhaps the most critical challenge facing mature open-source scientific software is long-term sustainability.
    Unlike commercial software with dedicated development teams, \pmg has evolved primarily through volunteer contributions from academic researchers, where software development competes with research publications, teaching, and other career pressures.
    This model, while fostering community ownership, introduces structural tensions in maintenance, code review capacity, and long-term feature development.
    
    The Materials Project has provided crucial institutional support, hiring dedicated development and engineering staff, computational resources, and organisational infrastructure.
    By providing infrastructure and maintenance, \pmg has established itself as a host for analysis- and application-specific materials science code that might otherwise never have been made public or reusable.
    This hybrid model (combining institutional backing with community contributions)
    has proven essential for \pmg's longevity.
    Yet acquiring funding for software maintenance remains challenging, as grant agencies strongly favour novel research over infrastructure upkeep.
    Projects like FAIRmat \cite{fairmat_web_2025}, as part of the German national
    research data infrastructure, and the Molecular Sciences Software Institute
    (MolSSI) \cite{molssi_web_2025} represent important steps toward recognising
    computational and data infrastructure as crucial research infrastructure deserving of sustained investment.
    
\begin{figure*}[t]
      \centering
      \includegraphics[width=0.65\textwidth]{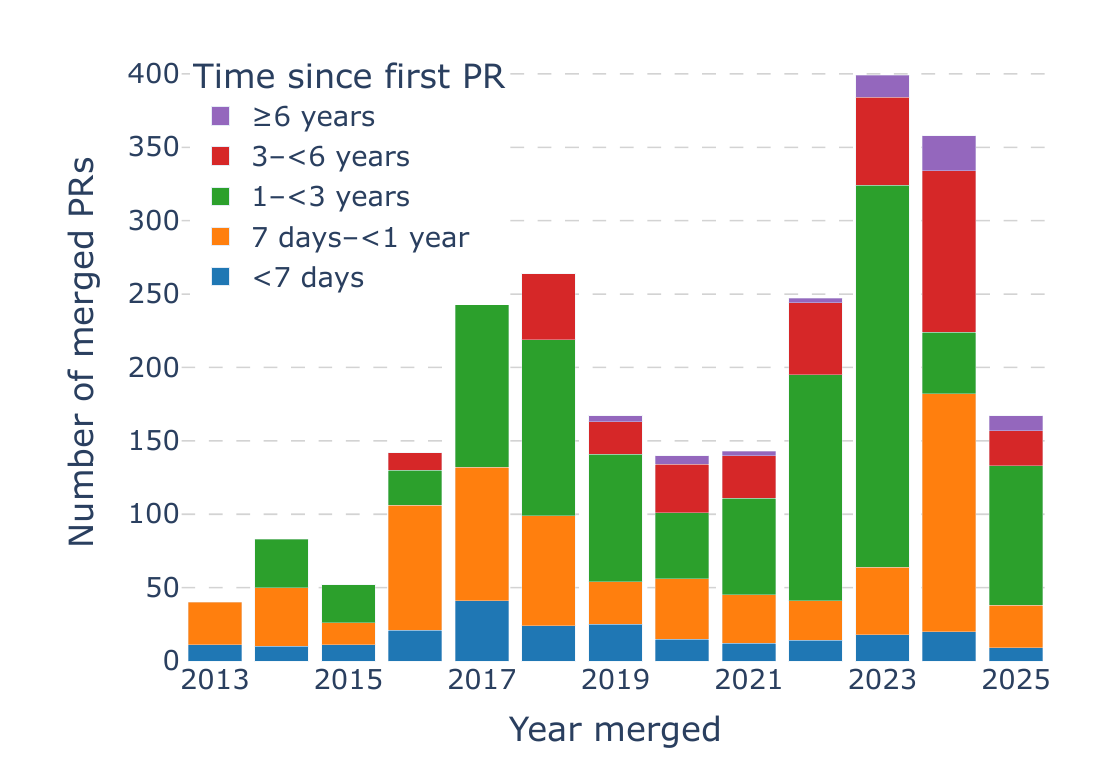}
      \caption{Annual merged non-bot \pmg pull requests grouped by contributor tenure at merger, measured from the author's first PR submission, including unmerged submissions. Years denote merger dates; all 2,494 PRs merged through 31 December 2025 are included, matching Fig.~\ref{fig:pr-topics}. Tenure bins are disjoint, with one year defined as 365.25 days.}
      \label{fig:pr-since-1st}
    \end{figure*}

    \Pmg has cultivated a globally distributed contributor base (Fig.~\ref{fig:contributors-worldmap}) with a continual influx of new contributors (Fig.~\ref{fig:pr-since-1st}), which affords
    some amount of resilience to changing resource situations at any one institution.
    However, the concentration of maintenance burden on a small number of core developers remains an ongoing concern.
    We hope that clear contribution guidelines, responsive code review, and explicit recognition of contributors (through authorship opportunities, acknowledgments in papers, and community visibility) can help sustain engagement and grow the pool of active maintainers.
    
    This incremental-development perspective aligns with Digital Discovery's recent Commit article concept \cite{aspuru_commit_2025}, which formalises citable reporting of improvements to prior software, data, and hardware
    publications.
    While the present manuscript contains a full retrospective, the underlying motivation is similar: incentivise and reward sustainable scientific software development through transparent, community-driven iteration.
    Sustained progress still depends on stable funding, explicit onboarding
    pathways, and incentive structures that recognize maintenance work. We will outline some strategies as part of this perspective.
    
    \subsection{Backwards Compatibility and API Evolution}
    
    A fundamental tension in mature software libraries is balancing API stability against the need for improvements.
    \Pmg has accumulated technical debt from early design decisions, e.g., inconsistent syntax for parsing Python objects from files, inefficient JSON representations of Python objects, complex object-oriented inheritance, and proliferation of similar objects.
    These are challenges currently known to \pmg developers and would require a careful assessment to ensure backwards compatibility without impeding further development.
    Since \pmg uses Calendar Versioning with the format \texttt{YYYY.MM.DD}, users currently need to consult the release notes to assess the upgrade impact.
    
    With thousands of users and hundreds of dependent packages, \pmg bears
    significant responsibility for API stability.
    Breaking changes can cascade through the ecosystem, disrupting research
    workflows and requiring coordinated updates across multiple projects.
    To manage this, \pmg is currently adopting established deprecation best practices: deprecated features emit a \texttt{FutureWarning} with clear removal timelines (typically 6--12 months), giving downstream maintainers time to adapt.
    Major changes are communicated through release notes, GitHub discussions, and the matsci.org forum.
    The namespace package architecture (\texttt{pymatgen-io-*},
    \texttt{pymatgen-analysis-*}) further enables specialised APIs to evolve
    independently, reducing the impact of breaking changes on the core library.
    
    For mature infrastructure libraries, backward compatibility is a community contract. Incremental deprecations with predictable removal timelines are imperative for evolving a codebase with a user base as large as \pmg while minimising the impact on downstream research.
    
    \subsection{Test Coverage and Code Quality}
    
\begin{figure*}[h!]
      \centering
      \includegraphics[width=\linewidth]{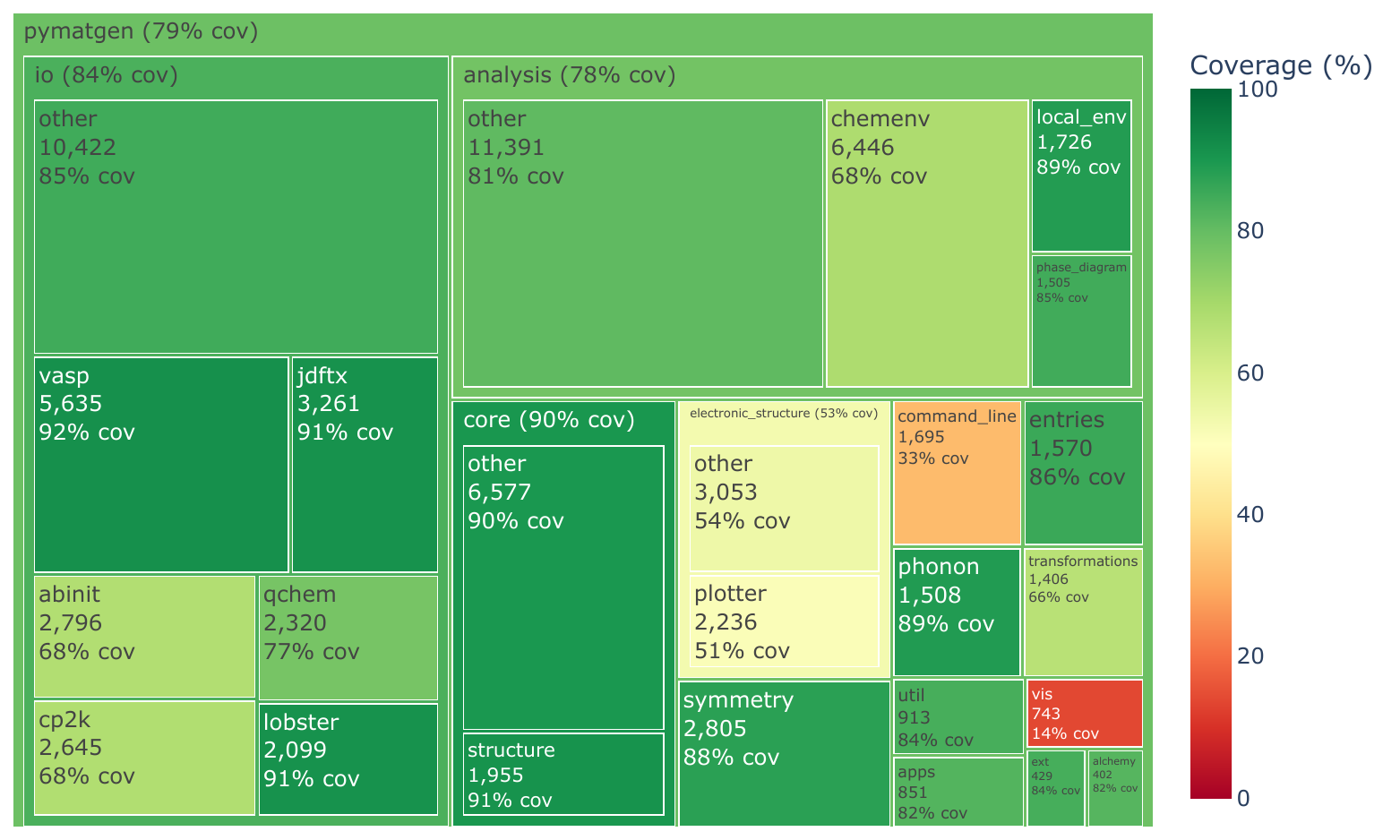}
      \caption{\Pmg Python code structure and line coverage in the 31 July 2025 coverage.py report. Cell area represents executable statements and colour represents the fraction covered by tests. Parent percentages use summed covered and executable statements. Modules are grouped to at most three path components, with groups below 1,500 statements combined into other cells and subpackages below 5,000 statements shown as single cells.}
      \label{fig:pmg-code-structure}
    \end{figure*}

    As shown in Fig.~\ref{fig:pmg-code-structure}, the 31 July 2025 coverage report records 60,242 covered statements out of 76,389 executable Python statements (78.9\%), though coverage varies significantly by module. Coverage is 90.2\% for \texttt{core}, 85.8\% for \texttt{entries}, and 83.6\% for \texttt{io}, while visualisation (\texttt{vis}) has 13.9\% coverage. Both cell areas and coverage percentages are derived from this report, with package coverage calculated from summed covered and executable statements.
    This heterogeneity reflects genuine trade-offs: some modules require licensed
    external executables (VASP, Gaussian) for meaningful testing, others involve
    long-running operations, and visualisation code requires rendering backends
    that complicate headless CI\@.
    Testing resources have been strategically focused on core modules serving the largest user bases.
    
    CI via GitHub Actions has been an important
    quality-control mechanism.
    The specifics of the CI have varied over time, but typically have included running the pytest test suite (or a comparable test suite) on every pull request across multiple Python versions.
    Test coverage is also tracked and reported.
    CI has benefited from the Python ecosystem's rapid migration of developer tools to Rust, which allows significant acceleration of \pmg's CI pipelines and local development workflows.
    In particular, static type analysis (e.g., mypy, Pyright) has helped catch subtle bugs during development
    time that would otherwise only surface during testing or production.
    For future advances, using the Rust-based type checker ty (once a stable release is minted) would be fast enough to run in tight feedback loops with LLM coding agents, catching type errors in seconds rather than minutes. 
    This is even true for large codebases like \pmg. If implemented, this could enable rapid iteration toward the correct code without waiting for much slower test suites to complete.
    Pre-commit hooks enforce these checks locally.
    However, unit tests are still necessary and the test suite's slow execution time, currently at over half an hour despite parallelisation, still creates friction for contributors.
    Improving test efficiency remains an ongoing priority.
    
    The emergence of frontier large language models (LLMs) is significantly reducing \pmg's maintenance burden.
    Tasks like writing tests, adding type annotations, improving docstrings, and refactoring legacy code can now be accelerated with LLM assistance.
    This is particularly valuable given the backlog of technical debt accumulated over 15 years.
    Incorporating AI-powered code review tools like
    \href{https://coderabbit.ai}{CodeRabbit},
    \href{https://cursor.com/bugbot}{Cursorbot} or \href{https://code.claude.com/docs/en/code-review}{Claude Code}
    can further reduce maintainer workload by providing automated first-pass reviews on pull requests.
    While early attempts at LLM-based code review on large codebases often introduced noise (e.g., overly verbose reviews) due to limited context windows and model capabilities, these tools have
    matured significantly and are now handling complex change sets reliably and catching real issues that would otherwise require slow and arduous manual inspection.
    LLMs also lower contribution barriers by helping newcomers navigate unfamiliar code.
    These AI-assisted tools have
    meaningfully increased throughput for \pmg's small maintainer team. At the same time, careful review remains critical to ensure that AI-assisted contributions meet community standards for correctness, maintainability, scientific reliability, and that contributors retain sufficient understanding of the resulting code to support its long-term stewardship.
    
    Quality at scale is best achieved through layered safeguards rather than any single metric.
    Fast static type and format checks, targeted issue-driven unit tests, and human-reviewed AI assistance together provide better
    reliability-per-maintainer-hour than focusing on any subset of these tools.
    
    \subsection{Governance and Contributor and User Onboarding}
    
    As a core component of the Materials Project's analysis tools, \pmg is
    governed according to principles established by the Materials Project
    stakeholders, which are in turn inspired by the best practices adopted by established community organisations like NumPy \cite{harris_array_2020} and NumFOCUS.
    Collectively, the MP members and software stakeholders responsible for adopting and enforcing codes of conduct were known as the Materials Project Software Foundation (MPSF). The MPSF worked to establish a democratic governance structure for the MP software ecosystem and, for example, facilitated the inclusion of MP software in the NumFOCUS foundation index.
    
\begin{figure*}[htbp]
      \centering
      \includegraphics[width=0.8\textwidth]{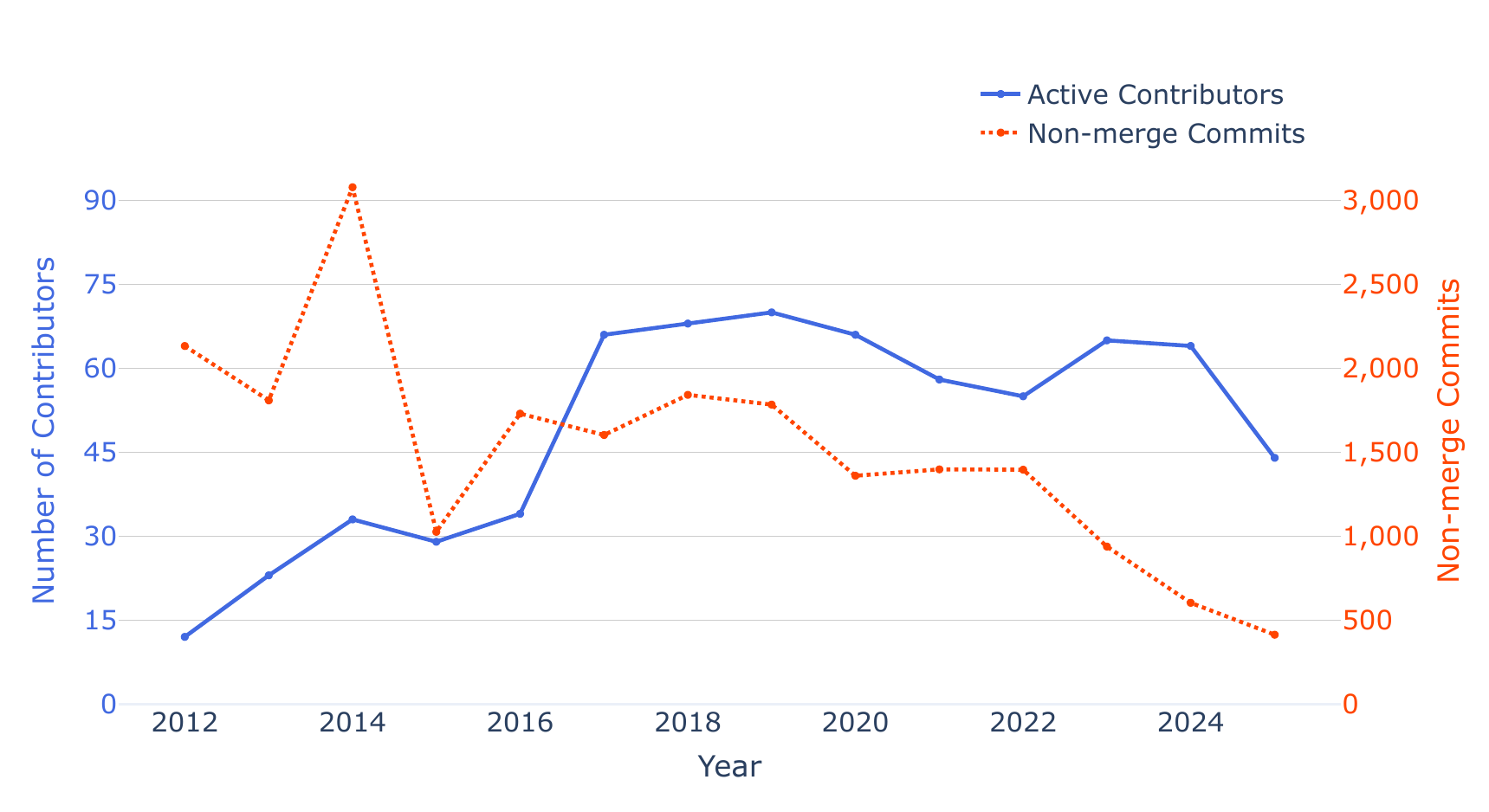}
      \caption{Number of active contributors and non-merge commits per year to \pmg from 2012 through 2025.}
      \label{fig:active-contributors}
    \end{figure*}

    These codes of conduct foster a welcoming environment for new contributors, aiming to cultivate longer-term community engagement that, in turn, benefits \pmg.
    As \pmg continues to expand in scope beyond the means of a small team of core maintainers, community engagement in maintaining its more niche feature sets is imperative. 
    Fostering a welcoming environment for new contributors is vital to ensuring \pmg's long-term sustainability. Over the last 10 years, the number of active contributors has generally ranged between approximately 30 and 75 per year, with the highest number of contributors in the year 2019 (Fig. \ref{fig:active-contributors}). The last 5 years have shown a slight decline in active contributors, making governance and onboarding strategies even more important. The same is true for the annual number of commits, even though commit sizes can vary significantly.
    
    Our experience suggests that successful community growth depends not only on attracting contributors but also on establishing processes that make it easy to submit, review, reproduce, and maintain contributions. We believe that \pmg's broad adoption cannot be attributed solely to its technical capabilities. A comparatively low barrier to contribution and a culture of responsive code review have played important roles in growing and sustaining the contributor community. Structured issue reporting, contribution templates, automated testing, and layered code review have all been important in enabling \pmg to scale beyond what could be supported by a small number of maintainers. Together, these practices help ensure that new functionality developed by the community can be incorporated efficiently, reducing friction for contributors and encouraging long-term engagement.
    
    As more junior community maintainers contribute to \pmg, they will eventually
    be invited to take on maintainer roles for sections of \pmg.
    Junior maintainers who express interest in maintaining the core feature set will be deputised by more experienced maintainers to handle quotidian maintenance.
    This simultaneously ensures that community-submitted issues are addressed in a timely fashion, and that more experienced maintainers can address larger questions of software direction, optimisation, etc.
    Because the analysis tools of \pmg are integral to MP's workflows and data pipelines, these features are virtually guaranteed to be maintained as long as MP continues its mission.
    However, the more peripheral features of \pmg will require this kind of community-driven maintenance structure to prevent them from becoming unmaintained and nonfunctional.
    
    An important lesson from \pmg's growth is that a healthy open-source ecosystem extends beyond code contributors. Community discussion forums and support channels such as the matsci.org forum play a critical role in user onboarding, dissemination of best practices, and the identification of future contributors and maintainers. These resources help scale support beyond the core development team's capacity while strengthening community engagement.
    
    Equally important has been the transfer of expertise through academic mentorship and research networks. Many new contributors and users were trained in groups that already had experience with \pmg, allowing knowledge, development practices, and community norms to be passed on to students, postdoctoral researchers, and collaborators. As these researchers moved to new institutions and established their own groups, they acted as multipliers, broadening adoption and cultivating the next generation of contributors and maintainers. This combination of formal community support structures and person-to-person knowledge transfer has been a key factor in sustaining \pmg's growth and impact.
    
    \subsection{Dependencies and Dependents}
    
    As discussed in previous sections, a large ecosystem of materials science codes
    depends on \pmg (see Fig.~\ref{fig:dependent-usage-of-pmg}), while \pmg itself depends on the open-source ecosystem for scientific
    computing, plotting, and data science
    (Fig.~\ref{fig:pmg-dependency-usage}).
    As a core component of the base layer of the computational materials science infrastructure that sits between foundational packages and hundreds of
    downstream dependents, \pmg's dependency management has ecosystem-wide
    implications.
    \Pmg follows the Scientific Python community's SPEC~0 policy
    \cite{spec0_minimum_supported_dependencies}, specifying minimum-version
    constraints, but avoiding upper-version caps to prevent resolver conflicts in
    downstream packages.
    When upstream libraries introduce breaking changes---as occurred with the
    NumPy~2.0 C~API transition \cite{harris_array_2020}---\pmg must adapt
    promptly, often ahead of its own dependents, to avoid blocking the wider
    ecosystem.
    This bidirectional responsibility is a recurring theme in maintaining community infrastructure: changes propagate both up and down the dependency graph, and
    delayed adaptation in either direction compounds across the ecosystem,
    underscoring that widely adopted software like \pmg requires a responsive maintainer team to avoid becoming a bottleneck for the broader ecosystem.

    \begin{figure*}[h]
      \centering
      \includegraphics[width=0.9\linewidth]{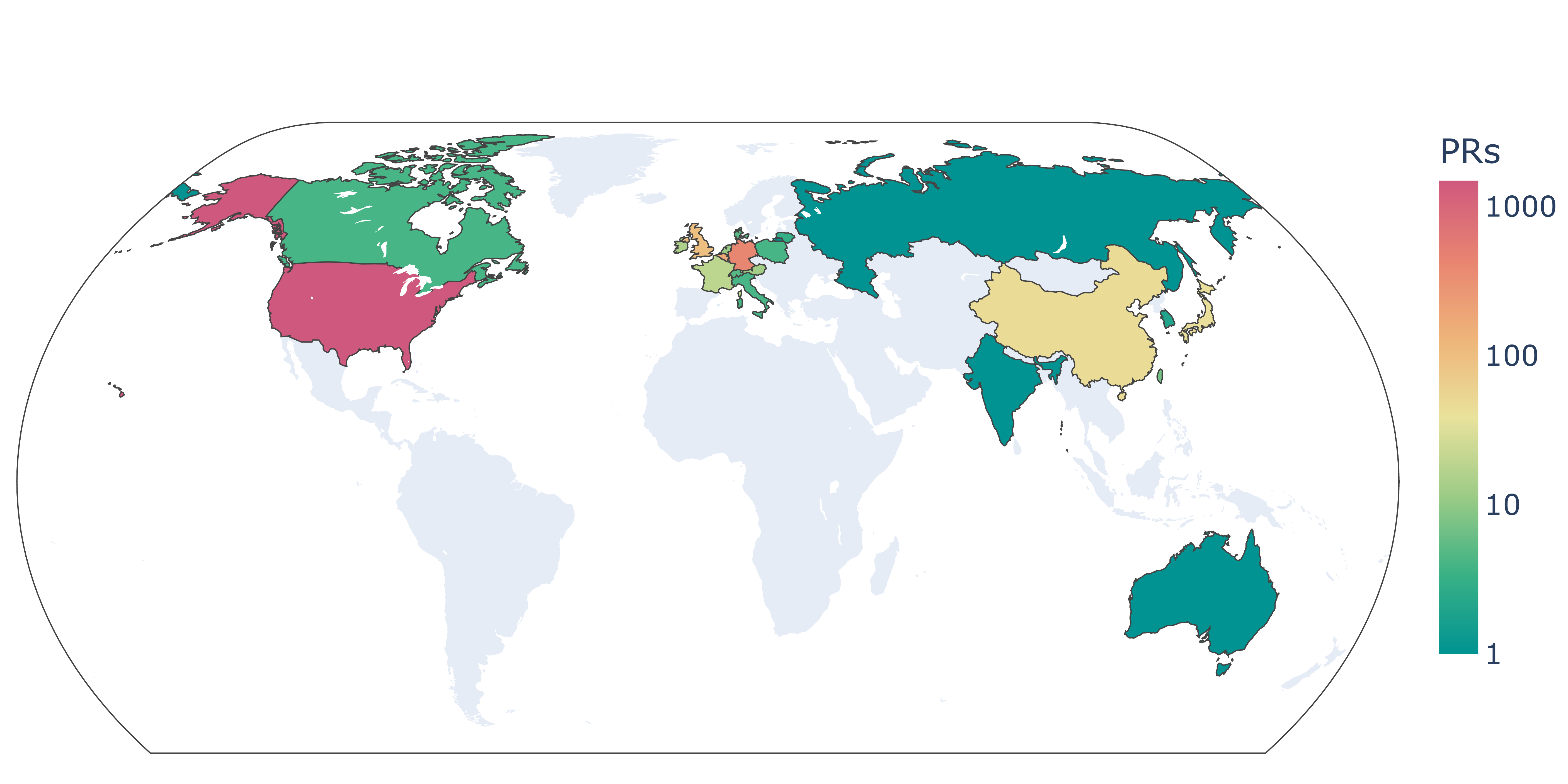}
      \caption{Geographic distribution of \pmg pull requests by contributor country. Countries were inferred from GitHub profile and affiliation snapshots, supplemented by manually curated mappings; these do not necessarily represent locations at the time of each contribution. The map omits 67 of 276 contributors with no country assignment, accounting for 136 of 2,498 merged PRs. See Supporting Information Section S6 for data sources and collection dates.}
      \label{fig:contributors-worldmap}
    \end{figure*}
    
    \section{Outlook and Future Developments}
    
    \Pmg has evolved from a specialised tool into a cornerstone of computational materials science.
    Its growth reflects the collaborative spirit of the materials informatics community and the importance of open-source software in scientific research.
    
    While the core features of \pmg are relatively stable, more peripheral
    features require constant community review beyond the stewardship of a small team of maintainers.
    Improving performance of the most essential \pmg libraries is key to its
    growth. \Pmg introduced namespace packages in
    \href{https://github.com/materialsproject/pymatgen/commit/f462d35c}{v2022.0.0} (March 2021), which allowed the independent installation and maintenance of \pmg ``add-on'' packages within the \pmg ``namespace'', i.e.\ importable
    from \texttt{pymatgen.io} and \texttt{pymatgen.analysis}.
    Critically, this decouples the release cadence of add-on packages from that of the core library---enabling the higher release frequency often needed during a package's early development without being blocked by the slower, more deliberate release cycle of \pmg itself.
    This is further supported by the standalone \texttt{pymatgen-core} package, now a dependency of \pmg, which provides fundamental materials science data structures together with I/O and analysis functionality for other developers to build upon.
    
    At its current scope, expanding and refining the feature set of \pmg would benefit from a large development team to ensure effective stewardship.
    As \pmg continues to grow in scope and adoption, we frame the next steps as high-impact community opportunities.
    In particular, broader shared ownership of maintenance tasks (code review, issue triage, and bug fixes in niche submodules) can strengthen participation and long-term stewardship.
    As contributors gain experience in these roles, they can naturally grow into deputy reviewer and maintainer positions.
    
    There are many active areas of materials science that could benefit from extensions of \pmg's key feature set.
    These include a wider variety and greater depth of control over \texttt{io} extensions for popular electronic structure (e.g., Quantum ESPRESSO), atomistic (e.g., LAMMPS, GROMACS), and continuum, phase-field and multiphysics codes.
    However, as noted previously, any extension of functionality that is not in a separately-installable namespace would require concomitant community review.
    
    While community forums such as \url{https://matsci.org} and the issue
    tracker for \pmg provide important support, there is significant opportunity to scale the documentation to match the library's impressive breadth. A promising direction for this expansion is a community-reviewed, LLM-assisted workflow,  which could improve coverage without increasing the burden on core maintainers.
    
    High-impact development opportunities informed by community needs and emerging trends in materials science include:
    
    \textbf{Machine learning integration}
    is likely to remain important as ML methods increasingly enter materials
    discovery and design \cite{butler_machine_2018}.
    High-value contributions include native support for modern frameworks (PyTorch, JAX), efficient structure-to-graph conversions for neural network architectures, and stronger integration with benchmarking frameworks like Matbench Discovery
    \cite{riebesell_framework_2025}.
    
    \textbf{Visualisation and user experience}
    improvements through tighter integration with Crystal Toolkit
    \cite{horton_crystal_2023}, pymatviz \cite{riebesell_pymatviz_2022},
    and MatterViz \cite{riebesell_matterviz_2025} could lower barriers to
    entry.
    Useful community efforts include enhanced Jupyter notebook support, interactive
    structure manipulation, and publication-ready figure generation.
    
    \textbf{Performance optimisations}
    remain a major opportunity, especially via strategic use of compiled extensions.
    \Pmg now operates in a different regime compared to its inception: in early
    development, \pmg was used to orchestrate much-slower DFT calculations, and
    the additional overhead from Python and its heavily object-oriented software structure was negligible.
    However, today, with the rise of very rapid ML forcefield simulation capabilities, this overhead is significant, and deeper optimisation is increasingly valuable.
    A successful example is given by the moyo project \cite{moyo}, which involved a Rust migration of symmetry analysis tools in spglib while maintaining an easy-to-use Python API via bindings.
    Similar approaches could accelerate often-used compute-intensive operations such as structure matching, neighbour searching, convex hull calculations, and more.
    
    \textbf{Extension package ecosystem}
    growth will strengthen \pmg's federated architecture: a stable core
    supplemented by domain-specific namespace packages (e.g.,
    pymatgen-analysis-diffusion \cite{deng_datadriven_2017},
    pymatgen-io-openmm) maintained by experts.
    Increasing the number and quality of these extensions distributes maintenance burden while enabling rapid iteration in specialised domains. New functionality areas---such as multi-scale modelling, bridging atomistic and
    mesoscale simulations---are best developed as extension packages rather than expanding the core.
    
    \textbf{GPU compatibility:}
    A promising direction is broader support for tensor backends (e.g., PyTorch) in array-heavy code paths, enabling GPU acceleration and potential speedups for large-scale analyses.
    
    \textbf{Agentic AI tools:}
    \Pmg functions could be exposed to LLM-based agents through tool interfaces, including Model Context Protocol (MCP) servers. Monty serialisation can facilitate data exchange, while explicit tool implementations, schemas, and input validation support reliable tool use. These tools can provide access to chemistry, materials science, and physical analysis capabilities to support agent workflows.
    
    We invite contributions at all levels---from bug reports and documentation
    improvements to new features and extension packages.
    The project's GitHub repository, documentation site, and community forums
    provide entry points for new contributors.
    
    \section*{Use of AI tools}
    Large language models were used in parts of the analysis: OpenAI GPT-5 grouped OpenAlex topics into thematic branches for Fig.~\ref{fig:mindmap} (Supporting Information Section S1), and OpenAI GPT-4o mapped institutional affiliations to countries for contributors without a GitHub profile location in Fig.~\ref{fig:contributors-worldmap} (Supporting Information Section S6). Copilot GPT-5.6 was used to unify the author contribution statements drafted by each co-author. Copilot was also used for brainstorming, translation, language editing, text revision, and proofreading during manuscript preparation. The authors reviewed, edited where necessary, and verified all text. LLM coding assistants, including Anthropic's Claude, helped write the figure and analysis scripts and were used to check consistency between the main text, figures, and Supporting Information. All outputs were checked by the authors.

    \section*{Data availability}
    Code to reproduce the analysis and figures presented in this manuscript is available at:
    https://github.com/janosh/pymatgen-2-paper
    
    
    \section*{Author contributions}
    Writing -- Original Draft, Writing -- Review \& Editing: Janine George, Matthew K. Horton, Aaron D. Kaplan, Seán R. Kavanagh, Janosh Riebesell, Andrew S. Rosen, Haoyu Yang, Anubhav Jain, Shyue Ping Ong. Methodology and Software -- all authors. Software contributions were made to the openly accessible repository available on GitHub. For more information on detailed software contributions from all authors, please refer to the supporting information.
    
    \section*{Conflicts of Interest}
    Anthony Onwuli is currently an employee of MatNex and holds options in the company. His software contributions to \pmg were made prior to joining MatNex. Michael Wolloch is an employee of VASP Software GmbH. 
    
    \section*{Acknowledgements}
    
    We thank the numerous developers, researchers, and users who have contributed
    to \pmg over the years.
    We also acknowledge the funding agencies and institutions that have supported
    its development, including the U.S.\ Department of Energy, the National Science
    Foundation, and various academic institutions.
    
    Special thanks is given to the broader open-source scientific computing
    community for their invaluable tools and libraries, especially NumFOCUS and
    their efforts around NumPy \cite{harris_array_2020}, pandas
    \cite{mckinney_data_2010,team_pandasdev_2025} and Matplotlib
    \cite{hunter_matplotlib_2007}, all of which \pmg heavily relies on.
    
    This work was intellectually led and primarily supported by the Materials Project, funded by the U.S. Department of Energy under award DE-AC02-05CH11231  (Materials Project program KC23MP).
    Tom Demeyere, Janine George, Aakash Naik were supported by ERC Grant MultiBonds (grant agreement Nº 101161771; Funded by the European Union. Views and opinions expressed are, however, those of the author(s) only and do not necessarily reflect those of the European Union or the European Research Council Executive Agency. Neither the European Union nor the granting authority can be held responsible for them.) Furthermore, Tom Demeyere, Janine George, Aakash Naik, Katharina Ueltzen would like to acknowledge the Gauss Centre for Supercomputing e.V. (www.gausscentre.eu) for funding the code developments by providing generous computing time on the GCS Supercomputer SuperMUC-NG at Leibniz Supercomputing Centre (www.lrz.de) (Project No. pn73da). Janine George is also part of the NFDI consortium FAIRmat funded by the Deutsche Forschungsgemeinschaft (DFG, German Research Foundation) – project 460197019 .
    Seán R. Kavanagh acknowledges membership of the UK's HEC Materials Chemistry Consortium, which is funded by EPSRC (EP/X035859), which provided access to the ARCHER2 UK National Supercomputing Service (http://www.archer2.ac.uk) for code developments.
    Andrew S. Rosen acknowledges support from the Princeton Office for the Dean for Research (ODFR) and Meta. Gian-Marco Rignanese is Research Director of the Fonds de la Recherche Scientifique - FNRS. 
    Haoyu Yang was supported by the NFDI consortium FAIRmat - Deutsche Forschungsgemeinschaft (DFG) - Project 460197019.
    Noam Bernstein's work was supported by the U. S. Naval Research Laboratory's base program.
    Brandon Bocklund's work was performed under the auspices of the U.S. Department of Energy by Lawrence Livermore National Laboratory (LLNL) under Contract DE-AC52-07NA27344.
    Part of the work at LLNL was funded by the Laboratory Directed Research and Development Program under project tracking code 26-SI-001.
    Matthew C. Kuner was supported by the National Science Foundation Graduate Research Fellowship Program under Grant No. DGE-2146752.
    Ihor Radchenko was supported by KEYTECH grant (project number WPC3/2022/50/KEYTECH/2024, National Centre for Research and Development, Poland), and by Teaming Horizon 2020 programme (GA. No. 857543) of the European Commission.
    Nav Nidhi Rajput's work was supported by the U.S. National Science Foundation (CSSI-2514002).
    Kaifeng Zheng was supported as part of the Molten Salts in Extreme Environments Energy Frontier Research Center, funded by the U.S. Department of Energy (DOE), Office of Science, through a subcontract from Brookhaven National Laboratory, which is operated under DOE Contract No. DE-SC0012704.    
    Shyue Ping Ong also acknowledges funding support from the National University of Singapore Materialyze.AI Startup Grant under WBS No. A-0010359-00-00.
    
    
    
    \balance
    
    
    \bibliography{refs} 

@article{Ganose2021,
  author  = {Ganose, Alex M. and Park, Junsoo and Faghaninia, Alireza and Woods-Robinson, Rachel and Persson, Kristin A. and Jain, Anubhav},
  title   = {Efficient calculation of carrier scattering rates from first principles},
  journal = {Nature Communications},
  year    = {2021},
  volume  = {12},
  number  = {1},
  pages   = {2222},
  doi     = {10.1038/s41467-021-22440-5},
  url     = {https://doi.org}
}

@article{jain_high-throughput_2011,
	title = {A high-throughput infrastructure for density functional theory calculations},
	volume = {50},
	issn = {0927-0256},
	url = {https://www.sciencedirect.com/science/article/pii/S0927025611001133},
	doi = {10.1016/j.commatsci.2011.02.023},
	number = {8},
	urldate = {2026-03-17},
	journal = {Computational Materials Science},
	author = {Jain, Anubhav and Hautier, Geoffroy and Moore, Charles J. and Ping Ong, Shyue and Fischer, Christopher C. and Mueller, Tim and Persson, Kristin A. and Ceder, Gerbrand},
	month = jun,
	year = {2011},
	pages = {2295--2310},
}

@article{butler_machine_2018,
  title = {Machine Learning for Molecular and Materials Science},
  author = {Butler, Keith T. and Davies, Daniel W. and Cartwright, Hugh and Isayev, Olexandr and Walsh, Aron},
  year = {2018},
  journal = {Nature},
  shortjournal = {Nature},
  volume = {559},
  number = {7715},
  eprint = {30046072},
  eprinttype = {pubmed},
  pages = {547--555},
  issn = {1476-4687},
  doi = {10.1038/s41586-018-0337-2}
}

@article{cohen_torchsim_2025,
  title = {{{TorchSim}}: {{An}} Efficient Atomistic Simulation Engine in {{PyTorch}}},
  shorttitle = {{{TorchSim}}},
  author = {Cohen, Orion Archer and Riebesell, Janosh and Goodall, Rhys and Kolluru, Adeesh and Falletta, Stefano and Krause, Joseph and Colindres, Jorge and Ceder, Gerbrand and Gangan, Abhijeet S},
  year = {2025},
  journal = {AI for Science},
  shortjournal = {AI Sci.},
  issn = {3050-287X},
  doi = {10.1088/3050-287X/ae1799},
  url = {http://iopscience.iop.org/article/10.1088/3050-287X/ae1799},
  urldate = {2025-11-01},
  pages = {025003},
}

@article{curtarolo_highthroughput_2013,
  title = {The High-Throughput Highway to Computational Materials Design},
  author = {Curtarolo, Stefano and Hart, Gus L. W. and Nardelli, Marco Buongiorno and Mingo, Natalio and Sanvito, Stefano and Levy, Ohad},
  year = {2013},
  journal = {Nature Materials},
  shortjournal = {Nature Mater},
  volume = {12},
  number = {3},
  pages = {191--201},
  publisher = {Nature Publishing Group},
  issn = {1476-4660},
  doi = {10.1038/nmat3568},
  url = {https://www.nature.com/articles/nmat3568},
  urldate = {2025-05-21}
}

@article{deng_datadriven_2017,
  title = {Data-{{Driven First-Principles Methods}} for the {{Study}} and {{Design}} of {{Alkali Superionic Conductors}}},
  author = {Deng, Zhi and Zhu, Zhuoying and Chu, Iek-Heng and Ong, Shyue Ping},
  date = {2017-01-10},
  journal = {Chemistry of Materials},
  shortjournal = {Chem. Mater.},
  volume = {29},
  number = {1},
  pages = {281--288},
  publisher = {American Chemical Society},
  issn = {0897-4756},
  doi = {10.1021/acs.chemmater.6b02648},
  url = {https://doi.org/10.1021/acs.chemmater.6b02648},
  urldate = {2025-05-21},
  year = {2017}
}

@misc{fairmat_web_2025,
  author = {{FAIRmat}},
  note = {https://www.fairmat-nfdi.eu/fairmat/},
  urldate = {2025-10-30}
}

@article{gonze_abinit_2020,
title = {The Abinitproject: Impact,  environment and recent developments},
volume = {248},
ISSN = {0010-4655},
url = {http://dx.doi.org/10.1016/j.cpc.2019.107042},
DOI = {10.1016/j.cpc.2019.107042},
journal = {Computer Physics Communications},
publisher = {Elsevier BV},
author = {Gonze,  Xavier and Amadon,  Bernard and Antonius,  Gabriel and Arnardi,  Frédéric and Baguet,  Lucas and Beuken,  Jean-Michel and Bieder,  Jordan and Bottin,  Fran\c{c}ois and Bouchet,  Johann and Bousquet,  Eric and Brouwer,  Nils and Bruneval,  Fabien and Brunin,  Guillaume and Cavignac,  Théo and Charraud,  Jean-Baptiste and Chen,  Wei and C\^oté,  Michel and Cottenier,  Stefaan and Denier,  Jules and Geneste,  Grégory and Ghosez,  Philippe and Giantomassi,  Matteo and Gillet,  Yannick and Gingras,  Olivier and Hamann,  Donald R. and Hautier,  Geoffroy and He,  Xu and Helbig,  Nicole and Holzwarth,  Natalie and Jia,  Yongchao and Jollet,  Fran\c{c}ois and Lafargue-Dit-Hauret,  William and Lejaeghere,  Kurt and Marques,  Miguel A.L. and Martin,  Alexandre and Martins,  Cyril and Miranda,  Henrique P.C. and Naccarato,  Francesco and Persson,  Kristin and Petretto,  Guido and Planes,  Valentin and Pouillon,  Yann and Prokhorenko,  Sergei and Ricci,  Fabio and Rignanese,  Gian-Marco and Romero,  Aldo H. and Schmitt,  Michael Marcus and Torrent,  Marc and van Setten,  Michiel J. and Van Troeye,  Benoit and Verstraete,  Matthieu J. and Zérah,  Gilles and Zwanziger,  Josef W.},
year = {2020},
month = mar,
pages = {107042}
}

@article{harris_array_2020,
  title = {Array programming with NumPy},
  volume = {585},
  ISSN = {1476-4687},
  url = {http://dx.doi.org/10.1038/s41586-020-2649-2},
  DOI = {10.1038/s41586-020-2649-2},
  number = {7825},
  journal = {Nature},
  publisher = {Springer Science and Business Media LLC},
  author = {Harris,  Charles R. and Millman,  K. Jarrod and van der Walt,  Stéfan J. and Gommers,  Ralf and Virtanen,  Pauli and Cournapeau,  David and Wieser,  Eric and Taylor,  Julian and Berg,  Sebastian and Smith,  Nathaniel J. and Kern,  Robert and Picus,  Matti and Hoyer,  Stephan and van Kerkwijk,  Marten H. and Brett,  Matthew and Haldane,  Allan and del Río,  Jaime Fernández and Wiebe,  Mark and Peterson,  Pearu and Gérard-Marchant,  Pierre and Sheppard,  Kevin and Reddy,  Tyler and Weckesser,  Warren and Abbasi,  Hameer and Gohlke,  Christoph and Oliphant,  Travis E.},
  year = {2020},
  month = sep,
  pages = {357–362}
}

@misc{horton_crystal_2023,
  title = {Crystal {{Toolkit}}: {{A Web App Framework}} to {{Improve Usability}} and {{Accessibility}} of {{Materials Science Research Algorithms}}},
  shorttitle = {Crystal {{Toolkit}}},
  author = {Horton, Matthew and Shen, Jimmy-Xuan and Burns, Jordan and Cohen, Orion and Chabbey, François and Ganose, Alex M. and Guha, Rishabh and Huck, Patrick and Li, Hamming Howard and McDermott, Matthew and Montoya, Joseph and Moore, Guy and Munro, Jason and O'Donnell, Cody and Ophus, Colin and Petretto, Guido and Riebesell, Janosh and Wetizner, Steven and Wander, Brook and Winston, Donald and Yang, Ruoxi and Zeltmann, Steven and Jain, Anubhav and Persson, Kristin A.},
  date = {2023-02-27},
  eprint = {2302.06147},
  eprinttype = {arXiv},
  eprintclass = {cond-mat},
  doi = {10.48550/arXiv.2302.06147},
  url = {http://arxiv.org/abs/2302.06147},
  urldate = {2023-02-16},
  pubstate = {prepublished},
  note = {arXiv:2302.06147}
}

@article{horton_promises_2021,
  title = {Promises and Perils of Computational Materials Databases},
  author = {Horton, M. K. and Dwaraknath, S. and Persson, K. A.},
  year = {2021},
  journal = {Nature Computational Science},
  shortjournal = {Nat Comput Sci},
  volume = {1},
  number = {1},
  pages = {3--5},
  publisher = {Nature Publishing Group},
  issn = {2662-8457},
  doi = {10.1038/s43588-020-00016-5},
  url = {https://www.nature.com/articles/s43588-020-00016-5},
  urldate = {2023-04-07}
}

@article{hunter_matplotlib_2007,
title = {{Matplotlib: A 2D Graphics Environment}},
  volume = {9},
  ISSN = {1521-9615},
  url = {http://dx.doi.org/10.1109/MCSE.2007.55},
  DOI = {10.1109/mcse.2007.55},
  number = {3},
  journal = {Computing in Science \& Engineering},
  publisher = {Institute of Electrical and Electronics Engineers (IEEE)},
  author = {Hunter,  John D.},
  year = {2007},
  pages = {90–95}
}

@article{jain_commentary_2013,
  title = {Commentary: {{The Materials Project}}: {{A}} Materials Genome Approach to Accelerating Materials Innovation},
  shorttitle = {Commentary},
  author = {Jain, Anubhav and Ong, Shyue Ping and Hautier, Geoffroy and Chen, Wei and Richards, William Davidson and Dacek, Stephen and Cholia, Shreyas and Gunter, Dan and Skinner, David and Ceder, Gerbrand and Persson, Kristin A.},
  year = {2013},
  journal = {APL Materials},
  volume = {1},
  number = {1},
  pages = {011002},
  publisher = {American Institute of Physics},
  doi = {10.1063/1.4812323},
  url = {https://aip.scitation.org/doi/10.1063%2F1.4812323},
  urldate = {2022-02-21}
}

@article{jain_computational_2016,
  title = {Computational Predictions of Energy Materials Using Density Functional Theory},
  author = {Jain, Anubhav and Shin, Yongwoo and Persson, Kristin A.},
  year = {2016},
  journal = {Nature Reviews Materials},
  shortjournal = {Nat Rev Mater},
  volume = {1},
  number = {1},
  pages = {1--13},
  publisher = {Nature Publishing Group},
  issn = {2058-8437},
  doi = {10.1038/natrevmats.2015.4},
  url = {https://www.nature.com/articles/natrevmats20154},
  urldate = {2024-02-26},
  issue = {1}
}

@article{kresse_efficient_1996,
  title = {Efficient Iterative Schemes for Ab Initio Total-Energy Calculations Using a Plane-Wave Basis Set},
  author = {Kresse, G. and Furthmüller, J.},
  year = {1996},
  journal = {Physical Review B},
  shortjournal = {Phys. Rev. B},
  volume = {54},
  number = {16},
  pages = {11169--11186},
  doi = {10.1103/PhysRevB.54.11169},
  url = {https://link.aps.org/doi/10.1103/PhysRevB.54.11169},
  urldate = {2019-08-22}
}

@article{larsen_atomic_2017,
  title = {The Atomic Simulation Environment—a {{Python}} Library for Working with Atoms},
  author = {Larsen, Ask Hjorth and Mortensen, Jens Jørgen and Blomqvist, Jakob and Castelli, Ivano E. and Christensen, Rune and Du\textbackslash lak, Marcin and Friis, Jesper and Groves, Michael N. and Hammer, Bjørk and Hargus, Cory},
  year = {2017},
  journal = {Journal of Physics: Condensed Matter},
  volume = {29},
  number = {27},
  pages = {273002},
  publisher = {IOP Publishing},
  url = {https://doi.org/10.1088/1361-648X/aa680e},
  urldate = {2024-03-01}
}

@article{mathew_atomate_2017,
  title = {Atomate: {{A}} High-Level Interface to Generate, Execute, and Analyze Computational Materials Science Workflows},
  shorttitle = {Atomate},
  author = {Mathew, Kiran and Montoya, Joseph H. and Faghaninia, Alireza and Dwarakanath, Shyam and Aykol, Muratahan and Tang, Hanmei and Chu, Iek-heng and Smidt, Tess and Bocklund, Brandon and Horton, Matthew and Dagdelen, John and Wood, Brandon and Liu, Zi-Kui and Neaton, Jeffrey and Ong, Shyue Ping and Persson, Kristin and Jain, Anubhav},
  date = {2017-11-01},
  year = {2017},
  journal = {Computational Materials Science},
  shortjournal = {Computational Materials Science},
  volume = {139},
  pages = {140--152},
  issn = {0927-0256},
  doi = {10.1016/j.commatsci.2017.07.030},
  url = {https://www.sciencedirect.com/science/article/pii/S0927025617303919},
  urldate = {2022-02-21}
}

@inproceedings{mckinney_data_2010,
  booktitle = {Data {{Structures}} for {{Statistical Computing}} in {{Python}}},
  author = {McKinney, Wes},
  date = {2010},
  pages = {56--61},
  location = {Austin, Texas},
  doi = {10.25080/Majora-92bf1922-00a},
  url = {https://conference.scipy.org/proceedings/scipy2010/mckinney.html},
  urldate = {2023-11-01},
  eventtitle = {Python in {{Science Conference}}},
  year = {2010}
}

@misc{molssi_web_2025,
  title = {The {{Molecular Sciences Software Institution}}},
  note = {https://molssi.org/},
  urldate = {2025-10-30},
  organization = {MolSSI}
}

@article{ong_python_2013,
  title = {Python {{Materials Genomics}} (Pymatgen): {{A}} Robust, Open-Source Python Library for Materials Analysis},
  shorttitle = {Python {{Materials Genomics}} (Pymatgen)},
  author = {Ong, Shyue Ping and Richards, William Davidson and Jain, Anubhav and Hautier, Geoffroy and Kocher, Michael and Cholia, Shreyas and Gunter, Dan and Chevrier, Vincent L. and Persson, Kristin A. and Ceder, Gerbrand},
  year = {2013},
  journal = {Computational Materials Science},
  shortjournal = {Computational Materials Science},
  volume = {68},
  pages = {314--319},
  issn = {0927-0256},
  doi = {10.1016/j.commatsci.2012.10.028},
  url = {https://www.sciencedirect.com/science/article/pii/S0927025612006295},
  urldate = {2023-11-01}
}

@misc{pan_benchmarking_2021,
  title = {Benchmarking {{Coordination Number Prediction Algorithms}} on {{Inorganic Crystal Structures}}},
  author = {Pan, Hillary and Ganose, Alex and Horton, Matthew and Aykol, Muratahan and Persson, Kristin and Zimmermann, Nils and Jain, Anubhav},
  date = {2021-01-05},
  eprinttype = {ChemRxiv},
  doi = {10.26434/chemrxiv.12508229.v2},
  url = {https://doi.org/10.26434/chemrxiv.12508229.v2},
  urldate = {2024-03-01},
  pubstate = {prepublished},
  note = {ChemRxiv:10.26434/chemrxiv.12508229.v2}
}

@misc{riebesell_matterviz_2025,
  title = {Matterviz},
  author = {Riebesell, Janosh and Evans, Matthew},
  date = {2025-09},
  doi = {10.5281/zenodo.17094509},
  note = {https://github.com/janosh/matterviz},
  urldate = {2025-09-10},
  version = {0.1.9}
}

@misc{riebesell_pymatviz_2022,
  title = {Pymatviz: {{A}} Toolkit for Visualizations in Materials Informatics},
  author = {Riebesell, Janosh and Goodall, Rhys and Baird, Sterling G.},
  date = {2022-10},
  doi = {10.5281/zenodo.7486816},
  note = {https://github.com/janosh/pymatviz},
  urldate = {2023-12-02},
  version = {0.8.0}
}

@misc{team_pandasdev_2025,
    author       = {{The pandas development team}},
    title        = {pandas-dev/pandas: Pandas},
    month        = feb,
    year         = 2020,
    publisher    = {Zenodo},
    version      = {latest},
    doi          = {10.5281/zenodo.3509134},
    note          = {https://doi.org/10.5281/zenodo.3509134}
}

@article{thompson_lammps_2022,
  title = {{{LAMMPS}} - a Flexible Simulation Tool for Particle-Based Materials Modeling at the Atomic, Meso, and Continuum Scales},
  author = {Thompson, Aidan P. and Aktulga, H. Metin and Berger, Richard and Bolintineanu, Dan S. and Brown, W. Michael and Crozier, Paul S. and In 'T Veld, Pieter J. and Kohlmeyer, Axel and Moore, Stan G. and Nguyen, Trung Dac and Shan, Ray and Stevens, Mark J. and Tranchida, Julien and Trott, Christian and Plimpton, Steven J.},
  year = {2022},
  journal = {Computer Physics Communications},
  shortjournal = {Computer Physics Communications},
  volume = {271},
  pages = {108171},
  issn = {00104655},
  doi = {10.1016/j.cpc.2021.108171},
  url = {https://linkinghub.elsevier.com/retrieve/pii/S0010465521002836},
  urldate = {2025-04-24}
}

@article{togo_firstprinciples_2023,
  title = {First-Principles {{Phonon Calculations}} with {{Phonopy}} and {{Phono3py}}},
  author = {Togo, Atsushi},
  year = {2023},
  journal = {Journal of the Physical Society of Japan},
  shortjournal = {J. Phys. Soc. Jpn.},
  volume = {92},
  number = {1},
  pages = {012001},
  publisher = {The Physical Society of Japan},
  issn = {0031-9015},
  doi = {10.7566/JPSJ.92.012001},
  url = {https://journals.jps.jp/doi/full/10.7566/JPSJ.92.012001},
  urldate = {2023-12-09}
}

@article{phonopy-phono3py-JPCM,
  author  = {Togo, Atsushi and Chaput, Laurent and Tadano, Terumasa and Tanaka, Isao},
  title   = {Implementation strategies in phonopy and phono3py},
  journal = {J. Phys. Condens. Matter},
  volume  = {35},
  number  = {35},
  pages   = {353001},
  year    = {2023},
  doi     = {10.1088/1361-648X/acd831}
}

@article{groom2016cambridge,
  title={The Cambridge structural database},
  author={Groom, Colin R and Bruno, Ian J and Lightfoot, Matthew P and Ward, Suzanna C},
  journal={Structural Science},
  volume={72},
  number={2},
  pages={171--179},
  year={2016},
  publisher={International Union of Crystallography}
}

@article{gallego2016magndata,
  title={MAGNDATA: towards a database of magnetic structures. I. The commensurate case},
  author={Gallego, Samuel V and Perez-Mato, J Manuel and Elcoro, Luis and Tasci, Emre S and Hanson, Robert M and Momma, Koichi and Aroyo, Mois I and Madariaga, Gotzon},
  journal={Applied Crystallography},
  volume={49},
  number={5},
  pages={1750--1776},
  year={2016},
  publisher={International Union of Crystallography}
}

@article{zagorac2019recent,
  title={Recent developments in the Inorganic Crystal Structure Database: theoretical crystal structure data and related features},
  author={Zagorac, Dejan and M{\"u}ller, H and Ruehl, S and Zagorac, J and Rehme, Silke},
  journal={Applied Crystallography},
  volume={52},
  number={5},
  pages={918--925},
  year={2019},
  publisher={International Union of Crystallography}
}

@article{andersen2021optimade,
  title={OPTIMADE, an API for exchanging materials data},
  author={Andersen, Casper W and Armiento, Rickard and Blokhin, Evgeny and Conduit, Gareth J and Dwaraknath, Shyam and Evans, Matthew L and Fekete, {\'A}d{\'a}m and Gopakumar, Abhijith and Gra{\v{z}}ulis, Saulius and Merkys, Andrius and others},
  journal={Scientific data},
  volume={8},
  number={1},
  pages={217},
  year={2021},
  publisher={Nature Publishing Group UK London}
}

@article{togo2024spglib,
  title={Spglib: a software library for crystal symmetry search},
  author={Togo, Atsushi and Shinohara, Kohei and Tanaka, Isao},
  journal={Science and Technology of Advanced Materials: Methods},
  volume={4},
  number={1},
  pages={2384822},
  year={2024},
  publisher={Taylor \& Francis}
}

@article{spglibv2,
  author = {Shinohara, Kohei and Togo, Atsushi and Tanaka, Isao},
  title = {Algorithms for magnetic symmetry operation search and identification of magnetic space group from magnetic crystal structure},
  journal = {Acta Cryst. A},
  year = {2023},
  volume = {79},
  number = {5},
  pages = {390--398},
  month = {Sep},
  doi = {10.1107/S2053273323005016},
  url = {https://doi.org/10.1107/S2053273323005016}
}

@misc{moyo,
  author = {Kohei Shinohara},
  title  = {{moyo: A fast and robust crystal symmetry finder, written in Rust}},
  year   = {2026},
  month  = {1},
  doi    = {10.6084/m9.figshare.31081162.v1},
  url    = {https://figshare.com/articles/software/moyo_A_fast_and_robust_crystal_symmetry_finder_written_in_Rust_/31081162},
  note   = {Source code available at \url{https://github.com/spglib/moyo}}
}

@article{virtanen_scipy_2020,
  author  = {Virtanen, Pauli and Gommers, Ralf and Oliphant, Travis E. and
            Haberland, Matt and Reddy, Tyler and Cournapeau, David and
            Burovski, Evgeni and Peterson, Pearu and Weckesser, Warren and
            Bright, Jonathan and {van der Walt}, St{\'e}fan J. and
            Brett, Matthew and Wilson, Joshua and Millman, K. Jarrod and
            Mayorov, Nikolay and Nelson, Andrew R. J. and Jones, Eric and
            Kern, Robert and Larson, Eric and Carey, C J and
            Polat, {\.I}lhan and Feng, Yu and Moore, Eric W. and
            {VanderPlas}, Jake and Laxalde, Denis and Perktold, Josef and
            Cimrman, Robert and Henriksen, Ian and Quintero, E. A. and
            Harris, Charles R. and Archibald, Anne M. and
            Ribeiro, Ant{\^o}nio H. and Pedregosa, Fabian and
            {van Mulbregt}, Paul and {SciPy 1.0 Contributors}},
  title   = {{{SciPy} 1.0: Fundamental Algorithms for Scientific
            Computing in Python}},
  journal = {Nature Methods},
  year    = {2020},
  volume  = {17},
  pages   = {261--272},
  adsurl  = {https://rdcu.be/b08Wh},
  doi     = {10.1038/s41592-019-0686-2},
}

@article{ward_matminer_2018,
  title = {Matminer: {{An}} Open Source Toolkit for Materials Data Mining},
  shorttitle = {Matminer},
  author = {Ward, Logan and Dunn, Alexander and Faghaninia, Alireza and Zimmermann, Nils E. R. and Bajaj, Saurabh and Wang, Qi and Montoya, Joseph and Chen, Jiming and Bystrom, Kyle and Dylla, Maxwell and Chard, Kyle and Asta, Mark and Persson, Kristin A. and Snyder, G. Jeffrey and Foster, Ian and Jain, Anubhav},
  year = {2018},
  journal = {Computational Materials Science},
  shortjournal = {Computational Materials Science},
  volume = {152},
  pages = {60--69},
  issn = {0927-0256},
  doi = {10.1016/j.commatsci.2018.05.018},
  url = {https://www.sciencedirect.com/science/article/pii/S0927025618303252},
  urldate = {2023-12-06}
}

@article{waroquiers_chemenv_2020,
  title = {ChemEnv
                    : a fast and robust coordination environment identification tool},
  volume = {76},
  ISSN = {2052-5206},
  url = {http://dx.doi.org/10.1107/S2052520620007994},
  DOI = {10.1107/s2052520620007994},
  number = {4},
  journal = {Acta Crystallographica Section B Structural Science,  Crystal Engineering and Materials},
  publisher = {International Union of Crystallography (IUCr)},
  author = {Waroquiers,  David and George,  Janine and Horton,  Matthew and Schenk,  Stephan and Persson,  Kristin A. and Rignanese,  Gian-Marco and Gonze,  Xavier and Hautier,  Geoffroy},
  year = {2020},
  month = jul,
  pages = {683–695}
}

@article{broberg_pycdt_2018,
	title = {{PyCDT}: {A} {Python} toolkit for modeling point defects in semiconductors and insulators},
	volume = {226},
	issn = {0010-4655},
	shorttitle = {{PyCDT}},
	url = {http://www.sciencedirect.com/science/article/pii/S0010465518300079},
	doi = {10.1016/j.cpc.2018.01.004},
	language = {en},
	urldate = {2020-05-18},
	journal = {Computer Physics Communications},
	author = {Broberg, Danny and Medasani, Bharat and Zimmermann, Nils E. R. and Yu, Guodong and Canning, Andrew and Haranczyk, Maciej and Asta, Mark and Hautier, Geoffroy},
	month = may,
	year = {2018},
	pages = {165--179},
}

@article{kumagai_insights_2021,
	title = {Insights into oxygen vacancies from high-throughput first-principles calculations},
	volume = {5},
	issn = {2475-9953},
	url = {https://link.aps.org/doi/10.1103/PhysRevMaterials.5.123803},
	doi = {10.1103/PhysRevMaterials.5.123803},
	language = {en},
	number = {12},
	urldate = {2021-12-28},
	journal = {Physical Review Materials},
	author = {Kumagai, Yu and Tsunoda, Naoki and Takahashi, Akira and Oba, Fumiyasu},
	month = dec,
	year = {2021},
	pages = {123803},
}

@article{ogawa_extended_2022,
	title = {An extended computational approach for point-defect equilibria in semiconductor materials},
	volume = {8},
	issn = {2057-3960},
	url = {https://www.nature.com/articles/s41524-022-00756-0},
	doi = {10.1038/s41524-022-00756-0},
	language = {en},
	number = {1},
	urldate = {2022-07-01},
	journal = {npj Computational Materials},
	author = {Ogawa, Takafumi and Taguchi, Ayako and Kuwabara, Akihide},
	month = dec,
	year = {2022},
	pages = {79},
}

@article{mosquera-lois_shakenbreak_2022,
	title = {{ShakeNBreak}: {Navigating} the defect configurational landscape},
	volume = {7},
	issn = {2475-9066},
	shorttitle = {{ShakeNBreak}},
	url = {https://joss.theoj.org/papers/10.21105/joss.04817},
	doi = {10.21105/joss.04817},
	language = {en},
	number = {80},
	urldate = {2022-12-25},
	journal = {Journal of Open Source Software},
	author = {Mosquera-Lois, Irea and Kavanagh, Seán R. and Walsh, Aron and Scanlon, David O.},
	month = dec,
	year = {2022},
	pages = {4817},
}

@article{muy_aiida-defects_2023,
	title = {{AiiDA}-defects: an automated and fully reproducible workflow for the complete characterization of defect chemistry in functional materials},
	volume = {5},
	issn = {2516-1075},
	shorttitle = {{AiiDA}-defects},
	url = {https://iopscience.iop.org/article/10.1088/2516-1075/ace014},
	doi = {10.1088/2516-1075/ace014},
	language = {en},
	number = {2},
	urldate = {2024-02-20},
	journal = {Electronic Structure},
	author = {Muy, Sokseiha and Johnston, Conrad and Marzari, Nicola},
	month = jun,
	year = {2023},
	pages = {024009},
}

@article{shen_pymatgen-analysis-defects_2024,
	title = {pymatgen-analysis-defects: {A} {Python} package for analyzing point defects in crystalline materials},
	volume = {9},
	issn = {2475-9066},
	shorttitle = {pymatgen-analysis-defects},
	url = {https://joss.theoj.org/papers/10.21105/joss.05941},
	doi = {10.21105/joss.05941},
	language = {en},
	number = {93},
	urldate = {2024-02-20},
	journal = {Journal of Open Source Software},
	author = {Shen, Jimmy-Xuan and Varley, Joel},
	month = jan,
	year = {2024},
	pages = {5941},
}

@article{kavanagh_doped_2024,
	title = {doped: {Python} toolkit for robust and repeatable charged defect supercell calculations},
	volume = {9},
	issn = {2475-9066},
	shorttitle = {doped},
	url = {https://joss.theoj.org/papers/10.21105/joss.06433},
	doi = {10.21105/joss.06433},
	language = {en},
	number = {96},
	urldate = {2024-05-20},
	journal = {Journal of Open Source Software},
	author = {Kavanagh, Seán R. and Squires, Alexander G. and Nicolson, Adair and Mosquera-Lois, Irea and Ganose, Alex M. and Zhu, Bonan and Brlec, Katarina and Walsh, Aron and Scanlon, David O.},
	month = apr,
	year = {2024},
	pages = {6433},
}

@article{riebesell_framework_2025,
	title = {A framework to evaluate machine learning crystal stability predictions},
	volume = {7},
	copyright = {2025 The Author(s)},
	issn = {2522-5839},
	url = {https://www.nature.com/articles/s42256-025-01055-1},
	doi = {10.1038/s42256-025-01055-1},
	language = {en},
	number = {6},
	urldate = {2025-07-07},
	journal = {Nature Machine Intelligence},
	author = {Riebesell, Janosh and Goodall, Rhys E. A. and Benner, Philipp and Chiang, Yuan and Deng, Bowen and Ceder, Gerbrand and Asta, Mark and Lee, Alpha A. and Jain, Anubhav and Persson, Kristin A.},
	month = jun,
	year = {2025},
	note = {Publisher: Nature Publishing Group},
	pages = {836--847},
}

@article{davies_smact_2019,
	title = {{SMACT}: {Semiconducting} {Materials} by {Analogy} and {Chemical} {Theory}},
	volume = {4},
	issn = {2475-9066},
	shorttitle = {{SMACT}},
	url = {https://joss.theoj.org/papers/10.21105/joss.01361},
	doi = {10.21105/joss.01361},
	language = {en},
	number = {38},
	urldate = {2025-10-29},
	journal = {Journal of Open Source Software},
	author = {Davies, Daniel W. and Butler, Keith T. and Jackson, Adam J. and Skelton, Jonathan M. and Morita, Kazuki and Walsh, Aron},
	month = jun,
	year = {2019},
	pages = {1361},
}

@article{ganose_atomate2_2025,
	title = {Atomate2: modular workflows for materials science},
	volume = {4},
	issn = {2635-098X},
	shorttitle = {Atomate2},
	url = {https://pubs.rsc.org/en/content/articlelanding/2025/dd/d5dd00019j},
	doi = {10.1039/D5DD00019J},
	language = {en},
	number = {7},
	urldate = {2025-10-29},
	journal = {Digital Discovery},
	author = {Ganose, Alex M. and Sahasrabuddhe, Hrushikesh and Asta, Mark and Beck, Kevin and Biswas, Tathagata and Bonkowski, Alexander and Bustamante, Joana and Chen, Xin and Chiang, Yuan and Chrzan, Daryl C. and Clary, Jacob and Cohen, Orion A. and Ertural, Christina and Gallant, Max C. and George, Janine and Gerits, Sophie and Goodall, Rhys E. A. and Guha, Rishabh D. and Hautier, Geoffroy and Horton, Matthew and Inizan, T. J. and Kaplan, Aaron D. and Kingsbury, Ryan S. and Kuner, Matthew C. and Li, Bryant and Linn, Xavier and McDermott, Matthew J. and Mohanakrishnan, Rohith Srinivaas and Naik, Aakash N. and Neaton, Jeffrey B. and Parmar, Shehan M. and Persson, Kristin A. and Petretto, Guido and Purcell, Thomas A. R. and Ricci, Francesco and Rich, Benjamin and Riebesell, Janosh and Rignanese, Gian-Marco and Rosen, Andrew S. and Scheffler, Matthias and Schmidt, Jonathan and Shen, Jimmy-Xuan and Sobolev, Andrei and Sundararaman, Ravishankar and Tezak, Cooper and Trinquet, Victor and Varley, Joel B. and Vigil-Fowler, Derek and Wang, Duo and Waroquiers, David and Wen, Mingjian and Yang, Han and Zheng, Hui and Zheng, Jiongzhi and Zhu, Zhuoying and Jain, Anubhav},
	month = jul,
	year = {2025},
	note = {Publisher: RSC},
	pages = {1944--1973},
}

@article{aspuru_commit_2025,
	title = {Commit: Mini article for dynamic reporting of incremental improvements to previous scholarly work},
	volume = {4},
	issn = {2635-098X},
	shorttitle = {Commit},
	url = {https://pubs.rsc.org/en/content/articlelanding/2025/dd/d4dd90053g},
	doi = {10.1039/D4DD90053G},
	language = {en},
	number = {2},
	urldate = {2026-02-17},
	journal = {Digital Discovery},
	author = {Aspuru-Guzik, Alan and Hein, Jason E. and Schrier, Joshua},
	month = jan,
	year = {2025},
	pages = {301--302},
	note = {Publisher: RSC},
}

@misc{rosen_quacc_2025,
	title = {quacc – {The} {Quantum} {Accelerator}},
	url = {https://zenodo.org/records/17373420},
	urldate = {2025-10-29},
	publisher = {Zenodo},
	author = {Rosen, Andrew},
	month = oct,
	year = {2025},
	doi = {10.5281/zenodo.17373420},
}

@article{ko_materials_2025,
	title = {Materials {Graph} {Library} ({MatGL}), an open-source graph deep learning library for materials science and chemistry},
	volume = {11},
	copyright = {2025 The Author(s)},
	issn = {2057-3960},
	url = {https://www.nature.com/articles/s41524-025-01742-y},
	doi = {10.1038/s41524-025-01742-y},
	language = {en},
	number = {1},
	urldate = {2025-10-29},
	journal = {npj Computational Materials},
	author = {Ko, Tsz Wai and Deng, Bowen and Nassar, Marcel and Barroso-Luque, Luis and Liu, Runze and Qi, Ji and Thakur, Atul C. and Mishra, Adesh Rohan and Liu, Elliott and Ceder, Gerbrand and Miret, Santiago and Ong, Shyue Ping},
	month = aug,
	year = {2025},
	note = {Publisher: Nature Publishing Group},
	pages = {253},
}

@article{zhangArtificialIntelligenceScience2025,
  title = {Artificial {{Intelligence}} for {{Science}} in {{Quantum}}, {{Atomistic}}, and {{Continuum Systems}}},
  author = {Zhang, Xuan and Wang, Limei and Helwig, Jacob and Luo, Youzhi and Fu, Cong and Xie, Yaochen and Liu, Meng and Lin, Yuchao and Xu, Zhao and Yan, Keqiang and Adams, Keir and Weiler, Maurice and Li, Xiner and Fu, Tianfan and Wang, Yucheng and Strasser, Alex and Yu, Haiyang and Xie, YuQing and Fu, Xiang and Xu, Shenglong and Liu, Yi and Du, Yuanqi and Saxton, Alexandra and Ling, Hongyi and Lawrence, Hannah and St{\"a}rk, Hannes and Gui, Shurui and Edwards, Carl and Gao, Nicholas and Ladera, Adriana and Wu, Tailin and Hofgard, Elyssa F. and Tehrani, Aria Mansouri and Wang, Rui and Daigavane, Ameya and Bohde, Montgomery and Kurtin, Jerry and Huang, Qian and Phung, Tuong and Xu, Minkai and Joshi, Chaitanya K. and Mathis, Simon V. and Azizzadenesheli, Kamyar and Fang, Ada and {Aspuru-Guzik}, Al{\'a}n and Bekkers, Erik and Bronstein, Michael and Zitnik, Marinka and Anandkumar, Anima and Ermon, Stefano and Li{\`o}, Pietro and Yu, Rose and G{\"u}nnemann, Stephan and Leskovec, Jure and Ji, Heng and Sun, Jimeng and Barzilay, Regina and Jaakkola, Tommi and Coley, Connor W. and Qian, Xiaoning and Qian, Xiaofeng and Smidt, Tess and Ji, Shuiwang},
  year = 2025,
  month = jul,
  journal = {Foundations and Trends{\textregistered} in Machine Learning},
  volume = {18},
  number = {4},
  pages = {385--912},
  publisher = {Now Publishers, Inc.},
  issn = {1935-8237, 1935-8245},
  doi = {10.1561/2200000115},
  urldate = {2025-10-29},
  langid = {english},
}

@article{fredericksPyXtalPythonLibrary2021,
  title = {{{PyXtal}}: {{A Python}} Library for Crystal Structure Generation and Symmetry Analysis},
  shorttitle = {{{PyXtal}}},
  author = {Fredericks, Scott and Parrish, Kevin and Sayre, Dean and Zhu, Qiang},
  year = 2021,
  month = apr,
  journal = {Computer Physics Communications},
  volume = {261},
  pages = {107810},
  issn = {0010-4655},
  doi = {10.1016/j.cpc.2020.107810},
  urldate = {2025-10-29},
}

@article{janssen_pyiron_2019,
	title = {pyiron: {An} integrated development environment for computational materials science},
	volume = {163},
	issn = {0927-0256},
	shorttitle = {pyiron},
	url = {https://www.sciencedirect.com/science/article/pii/S0927025618304786},
	doi = {10.1016/j.commatsci.2018.07.043},
	urldate = {2024-08-21},
	journal = {Computational Materials Science},
	author = {Janssen, Jan and Surendralal, Sudarsan and Lysogorskiy, Yury and Todorova, Mira and Hickel, Tilmann and Drautz, Ralf and Neugebauer, Jörg},
	month = jun,
	year = {2019},
	pages = {24--36},
}

@article{squires_guidelines_2026,
	title = {Guidelines for robust and reproducible point defect simulations in crystals},
	copyright = {2026 Springer Nature Limited},
	issn = {2058-8437},
	url = {https://www.nature.com/articles/s41578-025-00879-y},
	doi = {10.1038/s41578-025-00879-y},
	language = {en},
	urldate = {2026-01-20},
	journal = {Nature Reviews Materials},
	publisher = {Nature Publishing Group},
	author = {Squires, Alexander G. and Kavanagh, Seán R. and Walsh, Aron and Scanlon, David O.},
	month = jan,
	year = {2026},
	pages = {1--18},
}

@article{stier_materials_2024,
	title = {Materials {Acceleration} {Platforms} ({MAPs}): {Accelerating} {Materials} {Research} and {Development} to {Meet} {Urgent} {Societal} {Challenges}},
	volume = {36},
	copyright = {© 2024 The Author(s). Advanced Materials published by Wiley-VCH GmbH},
	issn = {1521-4095},
	shorttitle = {Materials {Acceleration} {Platforms} ({MAPs})},
	url = {https://onlinelibrary.wiley.com/doi/abs/10.1002/adma.202407791},
	doi = {10.1002/adma.202407791},
	language = {en},
	number = {45},
	urldate = {2024-12-17},
	journal = {Advanced Materials},
	author = {Stier, Simon P. and Kreisbeck, Christoph and Ihssen, Holger and Popp, Matthias Albert and Hauch, Jens and Malek, Kourosh and Reynaud, Marine and Goumans, T.p.m. and Carlsson, Johan and Todorov, Ilian and Gold, Lukas and Räder, Andreas and Wenzel, Wolfgang and Bandesha, Shahbaz Tareq and Jacques, Philippe and Garcia-Moreno, Francisco and Arcelus, Oier and Friederich, Pascal and Clark, Simon and Maglione, Mario and Laukkanen, Anssi and Castelli, Ivano Eligio and Carrasco, Javier and Cabanas, Montserrat Casas and Stein, Helge Sören and Ozcan, Ozlem and Elbert, David and Reuter, Karsten and Scheurer, Christoph and Demura, Masahiko and Han, Sang Soo and Vegge, Tejs and Nakamae, Sawako and Fabrizio, Monica and Kozdras, Mark},
	year = {2024},
	note = {\_eprint: https://onlinelibrary.wiley.com/doi/pdf/10.1002/adma.202407791},
	pages = {2407791},
}

@article{chen_universal_2022,
  title = {A universal graph deep learning interatomic potential for the periodic table},
  volume = {2},
  ISSN = {2662-8457},
  url = {http://dx.doi.org/10.1038/s43588-022-00349-3},
  DOI = {10.1038/s43588-022-00349-3},
  number = {11},
  journal = {Nature Computational Science},
  publisher = {Springer Science and Business Media LLC},
  author = {Chen,  Chi and Ong,  Shyue Ping},
  year = {2022},
  month = nov,
  pages = {718–728}
}

@article{shen_topological_2023,
  title = {Topological graph-based analysis of solid-state ion migration},
  volume = {9},
  ISSN = {2057-3960},
  url = {http://dx.doi.org/10.1038/s41524-023-01051-2},
  DOI = {10.1038/s41524-023-01051-2},
  number = {1},
  journal = {npj Computational Materials},
  publisher = {Springer Science and Business Media LLC},
  author = {Shen,  Jimmy-Xuan and Li,  Haoming Howard and Rutt,  Ann and Horton,  Matthew K. and Persson,  Kristin A.},
  year = {2023},
  month = jun,
  pages = {99}
}

@article{shen_charge_2020,
  title = {A charge-density-based general cation insertion algorithm for generating new Li-ion cathode materials},
  volume = {6},
  ISSN = {2057-3960},
  url = {http://dx.doi.org/10.1038/s41524-020-00422-3},
  DOI = {10.1038/s41524-020-00422-3},
  number = {1},
  journal = {npj Computational Materials},
  publisher = {Springer Science and Business Media LLC},
  author = {Shen,  Jimmy-Xuan and Horton,  Matthew and Persson,  Kristin A.},
  year = {2020},
  month = oct,
  pages = {161}
}

@article{deng_chgnet_2023,
  title = {CHGNet as a pretrained universal neural network potential for charge-informed atomistic modelling},
  volume = {5},
  ISSN = {2522-5839},
  url = {http://dx.doi.org/10.1038/s42256-023-00716-3},
  DOI = {10.1038/s42256-023-00716-3},
  number = {9},
  journal = {Nature Machine Intelligence},
  publisher = {Springer Science and Business Media LLC},
  author = {Deng,  Bowen and Zhong,  Peichen and Jun,  KyuJung and Riebesell,  Janosh and Han,  Kevin and Bartel,  Christopher J. and Ceder,  Gerbrand},
  year = {2023},
  month = sep,
  pages = {1031–1041}
}

@misc{simeon_tensornet_2023,
  doi = {10.48550/ARXIV.2306.06482},
  url = {https://arxiv.org/abs/2306.06482},
  author = {Simeon,  Guillem and de Fabritiis,  Gianni},
  title = {TensorNet: Cartesian Tensor Representations for Efficient Learning of Molecular Potentials},
  publisher = {arXiv},
  year = {2023},
  copyright = {Creative Commons Attribution 4.0 International}
}

@article{kresse_ultrasoft_1999,
	title = {From ultrasoft pseudopotentials to the projector augmented-wave method},
	volume = {59},
	doi = {10.1103/PhysRevB.59.1758},
	pages = {1758--1775},
	number = {3},
	author = {Kresse, G. and Joubert, D.},
	journal = {Phys. Rev. B},
    year =  {1999},
	date = {1999-01-15},
}

@article{blum_fhiaims_2009,
  title = {Ab initio molecular simulations with numeric atom-centered orbitals},
  volume = {180},
  ISSN = {0010-4655},
  url = {http://dx.doi.org/10.1016/j.cpc.2009.06.022},
  DOI = {10.1016/j.cpc.2009.06.022},
  number = {11},
  journal = {Computer Physics Communications},
  publisher = {Elsevier BV},
  author = {Blum,  Volker and Gehrke,  Ralf and Hanke,  Felix and Havu,  Paula and Havu,  Ville and Ren,  Xinguo and Reuter,  Karsten and Scheffler,  Matthias},
  year = {2009},
  month = nov,
  pages = {2175–2196}
}

@article{giannozzi_qe_2017,
  title = {Advanced capabilities for materials modelling with Quantum ESPRESSO},
  volume = {29},
  ISSN = {1361-648X},
  url = {http://dx.doi.org/10.1088/1361-648X/aa8f79},
  DOI = {10.1088/1361-648x/aa8f79},
  number = {46},
  journal = {Journal of Physics: Condensed Matter},
  publisher = {IOP Publishing},
  author = {Giannozzi,  P and Andreussi,  O and Brumme,  T and Bunau,  O and Buongiorno Nardelli,  M and Calandra,  M and Car,  R and Cavazzoni,  C and Ceresoli,  D and Cococcioni,  M and Colonna,  N and Carnimeo,  I and Dal Corso,  A and de Gironcoli,  S and Delugas,  P and DiStasio,  R A and Ferretti,  A and Floris,  A and Fratesi,  G and Fugallo,  G and Gebauer,  R and Gerstmann,  U and Giustino,  F and Gorni,  T and Jia,  J and Kawamura,  M and Ko,  H-Y and Kokalj,  A and K\"{u}\c{c}\"{u}kbenli,  E and Lazzeri,  M and Marsili,  M and Marzari,  N and Mauri,  F and Nguyen,  N L and Nguyen,  H-V and Otero-de-la-Roza,  A and Paulatto,  L and Poncé,  S and Rocca,  D and Sabatini,  R and Santra,  B and Schlipf,  M and Seitsonen,  A P and Smogunov,  A and Timrov,  I and Thonhauser,  T and Umari,  P and Vast,  N and Wu,  X and Baroni,  S},
  year = {2017},
  month = oct,
  pages = {465901}
}

@article{eastman_openmm_2023,
  title = {OpenMM 8: Molecular Dynamics Simulation with Machine Learning Potentials},
  volume = {128},
  ISSN = {1520-5207},
  url = {http://dx.doi.org/10.1021/acs.jpcb.3c06662},
  DOI = {10.1021/acs.jpcb.3c06662},
  number = {1},
  journal = {The Journal of Physical Chemistry B},
  publisher = {American Chemical Society (ACS)},
  author = {Eastman,  Peter and Galvelis,  Raimondas and Peláez,  Raúl P. and Abreu,  Charlles R. A. and Farr,  Stephen E. and Gallicchio,  Emilio and Gorenko,  Anton and Henry,  Michael M. and Hu,  Frank and Huang,  Jing and Kr\"{a}mer,  Andreas and Michel,  Julien and Mitchell,  Joshua A. and Pande,  Vijay S. and Rodrigues,  João PGLM and Rodriguez-Guerra,  Jaime and Simmonett,  Andrew C. and Singh,  Sukrit and Swails,  Jason and Turner,  Philip and Wang,  Yuanqing and Zhang,  Ivy and Chodera,  John D. and De Fabritiis,  Gianni and Markland,  Thomas E.},
  year = {2023},
  month = dec,
  pages = {109–116}
}

@misc{wortmann_fleur_2023,
  author       = {Wortmann, Daniel and Michalicek, Gregor and Baadji, Nadjib and Betzinger, Markus and Bihlmayer, Gustav and Br\"oder, Jens and Burnus, Tobias and Enkovaara, Jussi and Freimuth, Frank and Friedrich, Christoph and Gerhorst, Christian-Roman and Granberg Cauchi, Sabastian and Grytsiuk, Uliana and Hanke, Andrea and Hanke, Jan-Philipp and Heide, Marcus and Heinze, Stefan and Hilgers, Robin and Janssen, Henning and Kl\"uppelberg, Daniel Aaaron and Kovacik, Roman and Kurz, Philipp and Lezaic, Marjana and Madsen, Georg K. H. and Mokrousov, Yuriy and Neukirchen, Alexander and Redies, Matthias and Rost, Stefan and Schlipf, Martin and Schindlmayr, Arno and Winkelmann, Miriam and Bl\"ugel, Stefan},
  title        = {{FLEUR}},
  month        = may,
  year         = 2023,
  publisher    = {Zenodo},
  doi          = {10.5281/zenodo.7576163},
  url          = {https://doi.org/10.5281/zenodo.7576163},
  howpublished  = {Zenodo}
}

@article{nelson_lobster_2020,
  title = {LOBSTER: Local orbital projections,  atomic charges,  and chemical‐bonding analysis from projector‐augmented‐wave‐based density‐functional theory},
  volume = {41},
  ISSN = {1096-987X},
  url = {http://dx.doi.org/10.1002/jcc.26353},
  DOI = {10.1002/jcc.26353},
  number = {21},
  journal = {Journal of Computational Chemistry},
  publisher = {Wiley},
  author = {Nelson,  Ryky and Ertural,  Christina and George,  Janine and Deringer,  Volker L. and Hautier,  Geoffroy and Dronskowski,  Richard},
  year = {2020},
  month = jun,
  pages = {1931–1940}
}

@article{zimmermann_crystalnn_2020,
  title = {Local structure order parameters and site fingerprints for quantification of coordination environment and crystal structure similarity},
  volume = {10},
  ISSN = {2046-2069},
  url = {http://dx.doi.org/10.1039/C9RA07755C},
  DOI = {10.1039/c9ra07755c},
  number = {10},
  journal = {RSC Advances},
  publisher = {Royal Society of Chemistry (RSC)},
  author = {Zimmermann,  Nils E. R. and Jain,  Anubhav},
  year = {2020},
  pages = {6063–6081}
}

@article{ganose_robocrys_2019,
  title = {Robocrystallographer: automated crystal structure text descriptions and analysis},
  volume = {9},
  ISSN = {2159-6867},
  url = {http://dx.doi.org/10.1557/mrc.2019.94},
  DOI = {10.1557/mrc.2019.94},
  number = {3},
  journal = {MRS Communications},
  publisher = {Springer Science and Business Media LLC},
  author = {Ganose,  Alex M. and Jain,  Anubhav},
  year = {2019},
  month = sep,
  pages = {874–881}
}

@misc{verstraete_Abinit2025_2025,
  title = {Abinit 2025: {{New Capabilities}} for the {{Predictive Modeling}} of {{Solids}} and {{Nanomaterials}}},
  shorttitle = {Abinit 2025},
  author = {Verstraete, Matthieu J. and Abreu, Joao and Allemand, Guillaume E. and Amadon, Bernard and Antonius, Gabriel and Azizi, Maryam and Baguet, Lucas and Barat, Clementine and Bastogne, Louis and Bejaud, Romuald and Beuken, Jean-Michel and Bieder, Jordan and Blanchet, Augustin and Bottin, Francois and Bouchet, Johann and Bouquiaux, Julien and Bousquet, Eric and Boust, James and Brieuc, Fabien and {Brousseau-Couture}, Veronique and Bruneval, Nils Brouwer Fabien and Castellano, Alois and Castiel, Emmanuel and Charraud, Jean-Baptiste and Clerouin, Jean and Cote, Michel and Duval, Clement and Gallo, Alejandro and Gendron, Frederic and Geneste, Gregory and Ghosez, Philippe and Giantomassi, Matteo and Gingras, Olivier and {Gomez-Ortiz}, Fernando and Gonze, Xavier and Goudreault, Felix Antoine and Gruneis, Andreas and Gupta, Raveena and Guster, Bogdan and Hamann, Donald R. and He, Xu and Hellman, Olle and Holzwarth, Natalie and Jollet, Francois and Kestener, Pierre and Lygatsika, Ioanna-Maria and Nadeau, Olivier and MacEnulty, Lorien and Marazzi, Enrico and Mignolet, Maxime and O'Regan, David D. and Outerovitch, Robinson and Paillard, Charles and Petretto, Guido and Ponce, Samuel and Ricci, Francesco and Rignanese, Gian-Marco and {Rodriguez-Mayorga}, Mauricio and Romero, Aldo H. and Rostami, Samare and Royo, Miquel and Sarraute, Marc and Sasani, Alireza and Soubiran, Francois and Stengel, Massimiliano and Tantardini, Christian and Torrent, Marc and Trinquet, Victor and Vasilchencko, Vasilii and Waroquiers, David and Zabalo, Asier and Zadoks, Austin and Zhang, Huazhang and Zwanziger, Josef},
  year = 2025,
  month = jul,
  number = {arXiv:2507.08578},
  eprint = {2507.08578},
  primaryclass = {cond-mat},
  publisher = {arXiv},
  doi = {10.48550/arXiv.2507.08578},
  urldate = {2025-11-25},
  archiveprefix = {arXiv},
}

@article{naik_LobsterPypackage_2024,
  title = {{{LobsterPy}}: {{A}} Package to Automatically Analyze {{LOBSTER}} Runs},
  shorttitle = {{{LobsterPy}}},
  author = {Naik, Aakash Ashok and Ueltzen, Katharina and Ertural, Christina and Jackson, Adam J. and George, Janine},
  year = 2024,
  month = feb,
  journal = {Journal of Open Source Software},
  volume = {9},
  number = {94},
  pages = {6286},
  issn = {2475-9066},
  doi = {10.21105/joss.06286},
  urldate = {2025-11-25},
  langid = {english},
}

@article{kuhne_cp2k_2020,
	title = {{CP2K}: {An} electronic structure and molecular dynamics software package - {Quickstep}: {Efficient} and accurate electronic structure calculations},
	volume = {152},
	issn = {0021-9606},
	shorttitle = {{CP2K}},
	url = {https://doi.org/10.1063/5.0007045},
	doi = {10.1063/5.0007045},
	number = {19},
	urldate = {2026-02-17},
	journal = {The Journal of Chemical Physics},
	author = {Kühne, Thomas D. and Iannuzzi, Marcella and Del Ben, Mauro and Rybkin, Vladimir V. and Seewald, Patrick and Stein, Frederick and Laino, Teodoro and Khaliullin, Rustam Z. and Schütt, Ole and Schiffmann, Florian and Golze, Dorothea and Wilhelm, Jan and Chulkov, Sergey and Bani-Hashemian, Mohammad Hossein and Weber, Valéry and Borštnik, Urban and Taillefumier, Mathieu and Jakobovits, Alice Shoshana and Lazzaro, Alfio and Pabst, Hans and Müller, Tiziano and Schade, Robert and Guidon, Manuel and Andermatt, Samuel and Holmberg, Nico and Schenter, Gregory K. and Hehn, Anna and Bussy, Augustin and Belleflamme, Fabian and Tabacchi, Gloria and Glöß, Andreas and Lass, Michael and Bethune, Iain and Mundy, Christopher J. and Plessl, Christian and Watkins, Matt and VandeVondele, Joost and Krack, Matthias and Hutter, Jürg},
	month = may,
	year = {2020},
	pages = {194103},
}

@article{krylov_qchem_2020,
  title = {Q‐Chem: an engine for innovation},
  volume = {3},
  ISSN = {1759-0884},
  url = {http://dx.doi.org/10.1002/wcms.1122},
  DOI = {10.1002/wcms.1122},
  number = {3},
  journal = {WIREs Computational Molecular Science},
  publisher = {Wiley},
  author = {Krylov,  Anna I. and Gill,  Peter M.W.},
  year = {2012},
  month = nov,
  pages = {317–326}
}

@article{angqvist_icet_2019,
  title = {ICET – A Python Library for Constructing and Sampling Alloy Cluster Expansions},
  volume = {2},
  ISSN = {2513-0390},
  url = {http://dx.doi.org/10.1002/adts.201900015},
  DOI = {10.1002/adts.201900015},
  number = {7},
  journal = {Advanced Theory and Simulations},
  publisher = {Wiley},
  author = {Ångqvist,  Mattias and Muñoz,  William A. and Rahm,  J. Magnus and Fransson,  Erik and Durniak,  Céline and Rozyczko,  Piotr and Rod,  Thomas H. and Erhart,  Paul},
  year = {2019},
  month = may,
  pages = {1900015},
}

@article{wang_openff_2024,
  title = {The Open Force Field Initiative: Open Software and Open Science for Molecular Modeling},
  volume = {128},
  ISSN = {1520-5207},
  url = {http://dx.doi.org/10.1021/acs.jpcb.4c01558},
  DOI = {10.1021/acs.jpcb.4c01558},
  number = {29},
  journal = {The Journal of Physical Chemistry B},
  publisher = {American Chemical Society (ACS)},
  author = {Wang,  Lily and Behara,  Pavan Kumar and Thompson,  Matthew W. and Gokey,  Trevor and Wang,  Yuanqing and Wagner,  Jeffrey R. and Cole,  Daniel J. and Gilson,  Michael K. and Shirts,  Michael R. and Mobley,  David L.},
  year = {2024},
  month = jul,
  pages = {7043–7067}
}

@article{sundararaman_jdftx_2017,
  title = {JDFTx: Software for joint density-functional theory},
  volume = {6},
  ISSN = {2352-7110},
  url = {http://dx.doi.org/10.1016/j.softx.2017.10.006},
  DOI = {10.1016/j.softx.2017.10.006},
  journal = {SoftwareX},
  publisher = {Elsevier BV},
  author = {Sundararaman,  Ravishankar and Letchworth-Weaver,  Kendra and Schwarz,  Kathleen A. and Gunceler,  Deniz and Ozhabes,  Yalcin and Arias,  T.A.},
  year = {2017},
  pages = {278–284}
}

@misc{spec0_minimum_supported_dependencies,
  title = {{SPEC} 0 --- {Minimum} {Supported} {Dependencies}},
  author = {Mueller, Andreas C. and Sipőcz, Brigitta and Millman, Jarrod and Munk, Madicken and Haberland, Matt and Bussonnier, Matthias and Gommers, Ralf and Barnowski, Ross and van der Walt, Stéfan and Caswell, Thomas A.},
  year = {2023},
  url = {https://scientific-python.org/specs/spec-0000/},
  note = {Endorsed by NumPy, SciPy, matplotlib, networkx, scikit-image, xarray, and others}
}

@article{freysoldt_firstprinciples_2014,
	title = {First-principles calculations for point defects in solids},
	volume = {86},
	issn = {0034-6861, 1539-0756},
	url = {https://link.aps.org/doi/10.1103/RevModPhys.86.253},
	doi = {10.1103/RevModPhys.86.253},
	language = {en},
	number = {1},
	urldate = {2020-04-26},
	journal = {Reviews of Modern Physics},
	author = {Freysoldt, Christoph and Grabowski, Blazej and Hickel, Tilmann and Neugebauer, Jörg and Kresse, Georg and Janotti, Anderson and Van de Walle, Chris G.},
	month = mar,
	year = {2014},
	pages = {253--305},
}

@article{Yan2024,
  title = {The Case for a Defect Genome Initiative},
  volume = {36},
  ISSN = {1521-4095},
  url = {http://dx.doi.org/10.1002/adma.202303098},
  DOI = {10.1002/adma.202303098},
  number = {11},
  journal = {Advanced Materials},
  publisher = {Wiley},
  author = {Yan,  Qimin and Kar,  Swastik and Chowdhury,  Sugata and Bansil,  Arun},
  year = {2024},
  month = jan,
  pages = {2303098}
}

@article{horton_2025_mp,
  title = {Accelerated data-driven materials science with the Materials Project},
  volume = {24},
  ISSN = {1476-4660},
  url = {http://dx.doi.org/10.1038/s41563-025-02272-0},
  DOI = {10.1038/s41563-025-02272-0},
  number = {10},
  journal = {Nature Materials},
  publisher = {Springer Science and Business Media LLC},
  author = {Horton,  Matthew K. and Huck,  Patrick and Yang,  Ruo Xi and Munro,  Jason M. and Dwaraknath,  Shyam and Ganose,  Alex M. and Kingsbury,  Ryan S. and Wen,  Mingjian and Shen,  Jimmy X. and Mathis,  Tyler S. and Kaplan,  Aaron D. and Berket,  Karlo and Riebesell,  Janosh and George,  Janine and Rosen,  Andrew S. and Spotte-Smith,  Evan W. C. and McDermott,  Matthew J. and Cohen,  Orion A. and Dunn,  Alex and Kuner,  Matthew C. and Rignanese,  Gian-Marco and Petretto,  Guido and Waroquiers,  David and Griffin,  Sinead M. and Neaton,  Jeffrey B. and Chrzan,  Daryl C. and Asta,  Mark and Hautier,  Geoffroy and Cholia,  Shreyas and Ceder,  Gerbrand and Ong,  Shyue Ping and Jain,  Anubhav and Persson,  Kristin A.},
  year = {2025},
  month = jul,
  pages = {1522–1532}
}

@misc{kaplan2025matpes,
      title={A Foundational Potential Energy Surface Dataset for Materials}, 
      author={Aaron D. Kaplan and Runze Liu and Ji Qi and Tsz Wai Ko and Bowen Deng and Janosh Riebesell and Gerbrand Ceder and Kristin A. Persson and Shyue Ping Ong},
      year={2025},
      eprint={2503.04070},
      notes={arXiv:2503.04070},
      primaryClass={cond-mat.mtrl-sci},
      url={https://arxiv.org/abs/2503.04070}, 
}

@article{himanen_data-driven_2019,
	title = {Data-{Driven} {Materials} {Science}: {Status}, {Challenges}, and {Perspectives}},
	volume = {6},
	copyright = {© 2019 The Authors. Published by WILEY-VCH Verlag GmbH \& Co. KGaA, Weinheim},
	issn = {2198-3844},
	shorttitle = {Data-{Driven} {Materials} {Science}},
	url = {https://onlinelibrary.wiley.com/doi/abs/10.1002/advs.201900808},
	doi = {10.1002/advs.201900808},
	language = {en},
	number = {21},
	urldate = {2026-05-26},
	journal = {Advanced Science},
	author = {Himanen, Lauri and Geurts, Amber and Foster, Adam Stuart and Rinke, Patrick},
	year = {2019},
	note = {\_eprint: https://advanced.onlinelibrary.wiley.com/doi/pdf/10.1002/advs.201900808},
	pages = {1900808},
}

@article{flores-leonar_materials_2020,
	title = {Materials {Acceleration} {Platforms}: {On} the way to autonomous experimentation},
	volume = {25},
	issn = {2452-2236},
	shorttitle = {Materials {Acceleration} {Platforms}},
	url = {https://www.sciencedirect.com/science/article/pii/S2452223620300596},
	doi = {10.1016/j.cogsc.2020.100370},
	urldate = {2026-05-26},
	journal = {Current Opinion in Green and Sustainable Chemistry},
	author = {Flores-Leonar, Martha M. and Mejía-Mendoza, Luis M. and Aguilar-Granda, Andrés and Sanchez-Lengeling, Benjamin and Tribukait, Hermann and Amador-Bedolla, Carlos and Aspuru-Guzik, Alán},
	month = oct,
	year = {2020},
	pages = {100370},
}

@article{schmidt_recent_2019,
	title = {Recent advances and applications of machine learning in solid-state materials science},
	volume = {5},
	copyright = {2019 The Author(s)},
	issn = {2057-3960},
	url = {https://www.nature.com/articles/s41524-019-0221-0},
	doi = {10.1038/s41524-019-0221-0},
	language = {en},
	number = {1},
	urldate = {2019-08-09},
	journal = {Npj Comput. Mater.},
	author = {Schmidt, Jonathan and Marques, Mário R. G. and Botti, Silvana and Marques, Miguel A. L.},
	month = aug,
	year = {2019},
	pages = {1--36},
}

@article{epifanovsky2021software,
  title={Software for the frontiers of quantum chemistry: An overview of developments in the Q-Chem 5 package},
  author={Epifanovsky, Evgeny and Gilbert, Andrew TB and Feng, Xintian and Lee, Joonho and Mao, Yuezhi and Mardirossian, Narbe and Pokhilko, Pavel and White, Alec F and Coons, Marc P and Dempwolff, Adrian L and others},
  journal={The Journal of chemical physics},
  volume={155},
  number={8},
  pages={084801},
  year={2021},
  doi={10.1063/5.0055522},
  publisher={AIP Publishing}
}

@techreport{hagberg2007exploring,
  title={Exploring network structure, dynamics, and function using NetworkX},
  author={Hagberg, Aric and Swart, Pieter J and Schult, Daniel A},
  year={2007},
  institution={Los Alamos National Laboratory (LANL)}
}

@article{packmol,
author = {Martínez, L. and Andrade, R. and Birgin, E. G. and Martínez, J. M.},
title = {PACKMOL: A package for building initial configurations for molecular dynamics simulations},
journal = {Journal of Computational Chemistry},
volume = {30},
number = {13},
pages = {2157-2164},
doi = {https://doi.org/10.1002/jcc.21224},
url = {https://onlinelibrary.wiley.com/doi/abs/10.1002/jcc.21224},
eprint = {https://onlinelibrary.wiley.com/doi/pdf/10.1002/jcc.21224},
year = {2009}
}

@article{atat,
title = {The alloy theoretic automated toolkit: A user guide},
journal = {Calphad},
volume = {26},
number = {4},
pages = {539-553},
year = {2002},
issn = {0364-5916},
doi = {https://doi.org/10.1016/S0364-5916(02)80006-2},
url = {https://www.sciencedirect.com/science/article/pii/S0364591602800062},
author = {A. {van de Walle} and M. Asta and G. Ceder}
}

@article{zeopp,
title = {Algorithms and tools for high-throughput geometry-based analysis of crystalline porous materials},
journal = {Microporous and Mesoporous Materials},
volume = {149},
number = {1},
pages = {134-141},
year = {2012},
issn = {1387-1811},
doi = {https://doi.org/10.1016/j.micromeso.2011.08.020},
url = {https://www.sciencedirect.com/science/article/pii/S1387181111003738},
author = {Thomas F. Willems and Chris H. Rycroft and Michaeel Kazi and Juan C. Meza and Maciej Haranczyk}
}

@Article{open_comp_matsci,
author={Walsh, Aron},
title={Open computational materials science},
journal={Nature Materials},
year={2024},
month={Jan},
day={01},
volume={23},
number={1},
pages={16-17},
issn={1476-4660},
doi={10.1038/s41563-023-01699-7},
url={https://doi.org/10.1038/s41563-023-01699-7}
}

@misc{plotly,
  author    = {{Plotly Technologies Inc.}},
  title     = {Collaborative data science},
  publisher = {Plotly Technologies Inc.},
  address   = {Montr\'eal, QC},
  year      = {2015},
  url       = {https://plot.ly}
}
    \bibliographystyle{rsc} 
    
    \end{document}


\maketitle

\begin{quote}
\textbf{AI-use disclosure.} Drafts for the author contribution statements were initially contributed by all co-authors. GPT-5.6, as part of M365 Copilot, was then used to unify the author contribution statements, and they were then checked by the first-author team. The other parts of the Supporting Information were drafted by OpenAI Codex using the \texttt{gpt-5.6-sol} large language model with the \texttt{medium} reasoning level. The first-author team subsequently fact-checked and proofread the text and takes responsibility for its content.
\end{quote}

\section{Extended Author contribution statements}\label{extended-author-contribution-statements}

For all authors, additional contribution statements relevant specifically for the software development and validation are provided.

Janine George: Contributed an extensive part of the LOBSTER modules, developed local environment analysis based on LOBSTER analysis, and added functionalities for computing phonon properties, including Grüneisen parameters and thermal displacement parameters. Mentored Master's, PhDs, and postdocs during their pymatgen contributions, including code review; provided bug fixes and general user support within the pymatgen community.

Matthew K. Horton: Served as co-lead maintainer since 2018. Contributed to community engagement, onboarding and mentorship for new developers, pull-request review, quality control, maintenance, releases, continuous integration, as well as making individual code contributions for several features.

Aaron D. Kaplan: Contributed features and fixes to the VASP and CIF I/O packages; contributed features to the special quasi-random structure generation extension; assisted in migrating history of \texttt{pymatgen} after its partition into core and non-core functionality; co-developer of the \texttt{pymatgen-io-validation} add-on.

Seán R. Kavanagh: Contributed bug fixes and performance improvements across VASP I/O, symmetry and structural analysis tools, electronic structure analysis, and thermodynamics, together with defect correction fixes.

Janosh Riebesell: Served as a co-lead maintainer for a few years. Added features and bug fixes across core structure and composition classes, phase diagrams, phonon analysis, and simulation interfaces. Contributed to ongoing maintenance through pull-request review, making releases, codebase modernization, added static typing with CI checks, docs, refactors, expanded test coverage.

Andrew S. Rosen: Developed the interface between ASE Atoms objects and Pymatgen Structure/Molecule objects, fixed various VASP I/O parsing bugs, augmented the parsing of spin-polarized electronic structure properties, added catalysis-oriented density of states descriptors, developed the Chargemol interface, and other various bug fixes and feature additions.

Haoyu Yang: Contributed to pymatgen through improvements to VASP I/O, core functionality, and surfaces, as well as through testing, type annotations, code quality improvements, and code review.

Felix Adams: Contributed small fixes and improvements to serialization in the phase-diagram submodule.

Mahdi Amachraa: Contributed support for parsing VASP BSE optical-transition data.

Oxana Andriuc: Contributed to the development and testing of the adsorption module.

Omar A. Ashour: Conceived and developed pymatgen.io.espresso.

Saurabh Bajaj: Added and edited several property and analytical calculations.

Ella Banyas: Contributed to the development of the \texttt{pymatgen-io-espresso} add-on.

Luis Barroso-Luque: Contributed new features, performance enhancements, and bug fixes to core functionality such as neighbor lists and to analysis tools such as Ewald summation and structure matching.

Karlo Berket: Added an LRU cache to the StructureMatcher.

Noam Bernstein: Extended \texttt{pymatgen.io.vasp.outputs.Wavecar} to read files generated by the gamma-point-only executable.

Arpit Bhardwaj: Contributed updates to pymatgen's VASP surface-calculation input sets, aligning MVLSlabSet with the established MPSurfaceSet parameters from atomate1.

Ali Hussain Umar Bhatti: Improved pymatgen's LOBSTER parsing reliability by fixing the handling of ICOHPLIST.lobster files without trailing newlines, preventing incorrect bond counts and file-consistency errors.

Samuel M. Blau: Contributed to Q-Chem I/O and input sets and molecular graphs. 

Brandon Bocklund: Contributed improvements that added an anharmonic contribution to the quasiharmonic approximation and provided bug fixes for Quantum ESPRESSO support.

Alexander Bonkowski: Contributed to the phonon-analysis module, the LAMMPS I/O interface, visualization tools, and general codebase maintenance.

Danny Broberg: Was the primary author of PyCDT and the original point-defect library in pymatgen.

Guillaume Brunin: Contributed to the LAMMPS I/O interface, particularly to the definition of dedicated input sets and generators.

Kyle Bystrom: Contributed the \texttt{pymatgen.analysis.interface} module.

Charles A. Cardot: Updated the Potentials class in the FEFF module to consider radius.

Rees Chang: Added functionality for constructing approximately cubic supercells from arbitrarily shaped unit cells and for interfacing with the ShengBTE Boltzmann transport equation solver.

Chi Chen: Developed the Cython neighbor-finding optimization and contributed to the LAMMPS interface, \texttt{pymatgen-analysis-diffusion}, and various pymatgen ecosystem packages, including \texttt{MEGNet}, \texttt{M3GNet}, and \texttt{MAML}.

Xin Chen: Initiated and improved the pymatgen interfaces for ADF, NWChem, and Gaussian.

Wei Chen: Contributed to VASP input sets and the point-defect, elasticity, and surfaces modules.

Yuan Chiang: Fixed multiple I/O bugs for VASP and sped up the Bader and Chargemol population-analysis modules.

Kamal Choudhary: Added the JARVIS-Tools exporter and the SLME tool.

Orion A. Cohen: Developed the pymatgen-io-openmm add-on and made miscellaneous contributions to molecular dynamics and molecular graph handling.

Miguel D. Costa: Created a class to parse VASP's \texttt{WAVEDERF} file, which contains the dipole moments between states and is important in the context of dielectric properties.

Stephen T. Dacek III: Designed and implemented the StructureMatcher algorithm.

Maarten De Jong: Implemented the original versions of the elastic-constant and piezoelectric-computation modules in pymatgen and contributed to later versions.

Tom Demeyere: Refactored the LOBSTER I/O modules, including tests, and added type hints to specific modules.

Alan K. Dozier: Contributed to the initial development of code for the FEFF module that enables the efficient computation of X-ray absorption spectra (XAS) and electron energy-loss spectra (EELS) using FEFF.

Alexander Dunn: Improved Fermi-surface plotting, tests for bond-valence-sum methods, and the loading of pymatgen-relevant environment variables.

Shyam Dwaraknath: Contributed to several capabilities in pymatgen and its add-ons, including general analysis, I/O, molecules, defects, surfaces, interfaces, database interoperability, and integration into high-throughput workflows.

Maxwell T. Dylla: Contributed code and tests to the electronic-structure packages.

Kristjan Eimre: Contributed improvements and fixes to pymatgen's Gaussian I/O functionality.

Alexander R. Epstein: Contributed to the thermochemical analysis of quantum-chemistry data and interactions with the CREST software.

Christina Ertural: Implemented the NciCobiList class for parsing LOBSTER's multicenter ICOBI output files, including spin-polarized support and test coverage. Added the .to\_file() export alias for structures and molecules and improved the performance of the step-size search in CubicSupercellTransformation.

Marco Esters: Wrote the first iteration of the LOBSTER module.

Matthew L. Evans: Contributed maintenance, testing, continuous integration, and support for OPTIMADE.

Johannes Falke: Reported multiple issues and provided code fixes related to symmetry operations. Also wrote other libraries in the ecosystem that use pymatgen.

Adam Fekete: Improved MoleculeMatcher functionality for identifying geometrically equivalent molecular structures as part of pymatgen's molecular-analysis package.

Bernard Field: Extended VASP I/O capabilities to include reading data for VASP's new \texttt{KPOINTS\_OPT} feature, which is useful for band-structure calculations.

Alexandr Fonari: Reported bugs and issues discovered through the use of pymatgen in the Schrödinger Materials Science Suite and, where possible, provided fixes for review. Examples included a surface-generation bug fix, an improvement allowing phase diagrams to have an arbitrary number of terminal groups, and a correction to the rotational symmetry number for the D*h point group.

Nathan C. Frey: Contributed code for calculating and analyzing magnetic and topological order.

Dale W. Gaines II: Fixed errors in the symmetry points of Brillouin zones used for k-path generation.

Alex M. Ganose: Contributed to input-set design and VASP implementation, as well as bonding- and dimensionality-analysis methods (key contributions).

Matteo Giantomassi: Contributed to the development and maintenance of pymatgen's ABINIT integration, including tools for generating inputs, parsing outputs, and supporting ABINIT-related workflows.

Rhys E. A. Goodall: Patched phase diagrams, Wyckoff-position and prototype analysis, and contributed bug fixes and optimizations, including improvements to composition hash functions.

Petre-Flaviu Gostin: Contributed \texttt{XRDCalculator} bug fixes, annotation enhancements for diffraction plots, and maintenance updates.

Aayush Gupta: Fixed bugs related to use in upstream packages where pymatgen would otherwise raise errors.

Viet-Anh Ha: Contributed functionality for plotting projected bands on specific atoms, orbitals, and subshells defined by the azimuthal quantum number \emph{l} or magnetic quantum number \emph{m}.

Geoffroy Hautier: Mainly mentored students and postdoctoral researchers who contributed to pymatgen in recent years (including contributions to phonon and LOBSTER modules, among others).

Niclas Heinsdorf: Contributed methods for classifying bonds into symmetry-equivalent groups and explicitly identifying the symmetry operations that connect them.

Michael F. Herbst: Contributed bug fixes to the ABINIT output parser and band-structure plotter.

Tingzheng Hou: Contributed to the development and enhancement of pymatgen's LAMMPS data I/O functionality, including the implementation and optimization of the CombinedData framework.

Patrick Huck: Served as the Materials Project Technical Platform Architect and Infrastructure Lead.

Yuji Ikeda: Contributed an improvement to the VASP POTCAR parser.

Adam J. Jackson: Contributed various bug reports and a fix for loading compressed VASP band structures.

Saivenkataraman Jayaraman: Implemented the Pourbaix App in its original version.

Finn R. Kearns: Contributed a bug fix and unit test for VASP I/O.

Sunghyun Kim: Contributed a bug fix to the VASP I/O module, correcting the scientific-notation formatting of zero values when writing volumetric-data files such as CHGCAR and LOCPOT.

Ryan S. Kingsbury: Authored, tested, and debugged the initial MPScanRelaxSet for the SCAN/r2SCAN workflow; authored, tested, and debugged the MaterialsProjectDFTMixingScheme; refactored the Compatibility classes with a new abstract interface; supervised and tested the refitting of the GGA energy corrections for MP2020Compatibility; added support for advanced implicit-solvation models to Q-Chem I/O; contributed various bug fixes to the Ion class; and updated periodic-table data for ionic and van der Waals radii.

Adam M. Krajewski: Contributed new features and bug fixes to pymatgen's \texttt{core.Composition}, as well as community contributions.

Yu Kumagai: Contributed to identifying and fixing bugs and developing associated unit tests.

Matthew C. Kuner: Created the \texttt{pymatgen-io-validation} add-on package.

William T. Laderer: Contributed multiple output-file parsing improvements and bug fixes for the VASP and Critic2 modules.

Xiang-Guo Li: Contributed code development to \texttt{pymatgen.core} and to grain-boundary generation.

Weitang Li: Enhanced the \texttt{CovalentBond} class.

Yu Hsuan Liang: Contributed to the development and optimization of the TEM diffraction calculator.

Hanyu Liu: Developed and maintained the \texttt{pymatgen.io.pwmat} module, enabling pymatgen to parse \texttt{PWmat} input and output files and supporting compatibility with the latest \texttt{PWmat} versions.

Ke Liu: Contributed nine pull requests to pymatgen, primarily focused on BoltzTraP integration fixes and logging improvements in four pull requests, Python 3 compatibility patches in two pull requests, and fixes for Matplotlib deprecation warnings and CLI-tool bugs. These were primarily maintenance-level contributions addressing real-world usage pain points.

Miao Liu: Performed testing and bug resolution.

Ahmad Lutfi: Contributed speed and stability improvements to \texttt{ChemEnv} and coordination-environment functionality.

Olivier Marchand: Optimized the VolumetricData.from\_cube() implementation in pymatgen by redesigning file I/O and vectorizing volumetric-data parsing with NumPy, reducing .cube file parsing times by approximately 55--65\%. Validated output equivalence across benchmark datasets and contributed the implementation, documentation, benchmarking, and testing.

Matthew J. McDermott: Created the \texttt{chempot\_distance} module for chemical-potential diagrams and second-generation phase-diagram plotting. Contributed \texttt{GibbsComputedStructureEntry} and other entry updates, an updated reaction calculator and solving algorithm, and numerous bug fixes.

Bharat K. Medasani: Developed the initial defects module in pymatgen, followed by PyCDT.

Fanchen Meng: Implemented support for applying multiple pseudopotentials to the same element type, enabling more flexible and accurate treatment of chemically distinct environments.

Kyle D. Miller: Implemented functionality for modulating crystal-lattice distortion modes.

Henrique Miranda: Contributed the infrared-spectra module and phononwebsite export.

Rohith Srinivaas Mohanakrishnan: Worked on isomorphic comparison of MolecularGraph objects in pymatgen.

Joseph H. Montoya: Wrote the elasticity module and the more general tensor module, the adsorbate-site finder, and the current algorithm for multielement Pourbaix diagrams.

Guy C. Moore: Fixed a bug affecting the handling of noncollinear magnetic moments across different Python type initializations, improving the writing of MAGMOM in VASP INCAR input sets as well as the parsing of noncollinear magnetic output from VASP calculations.

Benjamin J. Morgan: Contributed to both \texttt{pymatgen.core} and the \texttt{pymatgen-analysis-diffusion} add-on through bug fixes to symmetry analysis, diffusion analysis, VASP output parsing, and \texttt{RadialDistributionFunction}, as well as API extensions to \texttt{PointGroupAnalyzer} and \texttt{ProbabilityDensityAnalysis}.

Jason M. Munro: As a postdoctoral researcher and subsequently a Materials Project staff member, wrote the mp-api client and drove its integration into pymatgen.ext.matproj, replacing the legacy MPRester. Rebuilt pymatgen's high-symmetry k-path infrastructure by adding the \texttt{KPath} hierarchy with three selectable conventions and continuous band-structure paths, and provided design input for the pymatgen-io-validation package.

Koki Muraoka: Enhanced lattice processing to support GULP and VESTA structural formats.

Aakash A. Naik: Contributed mainly to updating and developing LOBSTER-related modules in pymatgen and LobsterPy.

Willis O'Leary: Eliminated side effects related to ASE Atoms interoperability and contributed to the \texttt{pymatgen-io-espresso} add-on.

Anthony Onwuli: Fixed MPRester utilities, added a convenience route for retrieving phonon band structures, extended the Element class to enable users to check for radioactive elements and screen for radioactive Composition objects, and updated tests for the \texttt{io/zeopp} extension.

Jaesuk Park: Added a custom-precision-format feature, fixed critical parsing and conversion bugs, and added unit tests to restore and enhance pymatgen's interoperability with Quantum ESPRESSO and ASE.

Drew F. Parsons: Developed pymatgen packaging for Debian and Ubuntu Linux distributions and contributed patches that improved portability and testing, supporting the robustness of the software on diverse CPU architectures.

Kai Pei: Contributed to the conversion-battery module and the GGA(+U)/r2SCAN phase-diagram mixing scheme, resolving issues in electrochemical and thermodynamic analysis for solid-state batteries.

Guido Petretto: Made major contributions to phonon functionality and ABINIT I/O, as well as other minor contributions and bug fixes.

Thomas A. R. Purcell: Wrote and maintained the FHI-aims interface that was separated into the pymatgen-io-aims plugin.

Ji Qi: Contributed to MatPES-related functionality and constructed code for regional ionic-diffusion analysis.

Xiaohui Qu: Contributed to Q-Chem I/O and MoleculeMatcher.

Ihor Radchenko: Improved integration with ASE, contributed a more robust INCAR tag dictionary, reported multiple bugs, and developed the \texttt{IMDgroup-pymatgen} add-on.

Nav Nidhi Rajput: The Rajput group contributed a Custodian error handler for Gaussian-based DFT calculations of molecular systems, enabling automated error detection, correction, and job resubmission for high-throughput quantum-chemistry workflows within the pymatgen ecosystem.

Francesco Ricci: Contributed to the \texttt{BoltzTraP} and \texttt{electronic\_structure} modules and provided various bug fixes.

Benjamin S. Rich: Served as the lead developer for the construction and maintenance of the JDFTx I/O module.

William D. Richards: Was part of the initial team and contributed to structure transformations, structure matching, phase diagrams, diffusion analysis, the VASP interface, and other functionality.

Gian-Marco Rignanese: Provided scientific direction for the ABINIT interface with the PseudoDojo table handling, the OptimadeRester, the ChemEnv coordination environment identification, the phonon modules, the BoltzTraP2 transport interface, and the LOBSTER/COHP bonding analysis. Mentored PhDs and postdocs during their pymatgen contributions.

Xiaoxu Ruan: Contributed to the initial discussions, implementation, and testing of MatPESStaticSet in pymatgen.

Germain Salvato Vallverdu: Contributed to the Gaussian parser module and density-of-states analyses.

Jonathan Schmidt: Contributed a bug fix and a new test for the Chargemol caller, as well as a bug fix for magnetic-moment handling in phonon workflows.

Jimmy-Xuan Shen: Created pymatgen-analysis-defects and mp-pyrho and built the current iteration of the insertion-electrodes stack in pymatgen.

Ryuichi Shimogawa: Fixed the sorting in the FEFF I/O module used to create ATOMS input.

Kohei Shinohara: Contributed performance improvements to pymatgen's crystal-structure matching and prototype-identification functionality by eliminating redundant structure preprocessing.

Arunima K. Singh: Led the development of the Pourbaix decomposition-energy formalism for assessing the stability, passivation, and corrosion of metastable materials.

Eric Sivonxay: Contributed the Trajectory class, the MPMDSet VASP input set for AIMD simulations, and the Interface and InterfaceBuilder classes.

Stef Smeets: Improved structure-metadata handling and CIF compatibility.

Tess E. Smidt: Wrote code and tests from 2016 to 2019 for classes, methods, and functions in pymatgen related to polarization calculations for ferroelectric and other polar materials. Primary contributions included the implementation and testing of the \texttt{pymatgen.analysis.ferroelectricity.polarization} module and associated supporting files.

Andrei Sobolev: Co-authored the pymatgen FHI-aims plugin, pymatgen-io-aims, and contributed FHI-aims-specific code to pymatgen.

Evan Walter Clark Spotte-Smith: Contributed to input/output functionality for the Q-Chem electronic structure code, the initial implementation of molecular graph utilities, and some other minor contributions, for example, to local environment and bond detection algorithms.

Christopher Stihl: Added code through pull requests, most notably fixing a memory leak in pbc\_shortest\_vectors, implementing structure modification by rotating sites, and parsing plane-wave-basis-related quantities from VASP output.

Wenhao Sun: Developed algorithms for \texttt{pymatgen.core.surface} and the Pourbaix applet and developed visualization functionality for three-dimensional convex hulls in \texttt{pymatgen.analysis.phase\_diagram}.

Jack D. Sundberg: Contributed to pymatgen through small software enhancements and patches to core modules, including modules for composition, symmetry, and calculation input sets.

Yuta Suzuki: Developed and validated neutron-diffraction functionality in pymatgen's diffraction module.

Leopold Talirz: Improved the speed of nearest-neighbor routines.

Hanmei Tang: Developed and implemented the NEB-related analysis library in pymatgen-analysis-diffusion.

Eric Taw: Contributed a small bug fix for reading \texttt{vasprun.xml} when VASP output encountered a floating-point overflow.

Cooper Tezak: Developed and tested JDFTx parsers and schema.

Félix Therrien: Created a pull request to improve the retrieval of nearest-neighbor information.

Scott D. Thiel: Contributed the \texttt{pymatgen.io.res} module and its test coverage.

Steven B. Torrisi: Contributed improvements and bug fixes to parsers.

Mark E. Turiansky: Contributed the Wavecar reader implementation, the Wannier90 I/O module, and minor performance improvements, and maintained the Nonrad code that depends on pymatgen.

Katharina Ueltzen: Contributed code and tests on plotting, I/O functionality, bonding analysis, and symmetry.

Michiel J. van Setten: Contributed to the ABINIT interface and implementations required for GW workflows.

Josua Vieten: Improved the discretization of occupancies for disordered structures. 

Sudarshan Vijay: Contributed to parsing and analyzing Hamiltonian and overlap matrices and orbital coefficients from the Q-Chem electronic-structure program.

Nicholas Wagner: Developed the Global Instability Index (GII) calculator for matminer, a materials data mining library within the pymatgen ecosystem, as a structure featurizer in matminer.featurizers.structure.

Frank Wan: Developed the initial implementation of pymatgen's TEM diffraction calculator.

Zhenbin Wang: Added support for VASP optical-property calculations, including dielectric-function parsing, BSE input sets, and absorption-coefficient evaluation.

Jingyang Wang: Added lattice features and file-reading functionality in the Trajectory class.

Amanda Wang: Added uncertainty quantification for energy corrections used to improve formation-energy and phase-stability calculations according to experimental data.

Logan Ward: Implemented early versions of algorithms for analyzing coordination environments.

Thomas Warford: Contributed improvements to surface generation, structure matching, VASP I/O, and Chargemol (DDEC6) handling.

David Waroquiers: Developed and applied pymatgen's ChemEnv tools for identifying ideal and distorted coordination environments, supervised contributions to the LAMMPS I/O interface, and made several additional fixes and enhancements.

Mingjian Wen: Extended the \texttt{Trajectory} class and fixed various bugs.

Oskar Weser: Improved the generation of symmetry groups from a minimal set of generators.

Nicholas Winner: Contributed to the original development of CP2K support and its integration into pymatgen.

Donald Winston: Helped ensure that programmatic interfaces to pymatgen functionality continued to require limited, if any, adaptation or translation for web-user interfacing through materialsproject.org functionality for applying and interpreting pymatgen operations on web-accessible data.

Michael Wolloch: Fixed bugs and contributed information regarding VASP functionality.

Brandon M. Wood: Contributed to Q-Chem I/O in pymatgen.

Rachel Woods-Robinson: Co-developed the pymatgen-alloys package, developed XRD analysis tools, and contributed testing and various bug fixes.

Yihan Xiao: Contributed to Python classes related to interfacial reactivity.

Yaoshu Xie: Contributed modifications to the interface module.

Firat Yalcin: Contributed fixes and enhancements to core structure and surface routines, focusing on primitive-cell reduction, Miller-index generation, and slab generation.

Ruo Xi Yang: Contributed bug fixes and a new optical-absorption-coefficient analysis module.

Weike Ye: Contributed term-symbol properties to the Element class and a correction to the lanthanide MAGMOM defaults in the VASP input sets.

Pezhman Zarabadi-Poor: Fixed the LOBSTER parser for bandOverlaps.

Jinzhe Zeng: Set up the Python wheel build workflow and fixed several bugs.

Hongsheng Zhao: Contributed bug fixes and feature enhancements.

Kaifeng Zheng: Identified and resolved multiple critical, long-standing issues in the pymatgen XAFS module's FEFF I/O layer, substantially enhancing its reliability, accuracy, and interoperability with external codes.

Zhuoying Zhu: Contributed AIMD- and NEB-related code, unit tests, and Jupyter Notebook tutorials to pymatgen-analysis-diffusion and pymatgen.

He Zhu: Enriched the visualization options for band-structure data.

Nils E. R. Zimmermann: Developed and implemented analysis tools for neighbor searching, coordination geometry, structure comparison, and charged-defect analysis.

Daniel Zügner: Fixed an issue in which an element could occur multiple times in a structure's chemical system and added performance improvements that sped up the creation of many Structure objects by a factor of five.

Kristin Persson: As the Director of Materials Project, made the decision to formally adopt and support pymatgen as its main I/O software. As part of her stewardship of Materials Project, she has provided the funding and mentorship for students, postdocs and staff since 2011 to use, develop and maintain pymatgen. This includes code contributions, supporting the user community, bug fixes, mediating code conflicts, and providing new scientific directions.

Anubhav Jain: Contributed code and design to the battery electrode, computed entry and compatibility correction, and reaction balancing modules. Authored the battery analyzer, cost analysis and HHI resource criticality modules, indeterminate formula parsing and oxidation state guessing in Composition, the CrystalNN near-neighbor algorithm, and the combined band structure/DOS plotter. Led a major refactoring of the BoltzTraP interface and contributed extensively to the VASP input sets and output parsing. Developed the StructureNL format for structures with provenance metadata. Mentored students and postdocs during their pymatgen contributions, and provided code review, bug fixes and user support.

Shyue Ping Ong: Conceptualization, software architecture, and project leadership. Initiated pymatgen in 2011 and has served as lead maintainer since inception, with contributions spanning all core modules (core, io, analysis, electronic\_structure, phase\_diagram, entries). Supervision, funding acquisition, and writing.

\section{Scope}\label{scope}

This Supporting Information describes how the data shown in the main-text figures were collected and processed.

Analysis source and cached intermediate data are available at \url{https://github.com/janosh/pymatgen-2-paper}. GitHub, OpenAlex, and the pymatgen repository continue to change; the archived data associated with the article define the snapshot used for the figures.

\section{S1. OpenAlex citation topics (main-text Fig. 1)}\label{s1.-openalex-citation-topics-main-text-fig.-1}

The original pymatgen article was identified by OpenAlex work ID \texttt{W2015197254}. Works citing this article were queried through the OpenAlex Works API and grouped by their OpenAlex primary topic. Up to 50 topic groups were retrieved, and groups containing fewer than 10 citing works were excluded. The OpenAlex data were retrieved around 5 July 2025. OpenAlex primary topics are algorithmically assigned; the subsequent language-model grouping therefore formed a second classification layer.

The retained topic names and OpenAlex counts were supplied to OpenAI's GPT-5 model (\texttt{gpt-5}), which was prompted to organize them into five thematic branches with concise labels. The branches were ordered by their summed subtopic counts, and the four highest-count subtopics in each branch were retained. The labels and counts returned by the model were cached as structured data; these cached values are preserved in the revised figure. The 20 displayed subtopics sum to 2,699 citing-work assignments. This is a selected thematic summary rather than an exhaustive partition of citations. Branch counts sum the displayed children, and node colours and legend ticks share a logarithmic Viridis scale from 1 to the largest branch count, 1,253.

\section{S2. Shells of influence (main-text Fig. 2)}\label{s2.-shells-of-influence-main-text-fig.-2}

The repository-level values were collected from the \href{https://github.com/materialsproject/pymatgen}{\texttt{materialsproject/pymatgen}} GitHub repository in August 2026. The contributor total of 276 counts distinct authors of non-bot PRs merged through 31 December 2025, the same population as main-text Fig. 6. The GitHub dependency graph supplied counts of 2,713 dependent repositories and 626 dependent packages. These values refer to the original pymatgen repository and package rather than \texttt{pymatgen-core}.

Merged PRs were retrieved through the GitHub REST API and counted by calendar year of merge, rather than creation year. Bot-authored PRs were identified by GitHub usernames ending in \texttt{{[}bot{]}}, matched case-insensitively. Annual counts with and without bot-authored PRs are listed below.

{\def\LTcaptype{none} 
\begin{longtable}[]{@{}lrr@{}}
\toprule\noalign{}
Year & Merged PRs excluding bots & Merged PRs including bots \\
\midrule\noalign{}
\endhead
\bottomrule\noalign{}
\endlastfoot
2013 (incomplete) & 40 & 40 \\
2014 & 83 & 83 \\
2015 & 52 & 52 \\
2016 & 142 & 142 \\
2017 & 292 & 292 \\
2018 & 264 & 264 \\
2019 & 167 & 228 \\
2020 & 140 & 141 \\
2021 & 143 & 143 \\
2022 & 247 & 268 \\
2023 & 399 & 423 \\
2024 & 358 & 373 \\
2025 & 167 & 186 \\
\textbf{Total, 2013--2025} & \textbf{2,494} & \textbf{2,635} \\
\textbf{Annual average, 2014--2025} & \textbf{204.5} & \textbf{216.25} \\
\end{longtable}
}

The first recorded PR merge occurred on 29 April 2013; 2013 is marked as an incomplete year of PR activity and excluded from the annual average. Over the 12 calendar years from 2014 through 2025, 2,454 non-bot PRs were merged, giving an average of 204.5 merged PRs per year.

PyPI download counts were obtained from the public PyPI download data in Google BigQuery. Counts from August 2018 through December 2025 summed to 23,239,954 and were displayed as approximately 23.2 million. August 2018 was used as the starting month because the PyPI download pipeline significantly under-reported totals before 26 July 2018, as described in the \href{https://packaging.python.org/en/latest/guides/analyzing-pypi-package-downloads/\#caveats}{Python Packaging User Guide}.

The Materials Project values came from a \href{https://newscenter.lbl.gov/2026/01/13/accelerating-discovery-how-the-materials-project-is-helping-to-usher-in-the-ai-revolution-for-materials-science/}{Berkeley Lab article published on 13 January 2026}, which reported approximately 5,000 uses per day and more than 650,000 registered users. The daily use rate was converted to approximately 1.8 million uses per year.

\section{S3. Upstream and downstream package usage (main-text Fig. 3)}\label{s3.-upstream-and-downstream-package-usage-main-text-fig.-3}

Both panels were derived by statically analyzing Python files and code cells in Jupyter notebooks. Calls to APIs imported from the package were counted and resolved through their import aliases. Dynamically constructed calls were not counted. Link widths represent the resulting function call counts.

\subsection{S3.1 APIs of dependencies used by pymatgen (Fig. 3a)}\label{s3.1-apis-of-dependencies-used-by-pymatgen-fig.-3a}

Required dependencies were read from pymatgen's \texttt{pyproject.toml} file. Source code at pymatgen commit \texttt{\seqsplit{ab34799d8ab5dee80756489cf2ca28a97de78121}} was analyzed. Calls to each dependency were counted in the \texttt{alchemy}, \texttt{analysis}, \texttt{core}, \texttt{electronic\_structure}, \texttt{entries}, \texttt{io}, \texttt{phonon}, \texttt{symmetry}, and \texttt{transformations} subpackages. Dependencies and subpackages with fewer than 15 total calls were omitted from the figure.

\subsection{S3.2 pymatgen APIs used by downstream repositories (Fig. 3b)}\label{s3.2-pymatgen-apis-used-by-downstream-repositories-fig.-3b}

Repository names were collected from GitHub's pymatgen dependents page, and star counts were obtained through the GitHub API. The resulting table contained 1,446 repositories. Repositories were taken in descending order of stars until 12 were retained, spanning the 28 highest-starred repositories (each with at least 127 stars in the archived table). Of the 16 excluded, 8 had no pymatgen usage outside tests or contained documentation, course or Materials Project internal material, and 8 had fewer than 10 pymatgen calls after the filtering described below (e.g.\ 8 of 17 calls retained for \texttt{megnet}).

Each retained repository was shallow-cloned from its default branch on 26 September 2026 at the commits listed below.

{\def\LTcaptype{none} 
\begin{longtable}[]{@{}
  >{\raggedright\arraybackslash}p{(\linewidth - 2\tabcolsep) * \real{0.5000}}
  >{\raggedright\arraybackslash}p{(\linewidth - 2\tabcolsep) * \real{0.5000}}@{}}
\toprule\noalign{}
\begin{minipage}[b]{\linewidth}\raggedright
Repository
\end{minipage} & \begin{minipage}[b]{\linewidth}\raggedright
Commit
\end{minipage} \\
\midrule\noalign{}
\endhead
\bottomrule\noalign{}
\endlastfoot
\texttt{materials\_discovery} & \texttt{\seqsplit{29cf9852b7e11327fb3eccff6620e600add65169}} \\
\texttt{AIRS} & \texttt{\seqsplit{4a16c68a7da707c521019067dec51c227c10de45}} \\
\texttt{matgl} & \texttt{\seqsplit{d0b6bf03cd6bc93c1e565db7516bb22f6b0977cb}} \\
\texttt{PaddleScience} & \texttt{\seqsplit{d57561b139df953a77a2e134b7646048e445e731}} \\
\texttt{PyXtal} & \texttt{\seqsplit{65d4872dbf39bb40a8fbfdb0aaa468e758b49799}} \\
\texttt{pymatviz} & \texttt{\seqsplit{873472b4a91af84b8095916d50c163d93f8d1513}} \\
\texttt{all-atom-diffusion-transformer} & \texttt{\seqsplit{b9ce505f170597a7c8ca50d13ce8e15df21cf8c9}} \\
\texttt{quacc} & \texttt{\seqsplit{2ee58ec742e5a3cccd26a748b616457b15cf3e84}} \\
\texttt{doped} & \texttt{\seqsplit{9f08c3e18e7326b4973402e41bdc97cfdd11b31e}} \\
\texttt{matbench-discovery} & \texttt{\seqsplit{71633e8bdfdfd41d56d64b1d777e5686d9eda3ec}} \\
\texttt{matsciml} & \texttt{\seqsplit{a475867cf56fec1a6c1d153b61ff84bcc38dfcc5}} \\
\texttt{CrystaLLM} & \texttt{\seqsplit{8e81a86f9e064f219f09312f29ecbfc867211784}} \\
\end{longtable}
}

Python files and notebook code were analyzed after excluding \texttt{tests}, \texttt{docs}, \texttt{.github}, and hidden paths. Detected calls were grouped by the first pymatgen subpackage in the import path. Detected deprecated paths beginning with \texttt{pymatgen.Structure}, \texttt{pymatgen.Molecule}, \texttt{pymatgen.MPRester}, \texttt{pymatgen.Composition}, or \texttt{pymatgen.Element} were discarded. Calls assigned to \texttt{pymatgen.phonon} (5 calls) and \texttt{pymatgen.command\_line} (2 calls) were also discarded because of very few detected calls. The 12 retained repositories account for 914 drawn calls.

\section{S4. Repository development by topic and package (main-text Fig. 4)}\label{s4.-repository-development-by-topic-and-package-main-text-fig.-4}

\subsection{S4.1 Pull-request topics over time (Fig. 4a)}\label{s4.1-pull-request-topics-over-time-fig.-4a}

Pull requests (PRs) in all states were collected from \texttt{materialsproject/pymatgen} through the paginated GitHub REST API on 8 September 2026. The figure includes all 2,494 non-bot PRs merged before 1 January 2026, including 167 merged during the complete 2025 calendar year. Accounts identified as bots by GitHub or whose usernames end in \texttt{[bot]} (case-insensitively) were excluded. Each PR number is retained once, including repeated titles, and years refer to merger dates.

Each PR title was classified directly using ordered, case-insensitive regular-expression rules. The first matching theme was assigned in this order: documentation; testing and code quality; performance; bug fixes and refactoring; I/O and parsing; structural and analysis functionality; and Other / unclassified for unmatched titles. Every PR is assigned exactly once, with no title deduplication or counts inferred from LLM-generated summaries. These are approximate title-based classifications rather than manually validated descriptions of PR contents. The archived data retain each PR's number, title, author, dates, and assigned theme for inspection.

\subsection{S4.2 Package-level commit activity (Fig. 4b)}\label{s4.2-package-level-commit-activity-fig.-4b}

Non-merge commits were collected from the Git history of \texttt{materialsproject/pymatgen}, beginning at the initial commit \texttt{\seqsplit{fa7f41d8bd769a04cca1f78242ebf072664c871d}} on 23 October 2011 and ending on 31 December 2025. The figure excludes commits before 1 January 2012 and therefore shows activity from 2012 through 2025. Commits were counted for 13 selected paths: \texttt{command\_line}, \texttt{symmetry}, \texttt{alchemy}, \texttt{core/structure.py}, \texttt{core/periodic\_table.py}, \texttt{io/vasp}, \texttt{transformations}, \texttt{analysis}, \texttt{optimization}, \texttt{electronic\_structure}, \texttt{phonon}, \texttt{vis}, and \texttt{entries}. Both \texttt{pymatgen/} and \texttt{src/pymatgen/} paths were matched over the full period (src layout adopted on 26 June 2024).

Commits touching each path were counted by month and, from January 2012 onward, aggregated into calendar half-years (January--June and July--December). A commit touching more than one selected path was counted once for each corresponding path.

\section{S5. Pull requests by contributor tenure (main-text Fig. 5)}\label{s5.-pull-requests-by-contributor-tenure-main-text-fig.-5}

The tenure figure uses the same 2,494 non-bot merged PRs and merger-year grouping as Fig. 4a. PRs in all states were retrieved so that each author's first submission date includes unmerged PRs. Tenure is the elapsed time from that first submission to the merger date of each included PR, rather than its creation date.

PRs were assigned to five disjoint tenure groups: less than 7 days; 7 days to less than 1 year; 1 to less than 3 years; 3 to less than 6 years; and at least 6 years. One year is defined as 365.25 days. All merges through 31 December 2025 are included. Only PR activity is represented; reviews, comments, and other contributions are not counted.

\section{S6. Geographic distribution of pull-request contributions (main-text Fig. 6)}\label{s6.-geographic-distribution-of-pull-request-contributions-main-text-fig.-6}

The GitHub API was used to count merged PRs by author and retrieve each author's public, self-reported profile location. Accounts whose GitHub username ended in \texttt{{[}bot{]}}, matched case-insensitively, were excluded. Free-text locations were geocoded to countries with OpenStreetMap Nominatim. For unresolved contributors listed on the public pymatgen team page, the listed affiliations were mapped to countries using OpenAI's GPT-4o model. Additional mappings were manually curated. Contributors without a country mapping were omitted from the country totals. The PR counts and profile locations were retrieved around 16 February 2026.

Of 276 contributors, 124 (44.9\%) provided no location information in their GitHub profile. All 152 contributors who provided location information were assigned a country, including through manually curated geocoding results. Institutional affiliations listed on the public pymatgen team page, together with manually curated mappings, supplied countries for 57 of the 124 contributors without a profile location. The final country mapping covered 209 contributors (75.7\%) and 2,362 of 2,498 merged PRs (94.6\%). The remaining 67 contributors (24.3\%), accounting for 136 merged PRs (5.4\%), had no country mapping.

PR counts were summed by mapped country. The resulting values are numbers of merged PRs, not numbers of unique contributors. A contributor's mapped country was based on the public profile location or listed affiliation used during data collection.

\section{S7. Source-code size and test coverage (main-text Fig. 7)}\label{s7.-source-code-size-and-test-coverage-main-text-fig.-7}

Both treemap areas and coverage percentages were derived from the coverage.py 7.6.12 \href{https://github.com/user-attachments/files/21545087/2025-07-31-pymatgen-coverage.json}{JSON snapshot} generated on 31 July 2025 with branch coverage disabled. Areas count executable statements in reported Python files under \texttt{src/pymatgen/}; files with no executable statements are omitted, while files with zero covered statements are retained. Colours represent summed covered statements divided by summed executable statements, including for parent packages. The report records 60,242 covered statements out of 76,389 executable statements (78.9\%). Coverage is 90.2\% for \texttt{core}, 85.8\% for \texttt{entries}, 83.6\% for \texttt{io}, and 13.9\% for \texttt{vis}. Cells are grouped to at most three path components; groups below 1,500 statements are combined into other cells, and subpackages below 5,000 statements are shown as single cells. No separately installed pymatgen source tree is used for sizing; the figure describes this dated test-suite snapshot.

\section{S8. Contributor and commit activity (main-text Fig. 8)}\label{s8.-contributor-and-commit-activity-main-text-fig.-8}

Non-merge commit metadata were collected from the Git history of \texttt{materialsproject/pymatgen} from 1 January 2012 through 31 December 2025. Merge commits were excluded. Authors whose names ended in \texttt{{[}bot{]}}, matched case-insensitively, were also excluded.

Author records with an identical name or lower-cased email address were assigned a common contributor identifier. Non-merge commit counts were aggregated by calendar year according to the Git author date. A contributor with at least one non-merge commit in a given year was counted once for that year. The solid line reports the number of active contributors, and the dotted line reports the total number of non-merge commits in the same year. Only activity recorded in the Git commit history was included.